\documentclass[]{SciPost}

\def\sfrac#1#2{{\textstyle{#1\over #2}}}
\newcommand{\be}{\begin{equation}}
\newcommand{\ee}{\end{equation}}
\newcommand{\ba}{\begin{array}}
\newcommand{\ea}{\end{array}}
\newcommand{\bea}{\begin{eqnarray}}
\newcommand{\eea}{\end{eqnarray}}
\newcommand{\sss}{\scriptscriptstyle}
\newcommand{\nn}{\nonumber}
\usepackage{amsmath,amssymb,wrapfig}

\begin{document}

\begin{center}{\Large \textbf{Dark Atoms and Composite Dark Matter
}}\end{center}

\begin{center}
James M.\ Cline
\end{center}

\begin{center}
McGill University, Dept.\ of Physics, Montr\'eal, Qu\'ebec, Canada
\\
jcline@physics.mcgill.ca
\end{center}

\begin{center}
\end{center}


\section*{Abstract}
{
I selectively review the theoretical properties and observational 
limits
pertaining to dark atoms, as well as
composite dark matter candidates bound by a confining gauge 
interaction: dark glueballs, glueballinos, mesons and baryons.
  Emphasis is given to cosmological, direct and indirect
detection constraints.  Lectures given at Les Houches Summer School
2021: Dark Matter.}

\bigskip
\bigskip
 \centerline{\includegraphics[width=\linewidth]{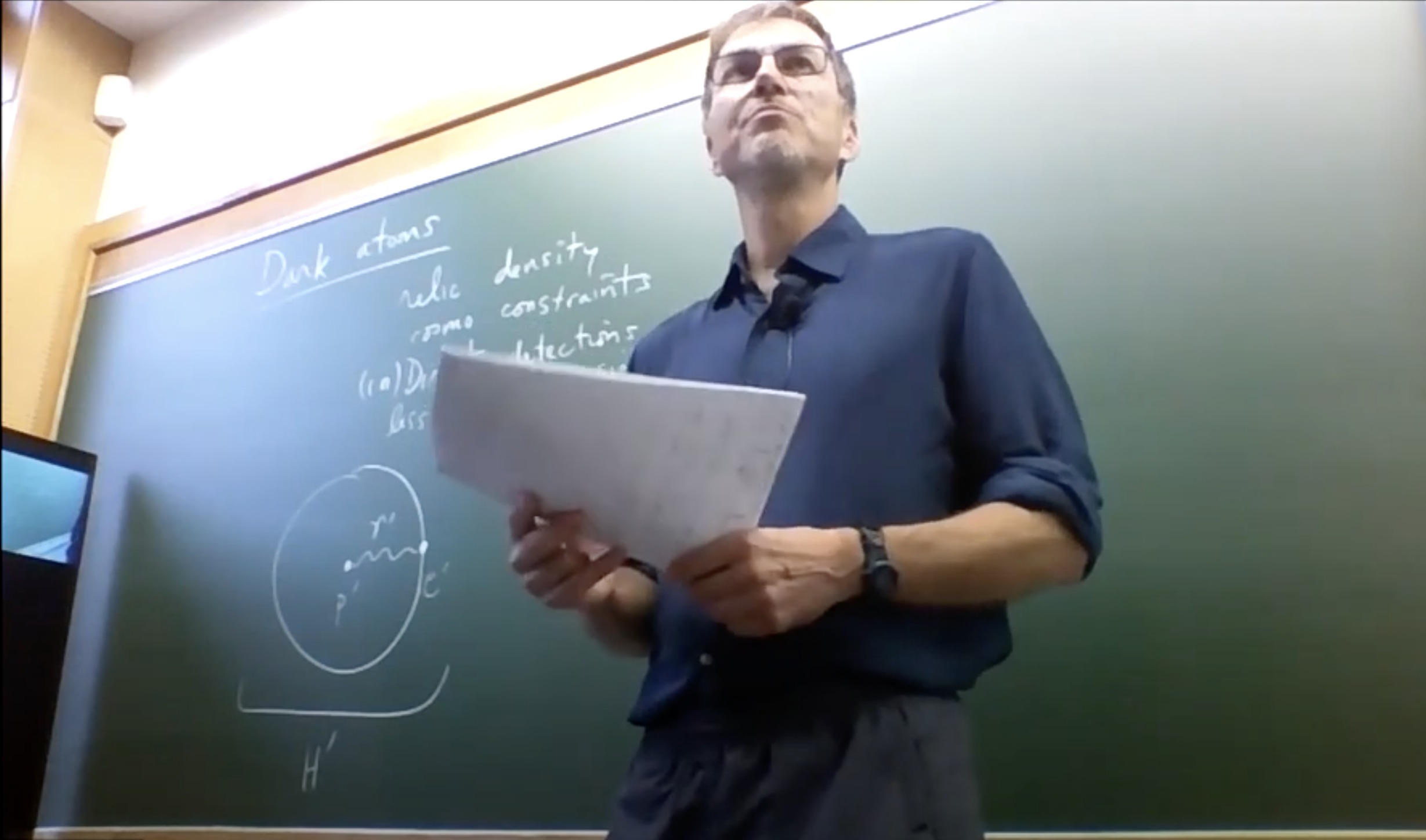}}

\vspace{10pt}
\noindent\rule{\textwidth}{1pt}
\tableofcontents\thispagestyle{fancy}
\noindent\rule{\textwidth}{1pt}
\vspace{10pt}
\newpage

\section{Introduction}
\label{sec:intro}
One of the earliest theoretical postulates for dark matter (DM), not to mention 
the now-popular framework of hidden sectors, was 
a mirror of the standard model (SM)
\cite{Kobzarev:1966qya,Okun:2006eb,Foot:2014mia}.  It provided two
examples of stable composite particles that could serve as dark
matter: atoms and baryons (nuclei).  Other early DM candidates
included magnetic monopoles, axions, massive neutrinos, sneutrinos
and photinos
\cite{Goodman:1984dc}.  Supersymmetric neutralinos seemed to enjoy a
favored status for many years.  The idea of a hidden sector with
complex structure---gauge interactions and DM multiplets---was
revitalized outside of the mirror context by Refs.\ 
\cite{Pospelov:2007mp,Arkani-Hamed:2008hhe}, inspired by cosmic ray anomalies
\cite{PAMELA:2008gwm}.  Around
the same time the ``hidden valley'' paradigm \cite{Strassler:2006im}
of a new confining gauge was proposed, mostly with signals for the
Large Hadron Collider in mind, but also with the awareness that a
stable bound state could serve as the DM.

In these lectures we review the various possibilities for
dark matter in the form of bound states, either in the case of a 
U(1)$'$
dark gauge group, leading to dark atoms, or of a confining
interaction, which could give rise to dark glueballs, mesons or
baryons.  Because atoms and baryons also exist in the visible sector,
it is reasonable to suppose they get their relic density in a similar
way as for visible matter---{\it i.e.}, we don't know!   In other
words, they could be asymmetric DM \cite{Kaplan:2009ag}, the origin of whose asymmetry
remains to be explained.  Beyond their relic density,
many interesting aspects can be addressed, in terms of direct and 
indirect signals.  

For some varieties of composite DM, it is not possible to have an asymmetry, 
in which case a calculation of 
the relic abundance is definitely called for.  A confining phase transition
can make the freezeout process more complex than in the standard thermal 
freezeout picture,
often depending on the relative temperatures of the two sectors.
There is the important model-dependent issue of which portals,
if any, exist between the hidden and visible sectors.  Of course
gravity always exists, and even it can play a role, as we will see for
glueballs.  Although in these lectures I will focus on hidden sectors,
it is interesting to note that DM could be a $\sim 25$\,TeV 
composite state of gluino-like fermions bound by 
QCD \cite{DeLuca:2018mzn,Gross:2018zha}.

\section{Dark atoms}
\label{sect:DA}
The simplest dark atom model outside of the mirror framework was
studied in Refs.\ \cite{Kaplan:2009de,Kaplan:2011yj}.  It consists
of a dark electron, proton and photon, $e'$, $p'$ and $\gamma'$
respectively, with a coupling strength of $\alpha' = g'^2/{4\pi}$.
In its minimal version, the only other fundamental parameters needed are the
masses $m_{e'}$ and $m_{p'}$.  Later we will consider the consequences
of also including a photon mass $m_{\gamma'}$ and kinetic mixing 
$\epsilon$ with
the SM hypercharge.  An important derived quantity is the binding
energy of the dark $H'$ atom,
\be
	B_{H'} = {\alpha'^2\over 2}\mu_{H'} = 
{\alpha'^2\over 2}{m_{e'} m_{p'}\over m_{e'} + m_{p'}}
	\cong {\alpha'^2\over 2} m_{H'}{R\over (1+R)^2}\,,
\label{BHeq}
\ee
where $\mu_{H'}$ is the reduced mass and $R = m_{p'}/m_{e'}>1$.
(Without loss of generality one can
assume that $R\ge 1$ since the sign of the U(1)$'$ charge is arbitrary.)  The mass of the $H'$ atom is
therefore $m_{H'} = m_{e'} + m_{p'} - B_{H'}$, which we usually
approximate as $m_{e'} + m_{p'}$, unless one is interested in the
regime of strong coupling.  The condition $m_{H'}>0$
implies the weak constraint 
\be
	\alpha' < \sqrt{2}(1+R) \,.
\ee
If it was violated, the mass of the atom is of course not negative; rather the perturbative calculation
(and possibly the nonrelativistic approximation) is breaking down.

\begin{figure}[t]
 \centerline{\includegraphics[width=0.7\linewidth]{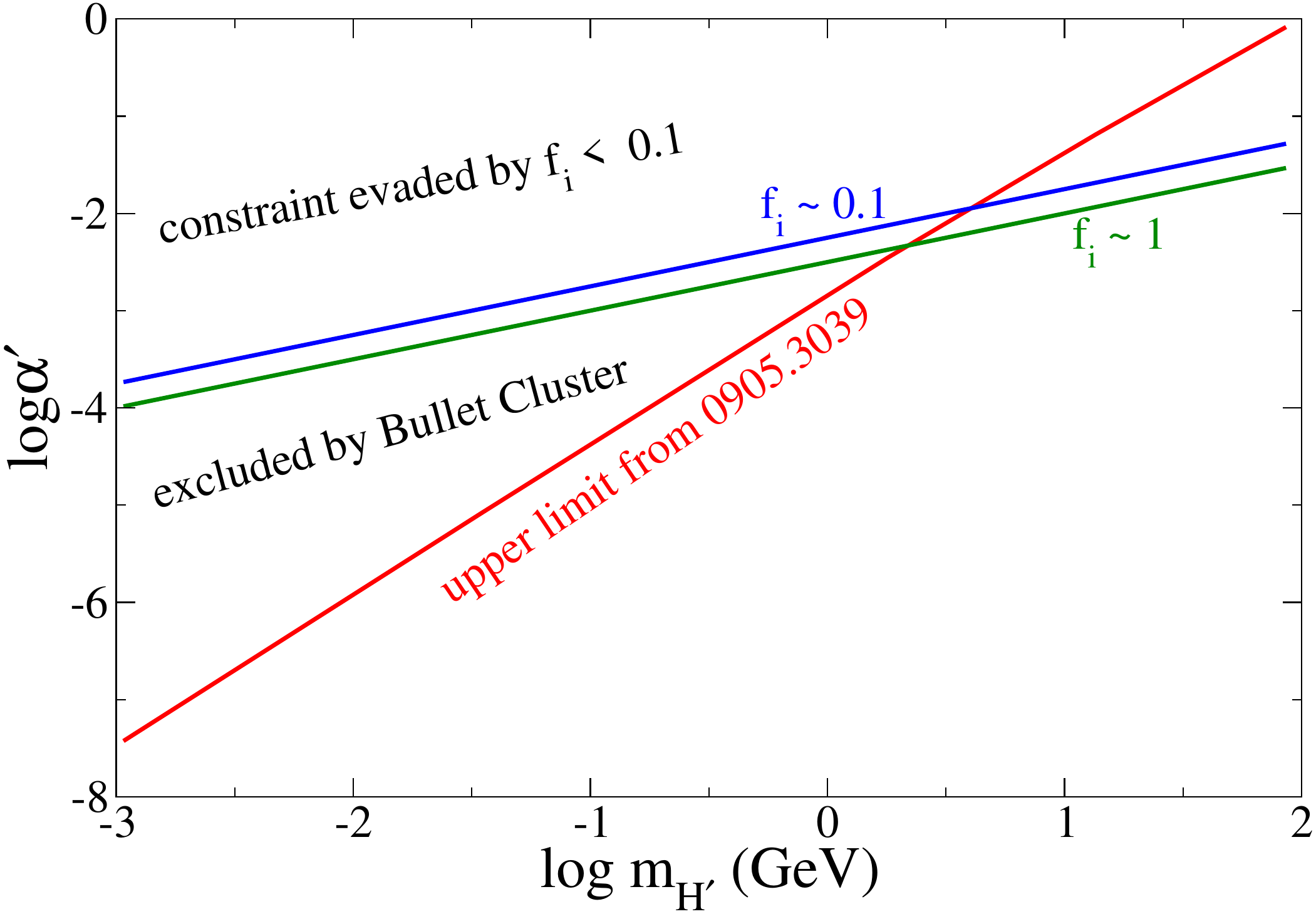}}
\caption{Region of dark atom parameter space excluded by Bullet
Cluster bounds for models with $R=1$ ($m_{e'} = m_{p'}$).  
}
\label{fig:fkty}
\end{figure}

\subsection{Cosmological evolution}
The cosmology of dark atom formation was initially worked out in
 Ref.\ \cite{Kaplan:2009de}, and later in more detail by Ref.\ 
\cite{Cyr-Racine:2012tfp}.  In the absence of a portal between the two
sectors, the dark sector will generally have a different temperature
$T'$ from the visible one $T$, and their ratio $\xi=T'/T$ can evolve with time.
One could imagine some initial ratio $\xi_i$ that is set by the
relative efficiency of reheating to the two sectors after inflation.

At temperatures $T'\sim m_{p'}/20$ and $T'\sim m_{e'}/20$
respectively, the symmetric components of $p'$ and $e'$ freeze out
through the annihilations $p'\bar p'\to\gamma'\gamma'$ and 
$e'\bar e'\to\gamma'\gamma'$.  This is followed by recombination at
$T' \lesssim B_{H'}$, below the binding energy due to the small
concentration of baryons relative to photons.  If $\xi_i \ll 1$, then all of these events occur 
when $T$ is significantly higher in the visible sector, {\it i.e.,}
at relatively early times compared to SM recombination.  

However it need not be the case that $\xi_i \ll 1$.  The dark photons 
are extra radiation species contributing to the Hubble expansion, 
conventionally parametrized as extra neutrino species,
\be
	\Delta N_{\rm eff} = {4/7}\left(11\over 4\right)^{4/3}
	g'_* \,\xi^4 < 0.45 \,,
\label{Neffbound}
\ee
where $g_*'=2$ if $\gamma'$ is the only dark radiation species,
$(11/4)^{1/3} = T_\gamma/T_\nu$ accounts for  the differential heating of
photons versus neutrinos after freezeout of the weak interactions, and $4/7 = \sfrac12(\rho_\gamma/\rho_\nu)$
for a single $\nu$ species.  The upper bound is from {\it Planck}
cosmic microwave background (CMB) constraints \cite{Planck:2018vyg}.  
Solving (\ref{Neffbound}) one finds
the modest constraint 
\be
	\xi ={T'\over T} < 0.57
\ee
at late times.  This would naturally result even if $\xi_i = 1$ after
inflation, if the two sectors remained decoupled, due to the much
larger entropy in the visible sector \cite{Fan:2013yva}.

An important quantity is the ionization fraction $f_i = n_{e'}/n_{H'}$
after recombination, the number of free $e'$ particles per dark atom.
It is determined by solving the appropriate Boltzmann equations
describing recombination.  Ref.\ \cite{Cline:2013pca} made an analytic fit to the 
numerical results of \cite{Kaplan:2009de},
\be
	f_i\cong {\rm min}\left[1,\, 
	10^{-10}\,\xi\,\alpha'^{-4}R^{-1}\left(m_{H'}\over{\rm
	GeV}\right)^2\right]\,.
\label{fiest}
\ee
Independently, Ref.\ \cite{Cyr-Racine:2012tfp} arrived at the estimate
\be
	f_i \sim 2\times 10^{-16}\,\xi\,\alpha'^{-6}
	\left(m_{H'}\over{\rm
	GeV}\right)\left(B_{H'}\over{\rm
	keV}\right)\,,
\ee
which is compatible with (\ref{fiest}) (using Eq.\ (\ref{BHeq})) for $R\gg 1$.

A large ionization fraction would be problematic because of the strong
Coulomb interactions between ions, in violation of Bullet Cluster
constraints on DM self-interactions 
\cite{Randall:2008ppe,Markevitch:2003at}.  The resulting bound on
$\alpha'$ was derived in Ref.\ \cite{Feng:2009mn} assuming that
$f_i=R=1$,
\be
	\alpha' < 10^{-2.9}\left(m_{H'}\over {\rm GeV}\right)^{1.5}\,,
\label{fengetal}
\ee
which is my fit to their numerical result (red line, Fig.\
\ref{fig:fkty}).  However one should combine
this with the estimate (\ref{fiest}) of $f_i$ as a function of $m_{H'}$ and
$\alpha'$ to see what region of parameter space is actually excluded.
I have done this exercise in Fig.\ \ref{fig:fkty}.  The actual
excluded region is the triangle at lower $m_{H'}$ between the red and
blue lines.  The blue line represents $f_i=0.1$, where the Bullet
Cluster bounds would be evaded.

More stringent bounds on $\alpha'$, by a factor of $10^4$, have been 
derived on the basis of observed DM halos with elliptical rather than
spherical morphology, since DM self-interactions would tend to erase
the ellipticity \cite{Miralda-Escude:2000tvu}. However, subsequent
analyses indicated that the ellipticity bound is not as stringent as
originally thought, but rather of the same order as the Bullet
Cluster constraint \cite{Peter:2012jh}.\footnote{Ref.\ 
\cite{Agrawal:2016quu} argues that the ellipticity bound is still more than two orders
of magnitude stronger than Eq.\ (\ref{fengetal}).}

Even if there is no significant ionization at early times, dark atoms
can reionize during structure formation, by shock heating as they
concentrate within galactic halos.  This causes the atoms to heat to
the virial temperature, which scales with redshift $z$ 
as \cite{Ghalsasi:2017jna}
\be
	T_{\rm vir} \sim G\, M_{\rm halo}^{2/3}\,\rho_m^{1/3}\, m_{p'} (1+z)
\ee
where $\rho_m$ is the present DM density.
If $T_{\rm vir}/\,B_{H'}\lesssim 0.1$, essentially no reionization
takes place.  This dimensionless ratio depends only upon $\alpha'$
and $R$ (apart from the dimensionless environmental parameter
$G M_{\rm halo}^{2/3}\rho_m^{1/3}$).  
From their Fig.\ 1, where contours of 
$T_{\rm vir}= 0.1\,B_{H'}$ in the plane of $m_{e'}$ versus $\alpha'$
are shown for a Milky-Way like galaxy with $m_{p'}= m_p$,
one can infer that 
\be
	\alpha' > 1.4\times 10^{-3}\,\sqrt{R}
\ee
is the condition to avoid reionization during structure formation
in our galaxy.

\begin{figure}[t]
 \centerline{\includegraphics[width=\linewidth]{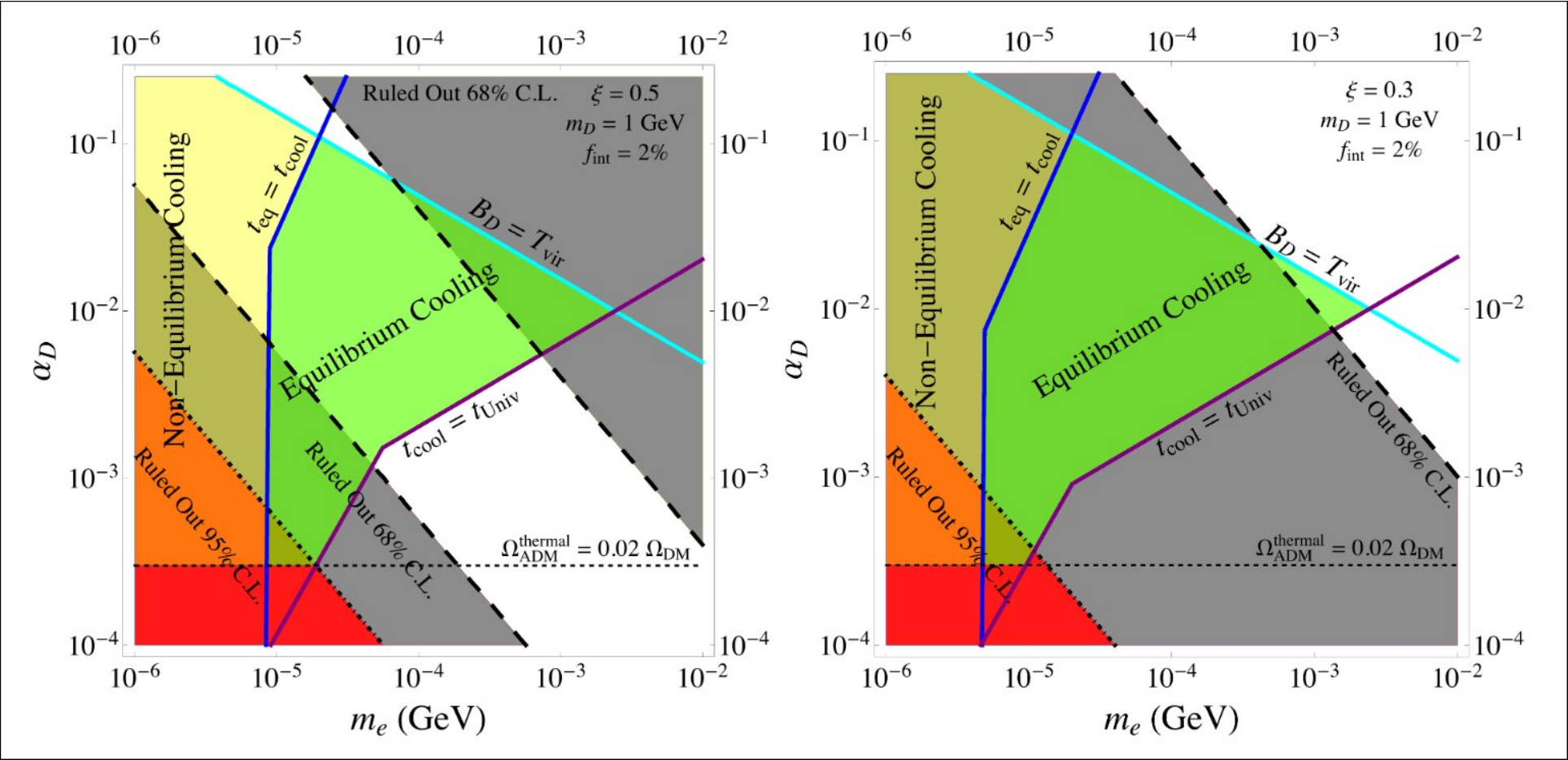}}
\caption{Constraints from DAO on dark atoms with $m_{H'}=1\,$GeV
assuming $\xi = 0.5$ (left) or $0.3$ (right) and $f_{int} = 0.02$.
Allowed regions are unshaded, and $\alpha_D = \alpha'$.  Taken from
Ref.\ \cite{Cyr-Racine:2013fsa}. 
}
\label{fig:DAO}
\end{figure}

\subsection{Dark acoustic oscillations}
If the dark sector is not too cold ($\xi$ not too small), and if the
ionization fraction is not too small, there can be
significant pressure waves in the dark sector at the surface of last
scattering for the CMB, analogous to
baryon acoustic oscillations.  These dark acoustic oscillations (DAO)
were studied in Ref.\ \cite{Cyr-Racine:2013fsa}, assuming that 
$\xi = 0.5$, and allowing for the possibility that dark atoms only
constitute a fraction $f_{int}$ of the total DM.  Under these
assumptions (for $m_{H'}=1\,$GeV), DAO rules out all
models in the remaining parameter space of  $\alpha'$ versus $R$
if $f_{\rm int}$ is as large as 0.05.  The constraints rapidly weaken
as $\xi$ or $f_{int}$ is decreased.  These results are illustrated 
in Fig.\ \ref{fig:DAO}.

\begin{figure}[t]
 \centerline{\includegraphics[width=0.54\linewidth]{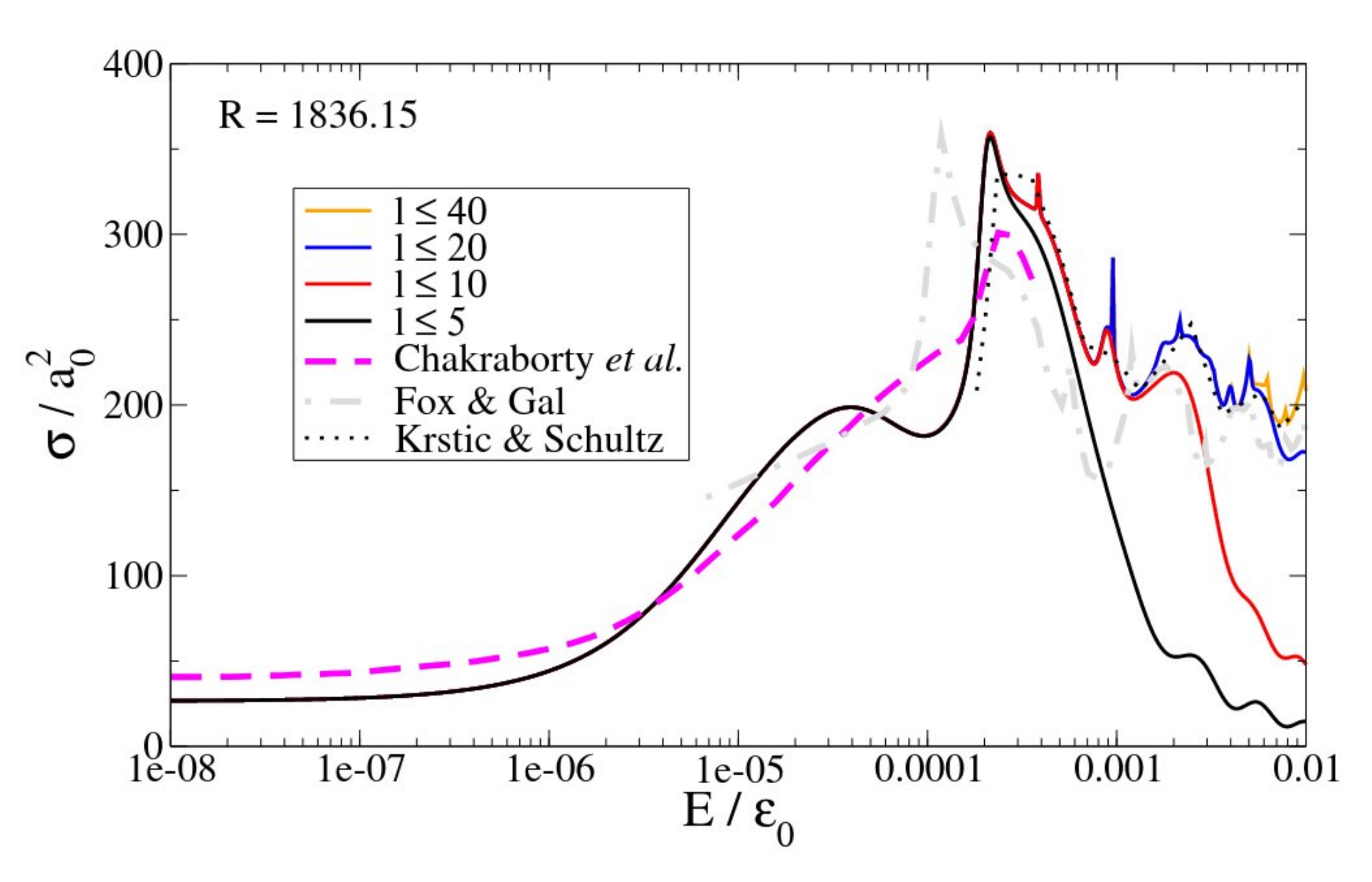}
\includegraphics[width=0.5\linewidth]{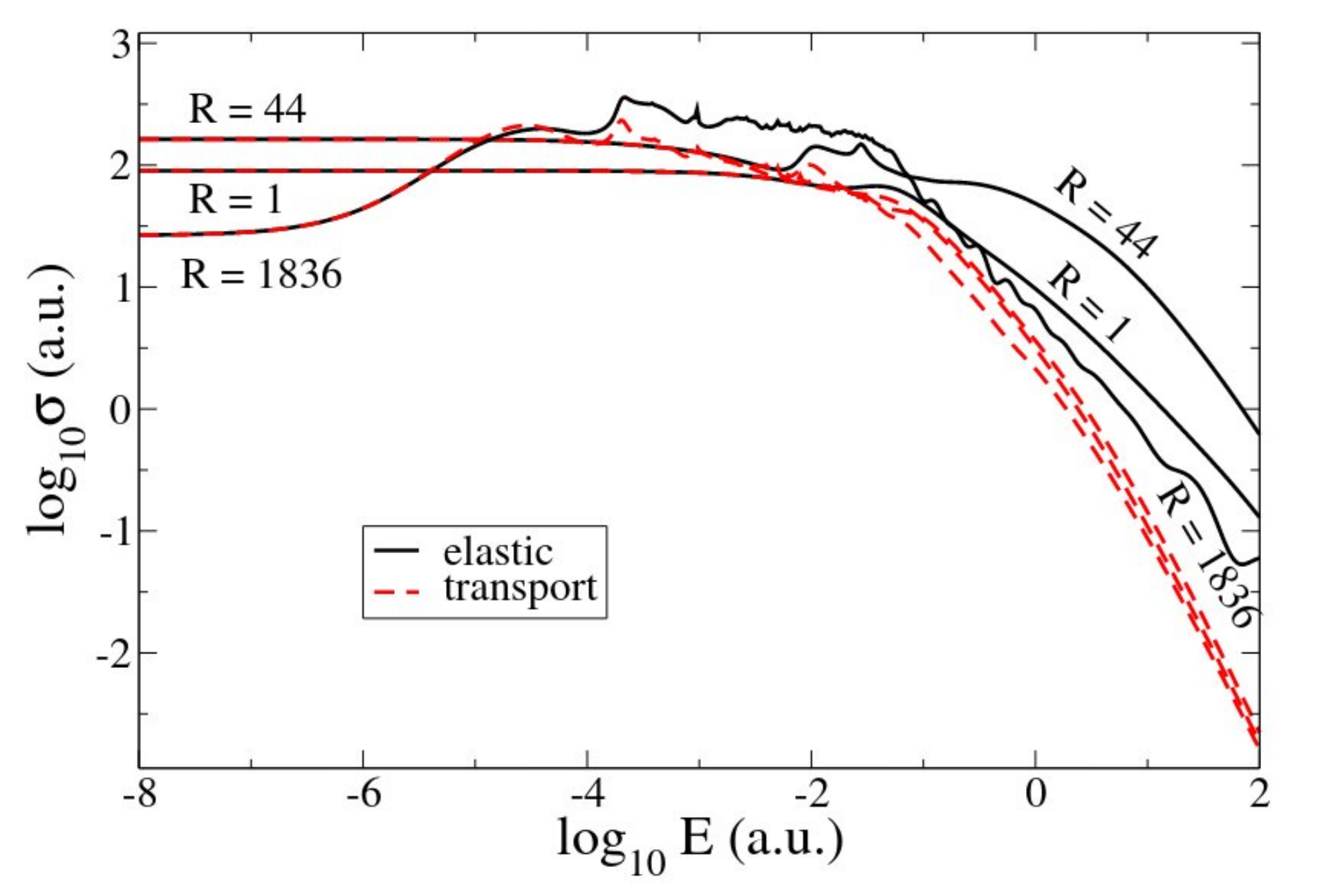}}
\caption{Left: Elastic scattering cross section of H atoms versus energy,
in atomic units, from Ref.\ \cite{Cline:2013pca}.  The convergence of the 
partial wave expansion is illustrated by summing different numbers of 
partial waves.
Right: elastic and and
transport cross sections versus energy, for several values of $R = m_p/m_e$.
}
\label{fig:sigatom}
\end{figure}

\subsection{Dark atom self-interactions}  
Although the Bullet Cluster puts an upper limit on the strength of DM
self-interactions, it is also known that nearly saturating the bound
by taking \cite{Rocha:2012jg}
\be
	{\sigma\over m}\sim 0.5\,{\rm cm^2\over g}
	\sim 0.9\,{\rm b\over GeV}
\label{sigm}
\ee
can have beneficial effects for ameliorating small-scale structure
problems of standard cold dark matter.  These include the cusp-core,
missing satellites and too-big-to-fail problems \cite{Weinberg:2013aya}.
Following that review article, the missing satellites seem to have been found \cite{Kim:2017iwr},
or the discrepancy may be a statistical fluctuation \cite{Fielder:2018szt}.
It is also possible that more realistic treatments of structure
formation including the effects of baryonic feedback can resolve some
of these
problems without the need for DM self-interactions 
\cite{2019MNRAS.487.1380G}.  However Ref.\ \cite{Kaplinghat:2019dhn}
notes that tuning the baryonic feedback to solve the cusp-core problem
results in discrepancies with the properties of high surface
brightness galaxies, and argues that self-interating DM still provides
a better fit halo profiles of diverse systems.

For an elementary DM particle of mass $m\sim 1\,$GeV, Eq.\ (\ref{sigm}) 
is a very large cross section, but with composite particles it is quite easy to
achieve.  For dark atoms we can expect a geometric cross section
governed by the dark Bohr radius, $a_0' = 1/(\alpha' \mu_{H'})$, with
$\sigma \sim \pi a_0'^2$.  In fact comparing to measured H atom scattering,
this is a significant underestimate: $\sigma/a_0^2 \sim 200-300$ at energies
above the atomic unit $E_0 = \alpha^2 \mu_H = 2 B_H$ (the Rydberg).
Moreover, $\sigma$ has complicated behavior as a function of energy, with
numerous resonances, as shown in Fig.\ \ref{fig:sigatom} (left).  Some of
these irregularities get smoothed out by considering the transport
cross-section $\sigma_t$ instead of the elastic cross section $\sigma$,
defined as 
\bea
	\sigma_t = \int d\cos\theta \,(1-\cos\theta){d\sigma\over d\cos\theta}
	\hbox{\quad or\quad}
	\sigma'_t = \int d\cos\theta\, (1-\cos^2\theta){d\sigma\over d\cos\theta}
\eea
This weights the cross section by the momentum transfer, which is the
physically relevant quantity since purely forward scattering does not have any
effect on DM structure formation.  For atom-atom scattering $\sigma_t'$ is the
appropriate choice, since exactly backward scattering is the same as forward
scattering for identical particles.  

It is amusing that, despite the complicated dependences on parameters,
these cross sections can be numerically computed for dark atoms for any values
of $\alpha'$, $m_{e'}$, $m_{p'}$, by working in atomic units and using results
from the atomic physics literature for the scattering potentials, which are
already determined in atomic units anyway.
  Then all dependences scale out of
the problem,\footnote{in the regime $R\gg 1$ where the Born-Oppenheimer approximation works.
The static H-H potential is computed assuming that the protons are immobile on the time scale
for the electron clouds to readjust themselves at a fixed proton-proton separation.}
 except for $R$, in the approximation that $B_{H'}\ll m_{H'}$.
One has to numerically solve the Schr\"odinger equation
\be
	\left[\partial_r^2 - {\ell(\ell+1)\over r^2} - f(R,\alpha')(V_{s,t}-E)\right]
	u_\ell^{s,t} = 0\,,
\ee
where $u_\ell = r\psi_\ell$,
for the partial waves in the spin singlet and triplet channels ($s,t$),
and sum over the orbital angular momentum $\ell$ and spins. 
  Here $f(R,\alpha') = R + 2 + R^{-1} - \alpha'^2/2$,
and usually one can neglect the $\alpha'^2/2$ correction.  
The same technique can be used for scattering of $H_2$ molecules.  Ref.\
\cite{Cline:2013pca} found that the energy-dependence  can be adequately
described by $\sigma'_t \cong (a_0 + a_1 E + a_2 E^2)^{-1}$ with $R$-dependent
coefficients.

\begin{figure}[t]
 \centerline{\includegraphics[width=\linewidth]{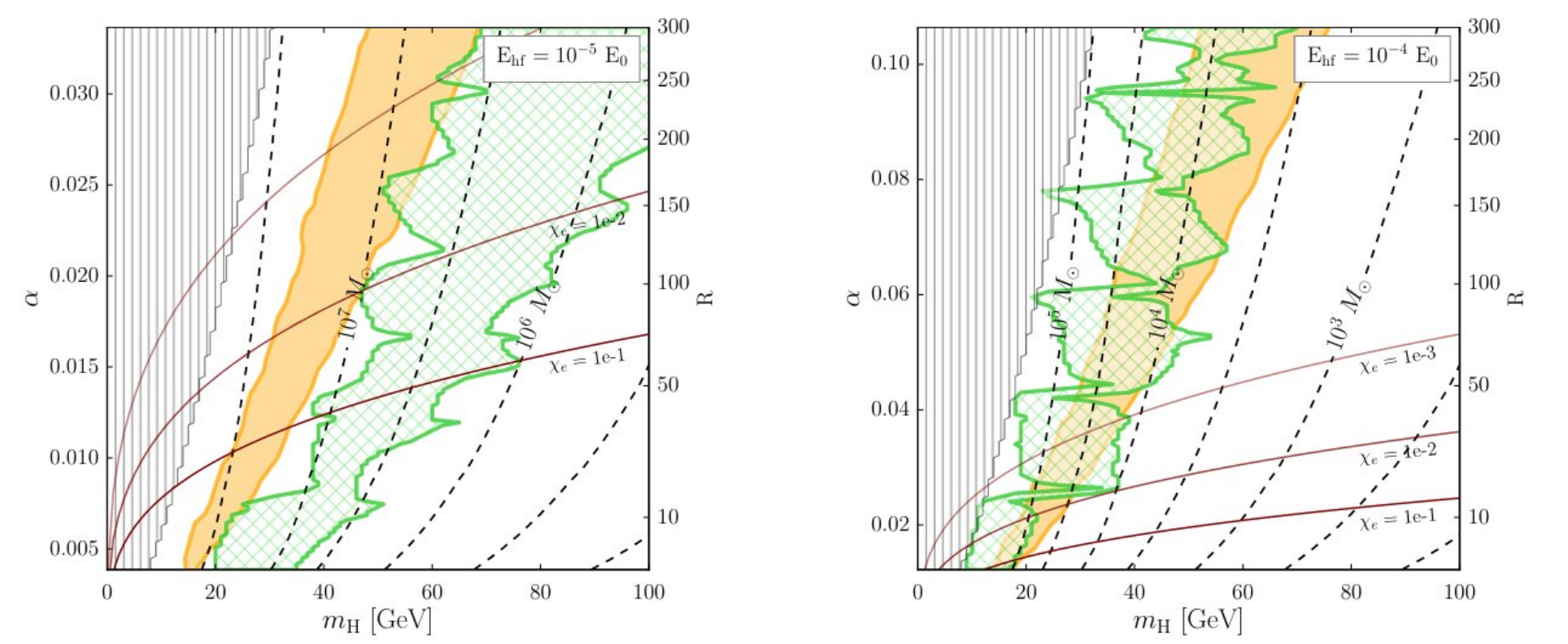}}
\caption{Regions of $\alpha'$ versus $m_{H'}$ where $\sigma'_t$ is compatible
with cluster halo profiles (orange) and lower mass halos (green), for
$E_{hf}/E_0 = 10^{-4}$ (left) and $10^{-5}$ (right), taken from Ref.\ 
\cite{Boddy:2016bbu}.  Contours of ionization fraction (here called $\chi_e$)
are shown, as well as contours of the minimum halo masses that can form, due
to DAO and the dark analog of Silk damping, assuming a dark temperature ratio
of $\xi = 0.6$. 
}
\label{fig:boddy}
\end{figure}

Interestingly, the reduced cross section at higher energies is compatible
with observations that galactic clusters, whose velocity dispersion is higher
than dwarf spheroidal or Milky Way-like galaxies, are also more cuspy and thus
require a smaller self-interaction cross section.  This was noted in Ref.\
\cite{Cline:2013zca} and studied in detail in Ref.\ \cite{Boddy:2016bbu}, which 
also took into account
the inelastic scatterings involving hyperfine transitions, whose energy is
\be
	E_{hf} = \sfrac83 \alpha'^4 \, {m_{e'}^2 m_{p'}^2\over m_{H'}^3} \cong 
	\sfrac83 \alpha'^2 E_0 {R\over (1+R)^2}\,.
\label{eq:ehf}
\ee
In Ref.\ \cite{Boddy:2016bbu} the parameter $R$ was traded for $E_{hf}$, and it
was shown that a good overlap between clusters and lower-mass halos could be
achieved if $E_{hf}/E_0 \cong 10^{-4}$, implying $\alpha'^2/R \cong 4\times
10^{-5}$ for $R\gg 1$.  This is illustrated in Fig.\ \ref{fig:boddy}.

\subsection{Relic density}
It is possible that dark atom constituents have equal and opposite aymmetries,
consistent with the universe having vanishing net U(1)$'$ charge.  Refs.\ 
\cite{Kaplan:2011yj,Choquette:2015mca} proposed UV completions in which the
dark atom asymmetry was directly linked to the baryon asymmetry through
leptogenesis.  One may ask whether it is possible to achieve the right relic
density without any asymmetric component, by the usual thermal freezeout via
annihilation into two photons, whose cross section is
\be
	\langle\sigma v \rangle_{\rm ann} = {\pi\alpha'^2\over m_{e',p'}^2}
	S
\ee 
respectively for the $e'$ and $p'$ components.  Here $S$ is a Sommerfeld
enhancement factor that is typically unimportant ($S\cong 1$ unless
$m_i\gtrsim$ TeV \cite{Agrawal:2016quu}).

\begin{figure}[t]
 \centerline{\includegraphics[width=0.75\linewidth]{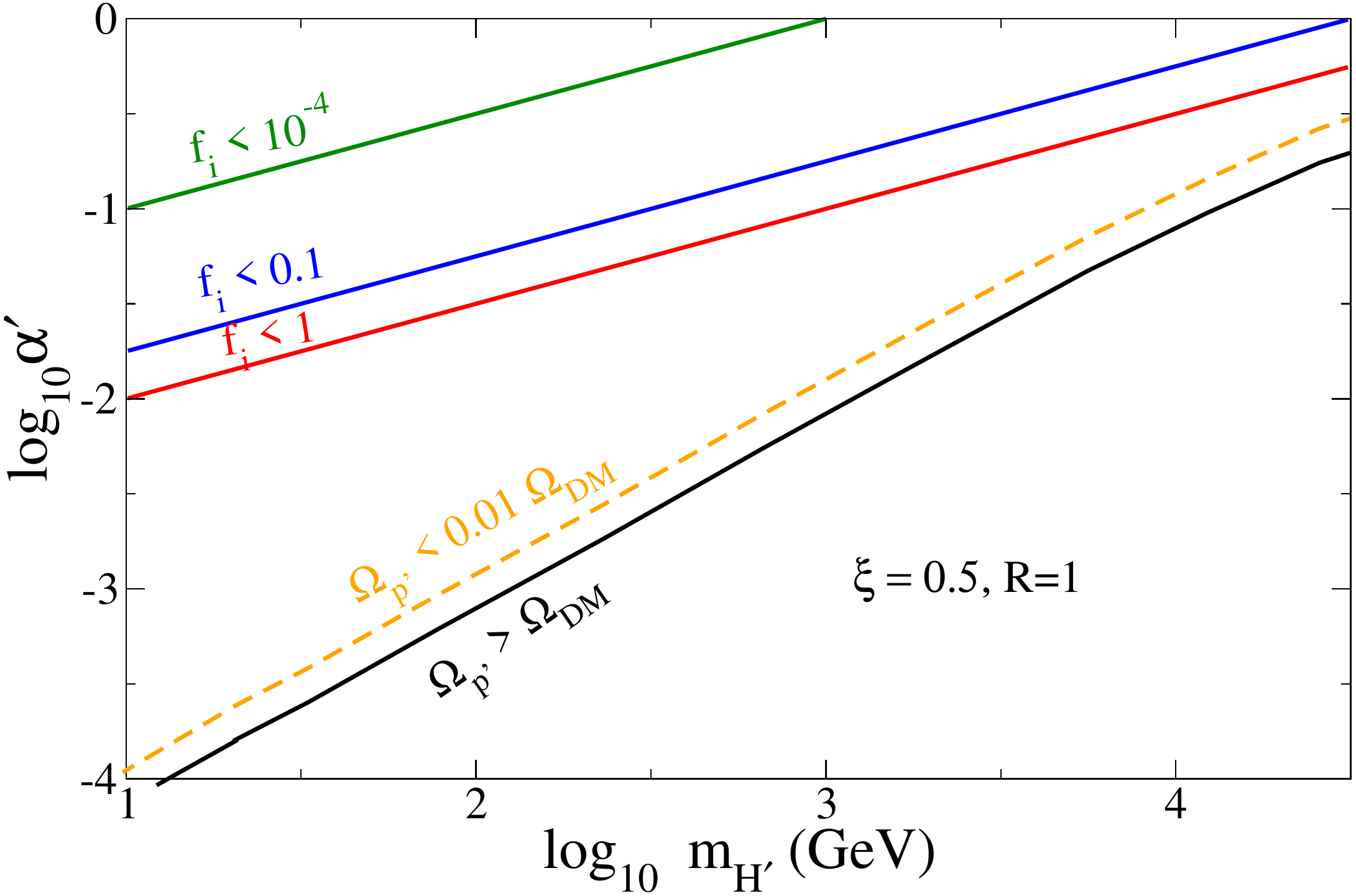}}
\caption{Black curve: contour of correct relic density for a model 
with only
$p'$ symmetric dark matter, reconstructed from  
Ref.\ \cite{Agrawal:2016quu}.  Dashed curve shows the suppression of
the symmetric component when the DM is assumed to be asymmetric.
Colored curves are contours of the
ionized fraction $f_i$.
}
\label{fig:relicden}
\end{figure}

Unless $R=1$ ($m_{e'}=m_{p'}$),
there will be more unannihilated $p'$s left over than $e'$s.
Hence it is natural to focus on the special case $R=1$
if atomic dark matter is symmetric.  Ref.\ \cite{Agrawal:2016quu} computed
the relic density in a model with only one constituent, which we could
identify as $p'$, and found the relationship between $\alpha'$ and $m_{p'}$
similar to the black curve in Fig.\ \ref{fig:relicden} to match the
observed DM density.  With two species of the same mass, this curve 
is adjusted for the fact that $m_{H'} = 2 m_{p'}$.  In this
scenario the DM remains fully ionized unless $\xi$ is very small.
On fig.\
\ref{fig:relicden} I have overlaid the contours of $f_i$ from Eq.\ 
(\ref{fiest}) for $\xi = 0.5$.  One would need $\xi$ to be smaller to comply
with the DAO constraints mentioned above, whereas $\alpha'$  is small enough
to satisfy Bullet Cluster and halo ellipticity constraints. In any case, the DM in this model
would not be in the form of atoms, but rather ions.  
It thus seems difficult to explain the relic density of dark atoms without an
asymmetry, unless $\xi$ is sufficiently small.

A related question is, given that dark atoms are a form of asymmetric
dark matter, how large of an unannihilated symmetric component can be
left over?  This question is answered for general asymmetric DM
models in the seminal reference \cite{Graesser:2011wi} (see their
Fig.\ 4).  For example,
if the relic density is mainly provided by the asymmetric component,
and the annihilation cross section is only 2.25 times greater than the
value needed for symmetric DM, then the symmetric component is
suppressed by a factor of 100.  This is illustrated by the dashed line
in Fig.\ \ref{fig:relicden}.

\subsection{Effect of dark photon mass}
A common elaboration of the model is to allow the dark photon to have
a small mass $m_{\gamma'}$.  It could come from a dark Higgs boson that is 
heavy enough to integrate out from our effective description, or it could come
from the St\"uckelberg mechanism through the interaction
\be
	\sfrac12 m_{\gamma'}^2\left( A'_\mu - \partial_\mu\theta\right)^2\,,
\label{eq:stck}
\ee
where $\theta \to \theta + \phi'$ under a U(1)$'$ gauge transformation
$A'_\mu \to A'_\mu + \partial_\mu\phi'$, so that the gauge symmetry is
maintained. The dark
Coulomb potential becomes a Yukawa potential with
a finite range $\lambda = 1/m_{\gamma'}$.  If this is still long compared to 
the Bohr radius $a_0' = (\alpha'\mu_{H'})^{-1}$, then the binding properties
of dark atoms will be slightly perturbed.  One can quantify this
effect by approximately solving the Schr\"odinger equation for the bound state,
\be
	\left(-{1\over 2\mu_{H'}}\partial_r^2 - 
\alpha'{e^{-r/\lambda}\over r}
	\right) u = E u\,,
\label{eq:sch}
\ee
where again $u = r\psi$.
If $\lambda \gg a_0$, we can expand $e^{-r/\lambda}\cong 1 - r/\lambda$ and
treat the extra term as a perturbation.  The shift in the binding energy
is
\be
	\Delta B_{H'} = -\left\langle {\alpha'\over r}\,{r\over\lambda}\right\rangle
	= -\alpha' m_{\gamma'}\,
\label{eq:dbh}
\ee
so one can estimate that dark atoms continue to exist as long as $m_{\gamma'}
\lesssim \alpha'\mu_{H'}/2$.  

More quantitatively, Ref.\ \cite{Petraki:2016cnz} solved Eq.\ (\ref{eq:sch})
numerically and their results for $B_{H'}$ can be fit by the formula
\be
	B_{H'} \cong \left(1-0.85 {m_{\gamma'}\over
	\alpha'\mu_{H'}}\right)^{2.16} {\mu_{H'}\alpha'^2\over 2}
\ee
which roughly agrees with Eq.\ (\ref{eq:dbh}) for small
$m_{\gamma'}$.\footnote{They should agree exactly at small $m_{\gamma'}$ since
the perturbative calculation is reliable; the discrepancy could be due to digitization
inaccuracies since Ref.\ \cite{Petraki:2016cnz} plots $B_{H'}^{1/2}$ 
with respect to $1/m_{\gamma'}$ rather than $m_{\gamma'}$.}\ \   This
indicates that
the more accurate constraint for having bound states is 
\be
	m_{\gamma'}\lesssim 1.2\, \alpha'\mu_{H'}\,.
\label{eq:mbound}
\ee

A nonnegligible $m_{\gamma'}$ affects the ionization fraction.  This was
studied in Ref.\ \cite{Kahlhoefer:2020rmg}, which noted that the
recombination interaction $e'+p'\to H' + \gamma'$ can be kinematically
blocked, leading to higher residual $f_i\sim 0.1$ when $m_{\gamma'} \sim
B_{H'}$.  Notice that this is still a small mass (by a factor of $\alpha'$)
compared to the constraint (\ref{eq:mbound}).

\subsection{Kinetic mixing}
So far we have not considered any portals between the dark and visible
sectors.  The most natural one for dark atoms is gauge kinetic mixing,
\be
	\sfrac12 \epsilon F'_{\mu\nu} F^{\mu\nu}\,,
\label{eq:kmix}
\ee
where $F'_{\mu\nu}$ is the U(1)$'$ field strength, and $F^{\mu\nu}$ is that of the
SM hypercharge, or U(1)$_{\sss EM}$ in an effective Lagrangian description.
To diagonalize the gauge boson kinetic terms for $\epsilon\ll 1$, we should
distinguish between the two cases $m_{\gamma'}=0$ or $m_{\gamma'}>0$ 
\cite{Holdom:1985ag}.
In the latter case, the field transformation that accomplishes this is
\be
	A_\mu\to A_\mu - \epsilon A_\mu',\qquad
	A'_\mu \to A'_\mu
\ee
so that SM particles acquire small couplings to the dark photon, for example
\be
	{\delta\cal L} = \epsilon e A'_\mu\left(\bar p\,\gamma^\mu\, p - 
	\bar e\,\gamma^\mu\, e\right)\,.
\label{epscoup}
\ee
This allows decays $\gamma'\to e^+e^-$ if $m_{\gamma'} > 2 m_e$.  For lighter
$\gamma'$, there is the decay $\gamma'\to 3 \gamma$ through an electron loop, 
with rate \cite{McDermott:2017qcg}
\be
	\Gamma \cong 1{\,\rm s}^{-1} \left(\epsilon\over 0.003\right)^2
	\left(m_{\gamma'}\over m_e\right)^9\,.
\ee

If $m_{\gamma'}=0$ then one has the freedom to choose arbitrary orthogonal linear
combinations of $A_\mu$ and $A'_\mu$ as the mass eigenstates.  The most
convenient choice is through the transformation
\be
	A_\mu \to A_\mu,\qquad A'_\mu \to A'_\mu + \epsilon A_\mu
\ee
which results in millicharges $q = \epsilon g'/e$ for the dark constituents
\cite{Cline:2012is},
\be
	{\delta\cal L} = \epsilon g'A_\mu \left(\bar p'\,\gamma^\mu\, p' - 
	\bar e'\,\gamma^\mu\, e'\right)
\ee
while the dark photon continues to couple only to the dark constituents.

Another option is St\"uckelberg mixing \cite{Kors:2004dx,Feldman:2007wj},
which uses the same Lagrangian (\ref{eq:stck}) but assumes that under the
combined SM U(1) and dark U(1)$'$ gauge transformations 
$A_\mu \to A_\mu + \partial_\mu\phi$ and $A'_\mu\to A'_\mu +
\partial_\mu\phi'$
the St\"uckelberg field transforms as
$\theta\to \theta + \phi' + \lambda\phi$.   The kinetic mixing term (\ref{eq:kmix})
can also be present.  Then after diagonalization of the kinetic terms, the
perturbed interaction Lagrangian is \cite{Kahlhoefer:2020rmg}
\be
	{\delta\cal L} = \lambda g' A_\mu\left(\bar p'\,\gamma^\mu\, p' - 
\bar e'\,\gamma^\mu\, e'\right)  - (\epsilon-\lambda)eA'_\mu 
\left(\bar p\,\gamma^\mu\, p - 
\bar e\,\gamma^\mu\, e\right)\,.
\ee
Notice that $m_{\gamma'}$ is assumed to be nonzero in this case.

\begin{figure}[t]
 \centerline{\includegraphics[width=0.75\linewidth]{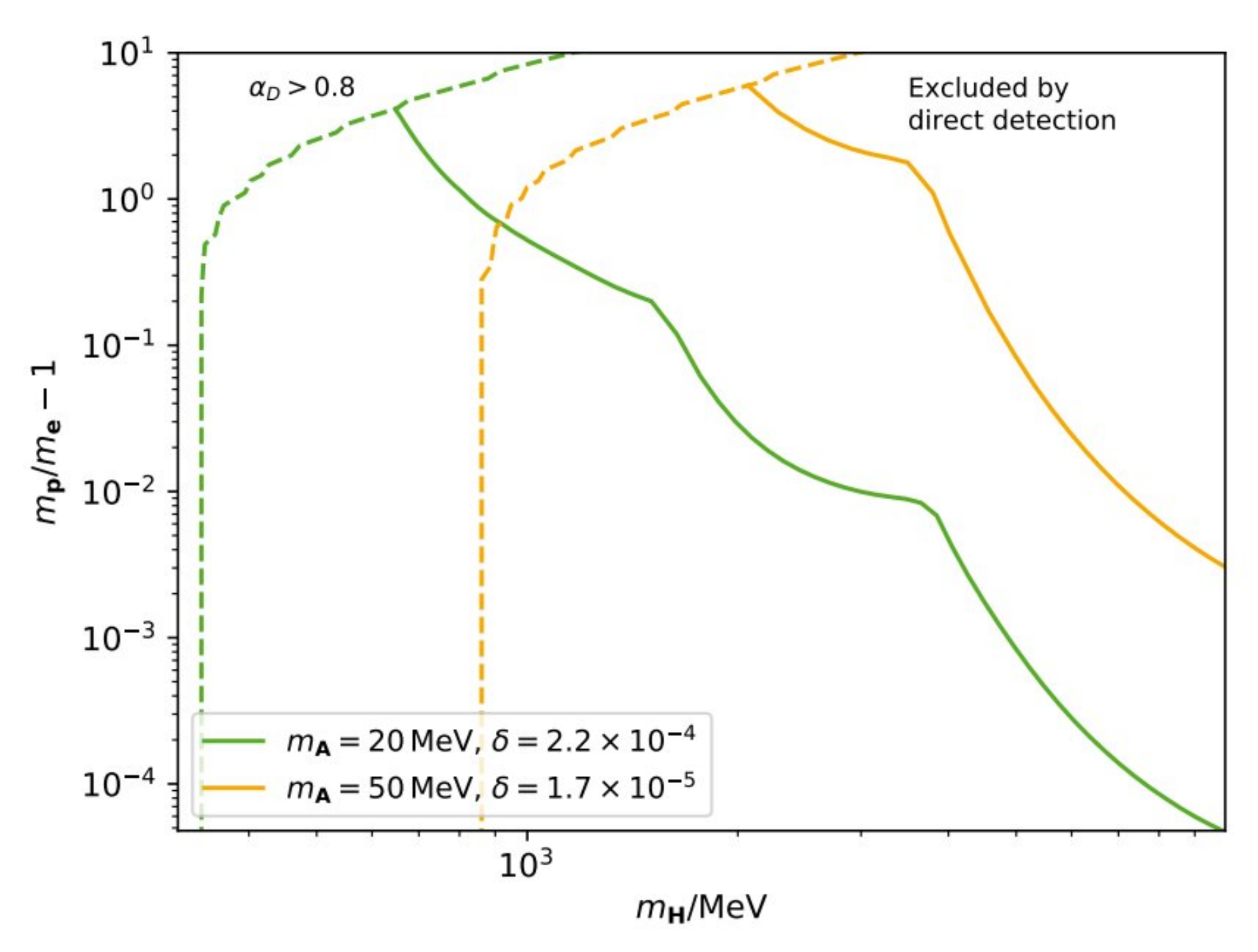}}
\caption{Constraints on $R-1$ versus $m_{H'}$ from direct detection of dark atoms in a model with
St\"uckelberg mixing and fraction $f_i=0.1$ of ionized millicharged dark
constituents, from Ref.\ \cite{Kahlhoefer:2020rmg}. The parameter $\delta$
is defined to be $\delta = \epsilon-\lambda$.
}
\label{fig:kahl}
\end{figure}

In general one could have scattering between dark H$'$ and visible
protons in direct detection experiments, mediated by both $\gamma$ and
$\gamma'$.  These interactions will be suppressed by the charge neutrality
of H$'$, but do not in general vanish because the charge distributions of
$p'$ and $e'$ do not exactly coincide, except in the special case $R=1$.
The Fourier transform of the charge distribution becomes a form factor
in the matrix element for scattering, that depends on the momentum $q$
transferred.  In the limit $q=0$, the photon (or dark photon) would be
sensitive to only the net charge of H$'$, which vanishes.  For small $q$,
the matrix element is suppressed by $q^2$ and one finds that
this cancels the $1/q^2$ photon propagator to give a contact interaction
in the case of massless $\gamma'$ \cite{Cline:2012is}.  The resulting
cross section for $p$-H$'$ scattering when $R\gg 1$ is
\be
	\sigma_p = 4\pi\alpha'^2\epsilon^2\left(m_p m_{H'}\over m_p + m_{H'} 
	\right)^2 a_0'^4\,.
\ee
If $R=1$, there is a
different (and weaker) velocity-suppressed contribution to direct detection from hyperfine 
transitions, of order
\be
	\sigma_p \sim 16 \alpha'^2\epsilon^2\left(m_p \over m_p + m_{H'} 
	\right)^2 \, {v^2\over q^2} \sim 16 
	{\alpha'^2\epsilon^2\over m_{H'}^2}\left(m_p \over m_p + m_{H'} 
	\right)^2\,.
\ee
This led to a weak limit $\epsilon g'/e \lesssim 10^{-2}$ in 2012, which at
that time was compatible with hints of direct detection by the CoGeNT
experiment \cite{CoGeNT:2010ols} for $m_{H'}\sim$\,6 GeV; the limit is significantly stronger now.

In principle, the ionized components $p'$ and $e'$ could lead to much stronger
limits unless $f_i\ll 1$, since there is no cancellation between charges when
they scatter on $p$, but this might not be the case if they are
millicharged.  It was shown that supernovae shock waves expel such particles
from the galactic disk \cite{Chuzhoy:2008zy,McDermott:2010pa}, making them
invisible to direct searches.  These studies however did not take into
account the possible effects of dark photon-mediated interactions
between the millicharged particles, which were shown to efficiently
randomize their directions in Ref.\ \cite{Foot:2010yz} (see also Ref.\
\cite{Chacko:2018vss}).  In this case,
ionized millicharged particles would not be expelled from the galaxy
by supernovae, unless the dark photon were sufficiently massive to
damp the self-interactions.

In the case of St\"uckelberg mixing, it is possible to have millicharged
constituents simultaneously with nonvanishing $m_{\gamma'}$.  Inspired by
the EDGES 21 cm anomaly \cite{Bowman:2018yin}, Ref.\ 
\cite{Kahlhoefer:2020rmg} constructed a model with enhanced ionization
fraction $f_i\sim 0.1$ by taking $m_{\gamma'} \sim 20-50$\,MeV, with a view
toward naturally explaining a subdominant component of millicharged DM through
the ionized fraction of atomic DM, without having to introduce it separately.
Achieving $f_i=0.1$ fixes $\alpha'$ in terms of the other model parameters,
and leads to constraints from direct detection in the $R$-$m_{H'}$ plane
shown in Fig.\ \ref{fig:kahl}.  The effect of hyperfine transitions has been
neglected in deriving these constraints.

\subsection{Further applications}
We have already discussed a number of observable effects of dark atoms:
DAO, self-interactions, direct detection.  The framework is very rich in 
possible phenomenological consequences.  Here we briefly describe
several more.

\begin{figure}[t]
 \centerline{\includegraphics[width=0.69\linewidth]{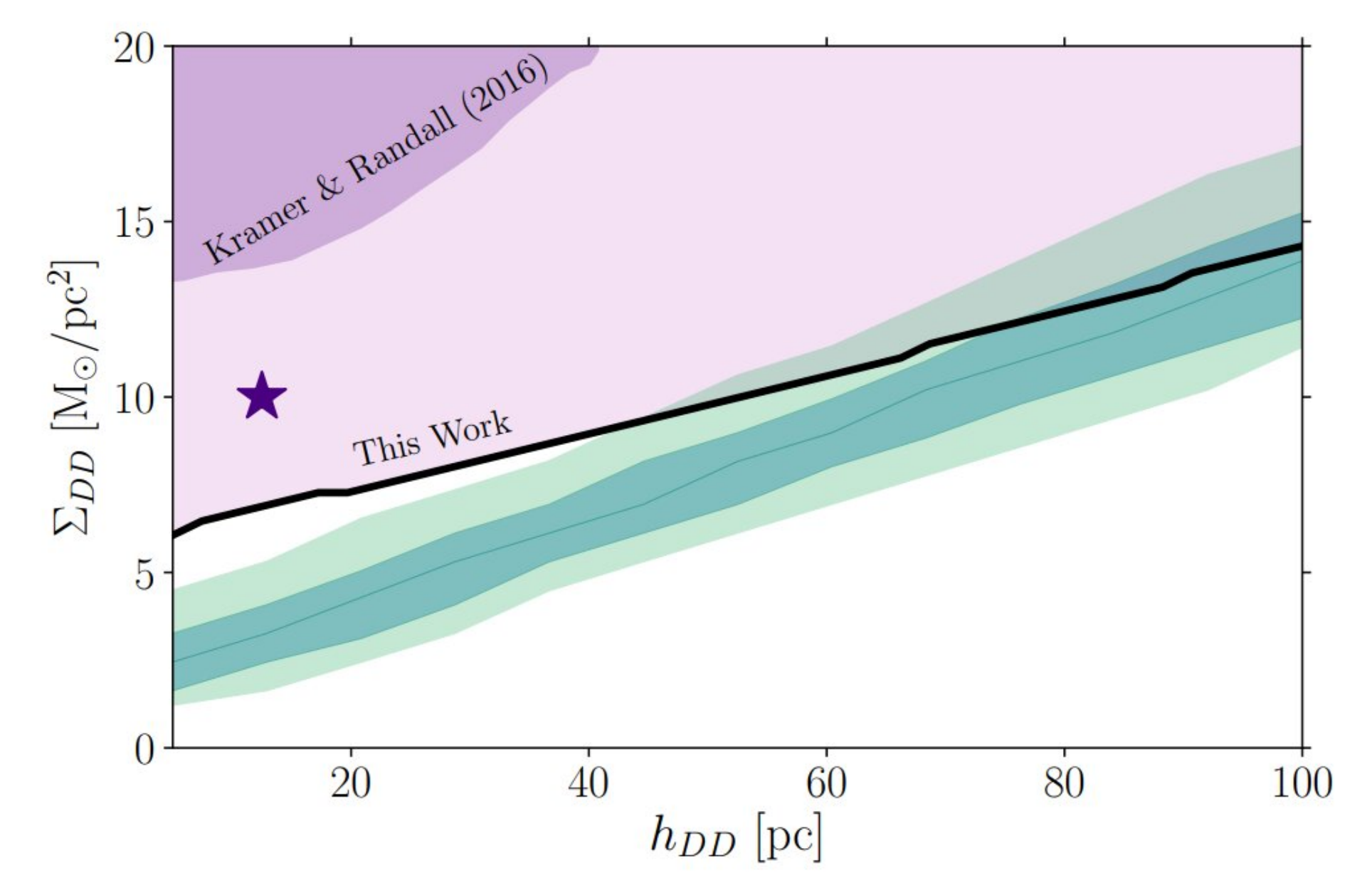}
\includegraphics[width=0.35\linewidth]{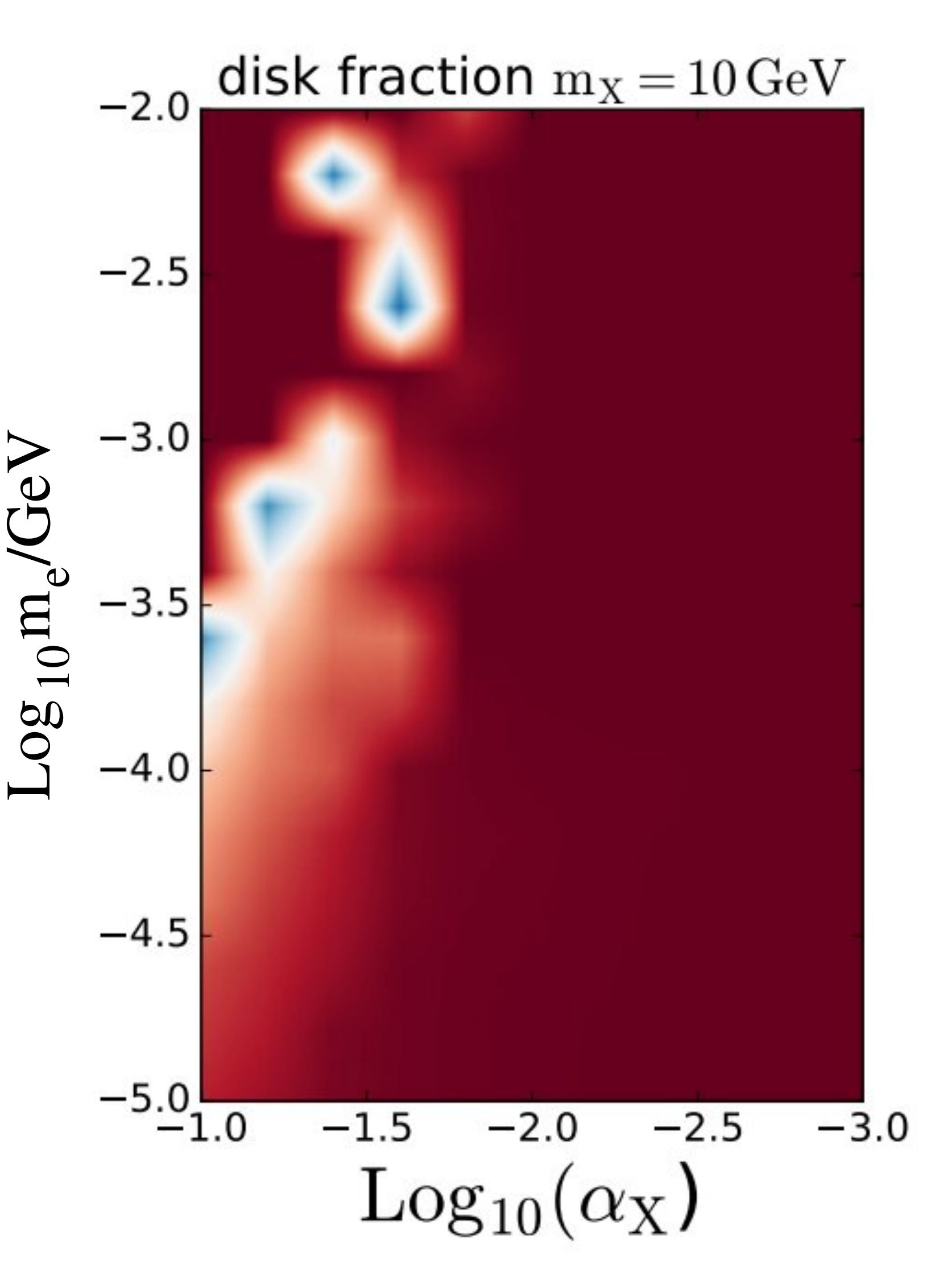}}
\caption{Left: constraints on a dark disk surface density $\Sigma_{DD}$
versus its thickness $h_{DD}$ from Ref.\ \cite{Schutz:2017tfp}.
The star denotes parameters preferred by Ref.\ \cite{Randall:2014lxa}
for explaining possibly enhanced periodic comet impacts on Earth.
Right: regions (blue/white) of $m_{e'}$ versus $\alpha'$ where a dark disk
forms, for $m_{H'} = 10\,$GeV, from Ref.\ \cite{Ghalsasi:2017jna}.
}
\label{fig:DD}
\end{figure}

\subsubsection{Dark disks}
In standard CDM structure formation, the DM halo is spheroidal and only
visible matter collapses to form the disk of a spiral galaxy.  However if some
fraction of DM has dissipative interactions similar to baryons, one might
expect it to collapse and form a disk that overlaps with the visible one.
This idea was explored in depth in Refs.\ 
\cite{Fan:2013tia,Fan:2013yva,McCullough:2013jma}.  In Ref.\ 
\cite{Fan:2013yva} it was argued that up to $\sim 10\%$ of DM could have
strong self-interactions while remaining consistent with Bullet Cluster
bounds, and that it could constitute up to 5\% of the mass in the Galactic
disk.  The formation mechanism is similar to that of the visible disk:
dark atoms fall toward the galactic center, virialize to temperatures
$T> B_{H'}$ through their dissipative interactions, becoming ionized and
then cooling via Brehmsstrahlung and Compton scattering on dark photons, that
are assumed to be present at the level $\xi \sim 0.5$.

This proposal has come under pressure from Gaia measurements, that are able to
constrain the surface density (mass per unit area) $\Sigma_{DD}$ of the 
dark disk \cite{Kramer:2016dqu,Schutz:2017tfp,Buch:2018qdr,Widmark:2021gqx}.
Constraints on $\Sigma_{DD}$ depend on its assumed thickness $h_{DD}$, as illustrated in Fig.\ \ref{fig:DD}
(left).  Ref.\
\cite{Buch:2018qdr} obtained stronger limits, similar to the green curves
in Fig.\ \ref{fig:DD} (left).

The previous works assume that a dark disk forms, but this need not be the
case.  Ref.\ \cite{Ghalsasi:2017jna} shows that with a subdominant component
(5\%) of atomic dark matter, cooling occurs too early for dark disks to
survive; they tend to be transformed into bulges by subsequent gravitational
torques from dense DM clumps.  The small (blue) parameter regions where disks form
are shown in Fig.\ \ref{fig:DD} (right) for $m_{H'} = 10\,$GeV; these regions
enlarge and merge to some extent for lighter $m_{H'} = 1\,$GeV atoms.

A thorough study of dark atomic structure formation within our galaxy 
was made in Ref.\ \cite{Chacko:2018vss}, in the context of a Twin
Higgs mirror sector that constitutes a fraction of the total dark
matter.  It includes kinetic mixing that induces nanocharges for the
dark constituents and enables direct detection.  Depending upon
details of the dark astrophysics, the mirror constituents may form a
disk or remain in a halo, and they may be in ionized or atomic form.

\begin{figure}[t]
 \centerline{\includegraphics[width=0.7\linewidth]{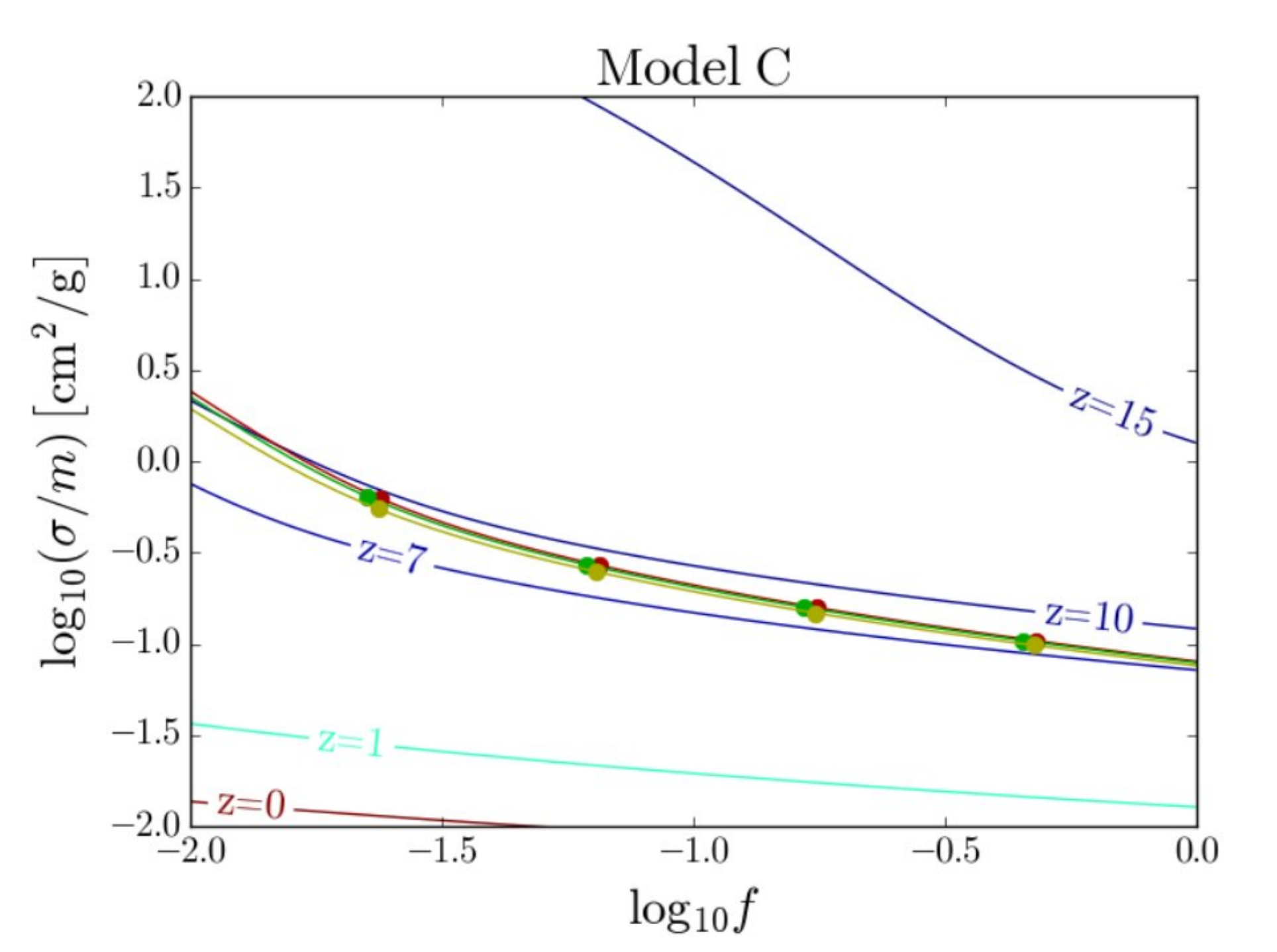}}
\caption{Trajectories in the plane of $\sigma/m$ versus fraction
of atomic DM $f$, with scattering assumed to be dissipative,
showing the degeneracy of DM parameters with the amount of accretion
subsequent to BH formation, for models 
consistent with redshifts of three observed SMBHs (taken from 
Ref.\ \cite{Choquette:2018lvq}).  Successive clusters represents steps
of 1 $e$-folding in mass growth, from right to left, and contours of 
constant redshift $z$ are shown.  The BHs corresponding to the rightmost
cluster must have undergone 1 or 2 $e$-foldings of growth to 
match the observations. 
}
\label{fig:smbh}
\end{figure}

\subsubsection{Early SMBH formation}
\label{sect:smbh}
Observations of supermassive black holes (SMBHs) at surprisingly high
redshifts \cite{2011Natur.474..616M,Banados:2017unc}  
have sparked interest in the possibility that a fraction $f$ of strongly
interacting dark matter could catalyze their formation \cite{Pollack:2014rja}.
The very large cross sections are naturally accommodated by dark atoms
constituting this part of the total DM \cite{DAmico:2017lqj}.  The dissipative
interactions of atomic DM can accelerate the process of gravothermal collapse
that would initiate formation of a black hole, at an earlier time than in 
standard CDM cosmology.  Subsequent accretion could then allow the BH to
reach its observed mass by redshifts $z\sim 7$.   This scenario was confirmed
using a modified $N$-body gravitational simulation in Ref.\ 
\cite{Choquette:2018lvq}, which found that dissipative scattering is much more
effective than elastic scattering for seeding SMBHs, and could 
allow for a fraction as large as $f=1$ of atomic DM while marginally
satisfying
Bullet Cluster constraints.  Fig.\ \ref{fig:smbh} shows the allowed parameters
assuming dissipative scattering.
For smaller $f$, a somewhat larger $\sigma/m$ and number of $e$-foldings of accretion would
be needed, as can be estimated from the curve by counting clusters. 

\subsubsection{3.5 keV X-ray line}
\label{kevsect}
The origin of a 3.5 keV X-ray signal \cite{Bulbul:2014sua,Boyarsky:2014jta}
in XMM-Newton observations of galactic clusters and M31 remains controversial,
but decays of 7 keV sterile neutrino DM into photons have been a highly
studied candidate.  In Ref.\ \cite{Cline:2014eaa} we considered the
alternative 
possibility that 3.5 keV corresponds to the dark hyperfine transition energy
(\ref{eq:ehf}), if $m_{\gamma'} > E_{hf}$ to kinematically 
block the decay of the triplet excited state H$_3'\to$ H$_1' + \gamma'$ into
dark photons, while introducing kinetic mixing
to allow the visible decay  H$_3'\to$ H$_1' + \gamma$.  The excited state
could either be primordial, with a lifetime similar to the age of the
universe, or it could be short-lived and result from late-time
self-interactions of H$'$.  The latter scenario requires relatively large
kinetic mixing, and is constrained by direct detection toward heavy dark
atoms, $m_{H'}> 350\,$GeV (in 2014, no doubt larger now in light of stronger
direct limits).

\begin{figure}[t]
 \centerline{\includegraphics[width=0.7\linewidth]{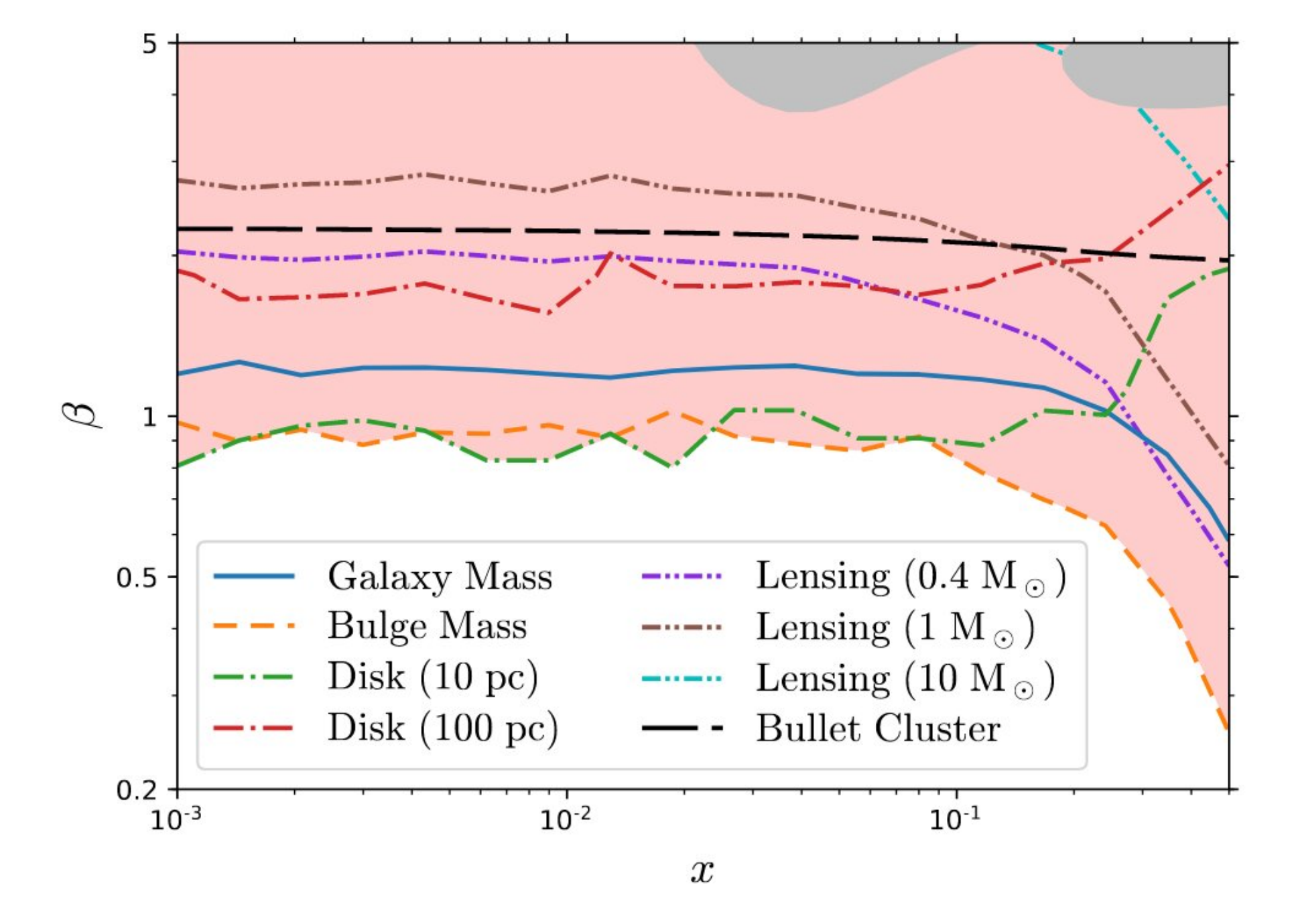}}
\caption{Constraints on the ratio $\beta$ 
of mirror baryons to visible baryons versus $x=\xi=T'/T$ from structure
formation in models with perfect mirror symmetry, from Ref.\
\cite{Roux:2020wkp}.  $\beta$ is related to $f$, the fraction of atomic
DM, by $f = \beta\, \Omega_b/\Omega_{DM} \cong 0.18\,\beta$.
}
\label{fig:JS}
\end{figure}

\subsubsection{Dark molecules, planets, stars \dots}

Apart from dark disks, other more complex structures beyond atoms can form,
depending upon the parameters in the dark sector, including the important
environmental ones $\xi = T'/T$ and $f$, the fraction of DM comprised by dark
atoms, versus conventional CDM.  In mirror models or other variants
having nontrivial chemistry, the abundance of He$'$ plays an important role in
structure formation.   
Ref.\ \cite{Ghalsasi:2017jna} studied structure formation in a simple dark sector
without chemistry, taking  $f=0.05$
(the limit from DAO assuming $\xi=0.5$) and $R\gg 1$, finding that much of the
parameter space is ruled out by the formation of MACHO-like structures or
a dark bulge in excess of constraints on the observed mass-to-light ratio of
the luminous part of the galaxy.

Ref.\ \cite{Roux:2020wkp} repeated this analysis for the model of exact mirror
symmetry, with $\xi$ and $f$ being the only free parameters.  Largely due to
the effects of He$'$, not present in simple atomic DM models, the structure
formation constraints found in the latter are significantly relaxed, and allow
for $f\lesssim 0.14$, as shown in Fig.\ \ref{fig:JS}.  He$'$ ions are
efficient for capturing free electrons, lowering the ionization fraction,
and impeding the formation of H$_2'$ molecules, which are important building
blocks for structure.  This ultimately reduces the
number of dense dark structures that are constrained by MACHO searches or
mass-to-light observations.

The previous studies were done using the extended Press-Schechter formalism
for simulating the merger history of DM halos.  Eventually it may be
interesting to repeat these using gravitational $N$-body simulations
including hydrodynamics, that could provide a closer to first-principles 
analysis.  Cooling rates in these complex dark sectors including molecules
with dissipative interactions have been computed
in Refs.\ \cite{Rosenberg:2017qia,Ryan:2021dis} as a necessary first step to enable such simulations.

\begin{figure}[t]
 \centerline{\includegraphics[width=\linewidth]{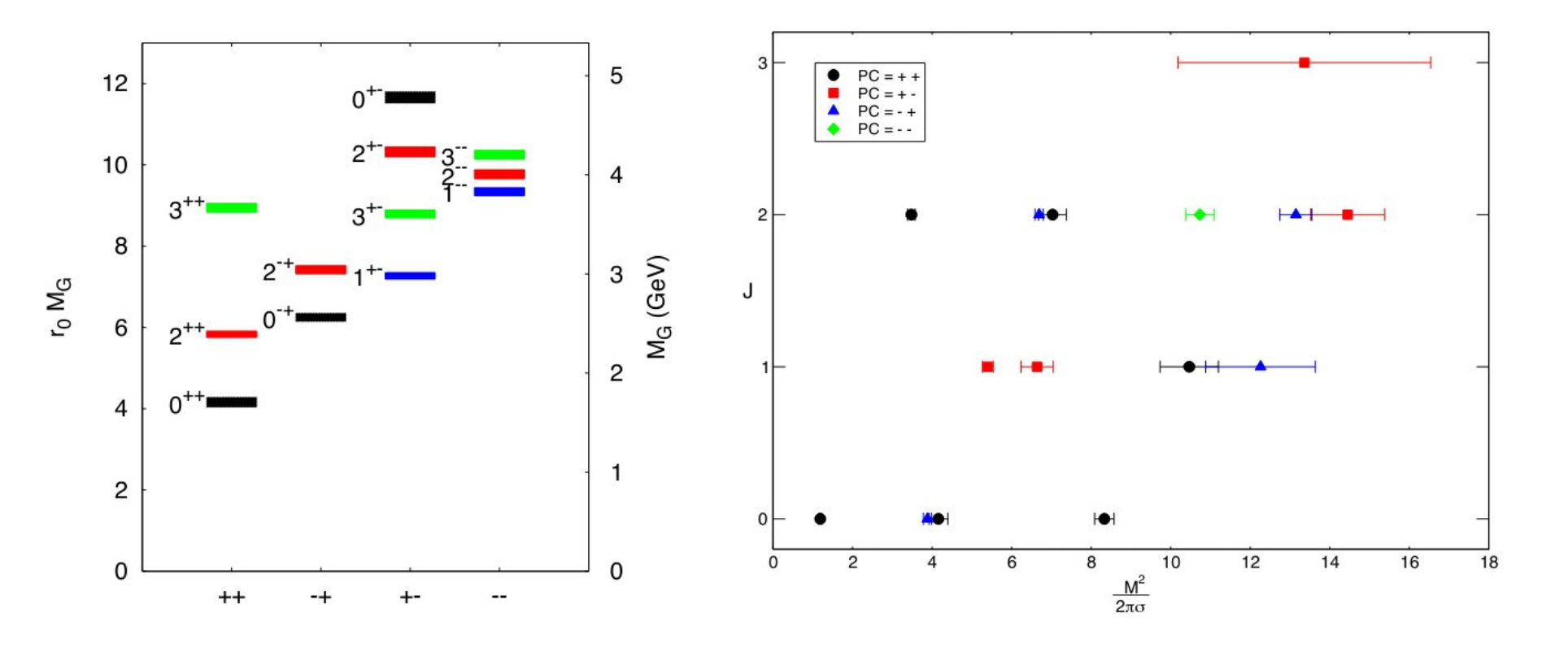}}
\caption{Spectrum of glueballs for pure SU(3) \cite{Chen:2005mg} (left) as a 
function of $J^{PC}$ quantum numbers and for large $N$ \cite{Lucini:2010nv}
showing $J$ versus $m^2$ (right).  The parameters $r_0\sim 1/\Lambda'$ and $
\sigma\sim\Lambda'^2$.
Figure from Ref.\ \cite{Kribs:2016cew}.
}
\label{fig:neil}
\end{figure}

\section{Dark Glueballs}
The simplest nonabelian confining sector is one consisting of gauge bosons
alone.  At temperatures below the confinement scale $\Lambda'$, there is a
lightest glueball state with quantum numbers $0^{++}$, and a spectrum of
excited states, that have been studied on the lattice for SU(3) and SU($N$) gauge
theories \cite{Ochs:2013gi,Kribs:2016cew}.  For example in real-world QCD,
but with quarks omitted, the lightest glueball mass is predicted to be
1750\,MeV \cite{Morningstar:1999rf}.  Taking $\Lambda_{QCD} = 260\,$MeV
\cite{Gockeler:2005rv}, we could expect the lightest glueball mass to scale as
$m_0 = 6.7\,\Lambda'$ for an SU(3) hidden sector with a different confinement
scale.  Glueball spectra for pure SU(3) and large-$N$ SU($N$) as determined by
lattice gauge theory are shown in Fig.\ \ref{fig:neil}.  The first proposal of
hidden sector glueballs as DM was as early as Ref.\ \cite{Faraggi:2000pv},
motivated by hints of DM self-interactions for cosmological structure 
formation and by string theory.

One can quickly be convinced that it is not possible to explain the relic 
density of dark glueballs using conventional thermal freezeout, even if there
is some portal to the SM such as $M^{-1}\phi^2 \bar f f$, where 
$\phi$ is the
effective glueball field, $f$ is a SM fermion and $M$ is a mass scale.
Since $\phi$ carries no global charge, it cannot be stabilized against decay
and the existence of such an operator would imply that $\phi\bar f f$ is 
also present, and generically more important.  Thus glueballs with portal interactions will be unstable,
and if their lifetime is longer than the age of the universe, their
annihilation rate will be even slower. This is borne out in real QCD, where
there is no stable glueball because it mixes with mesons of the same quantum
numbers.   The same argument implies that elastic scattering rates of
glueballs on visible baryons for direct detection are negligible 
\cite{Cline:2013zca}.

\subsection{Relic density}
Elaborating on the previous statements, suppose there is an effective coupling
\be
	{\cal O} = {1\over M^3}G_{\mu\nu}G^{\mu\nu} \bar f f
\ee
between the SU(N)$'$ field strength and a SM fermion, for example.  Then by
dimensional analysis we have matrix elements for decay and scattering of order
\be
	\langle f\bar f|{\cal O}|\phi\rangle \sim {\Lambda'^4\over M^3}\,,
	\quad \langle f\bar f|{\cal O}|\phi\phi\rangle 
\sim {\Lambda'^3\over M^3}\,,
\ee
leading to decay rate and scattering cross section
\be
	\Gamma\sim {\Lambda'^7\over M^6},\quad \langle\sigma v\rangle\sim
	{\Lambda'^4\over M^6}\,.
\ee
Equating $\langle\sigma v\rangle$ to the canonical cross section for thermal
freezeout $\sim 3\times 10^{-26}$\,cm$^3$/s gives $\Lambda'^2\sim 5\times
10^{-5} M^3$/GeV, while demanding that $1/\Gamma$ exceed the age of the
universe requires $\Lambda'^7 \lesssim 10^{-42} M^6$\,GeV.  The nontrivial
solution of these equations is $M\sim 1$\,keV, $\Lambda'\sim 0.01$\,eV, which
is too small for thermal freezeout.

If there is no portal to the SM, then glueballs will form from gluons
when the dark sector temperature $T'$ falls below $\Lambda'$.\footnote{or if
the dark sector has not yet thermalized, then when the density falls below
$n' \sim \Lambda'^3$.}  Their initial
density can be rougly estimated by equating the energy density of gluons to
that of glueballs at the time of the confinement phase transition.  Ref.\
\cite{Soni:2016gzf} first pointed out that $3\to 2$ scattering processes
mediated by an effective operator $\sim \phi^5/(5! N^3\Lambda)$  would determine the
subsequent relic abundance.  The $3\to 2$ process, originally studied for DM
evolution in Ref.\ \cite{Carlson:1992fn} and dubbed ``cannibalism,'' will come
back later in our discussion of dark mesons.  A notable feature of
the mechanism is that it causes the DM to cool more slowly by the conversion of mass into
kinetic energy, if the dark sector is secluded.  In the absence of a portal interaction for keeping the DM in
kinetic equilibrium with the SM, this may result in warm dark matter, which is
now disfavored by Lyman-$\alpha$ constraints \cite{Viel:2013fqw}, as
well as Milky Way satellite counts \cite{Nadler:2021dft} and 
measurements of the dark-matter subhalo mass function in the inner Milky
Way \cite{Banik:2019smi}.    Ref.\
\cite{Soni:2016gzf} finds that the glueballs are cold DM for masses above
1 MeV.

If the hidden sector starts out sufficiently cold, $\xi\ll 1$, the number
density of gluons is suppressed and $3\to 2$ processes may never come into
equilibrium.  In this case the relic glueball density can be estimated by
converting the energy density of gluons at the transition when $T'=\Lambda'$
and $T=T'/\xi$.  This estimate was made in Ref.\ \cite{Boddy:2014yra}, giving
\be
\label{eq:Ogb}
	\Omega_{gb} \sim 4\times 10^8\, {(N^2-1)\over g_{*}}{\Lambda'\over {\rm
	GeV} }\,\xi^3 = 2\times 10^8\, \left({s'\over s}\right){\Lambda'\over {\rm
	GeV} }\,,
\ee 
where $\xi$ and $g_*$ (counting SM degrees of freedom) are evaluated at the 
transition, and $s'/s = 2(N^2-1)\xi^3/g_*$ is the ratio of entropies in the two sectors.
 A small value of $\xi$ is thus also needed for getting the desired 
abundance $\Omega_{gb} = 0.27$, for reasonably large $\Lambda'$.
Interestingly, Refs.\ \cite{Forestell:2016qhc,Forestell:2017wov} get a very similar result
even taking into account $3\to 2$ scattering, 
$\Omega_{bg}= 3\times 10^8\,R\,\Lambda_x$/GeV,\footnote{from digitizing Fig.\ 1 of
their first paper} where $R=s'/s$ is the ratio of entropies
immediately {\it after} the phase transition, and $\Lambda_x$ is identified
with the glueball mass $m_0$.  The strength of the $3\to 2$
transition is estimated using large-$N$ \cite{Manohar:1998xv} and 
naive dimensional analysis (NDA)
arguments \cite{Manohar:1983md} giving the glueball potential
\be
	V(\phi) \sim \sum_{n=2} {1\over n!}\left(4\pi\over N\right)^{n-2}
	\Lambda_x^{4-n}\phi^n\,.
\label{eq:pot}
\ee
In addition to the lowest mass glueball state $0^{++}$, there are many other 
stable excited states, whose relic density has been shown to be much smaller
in Refs.\ \cite{Forestell:2016qhc,Forestell:2017wov}.

\begin{figure}[t]
 \centerline{\includegraphics[width=0.64\linewidth]{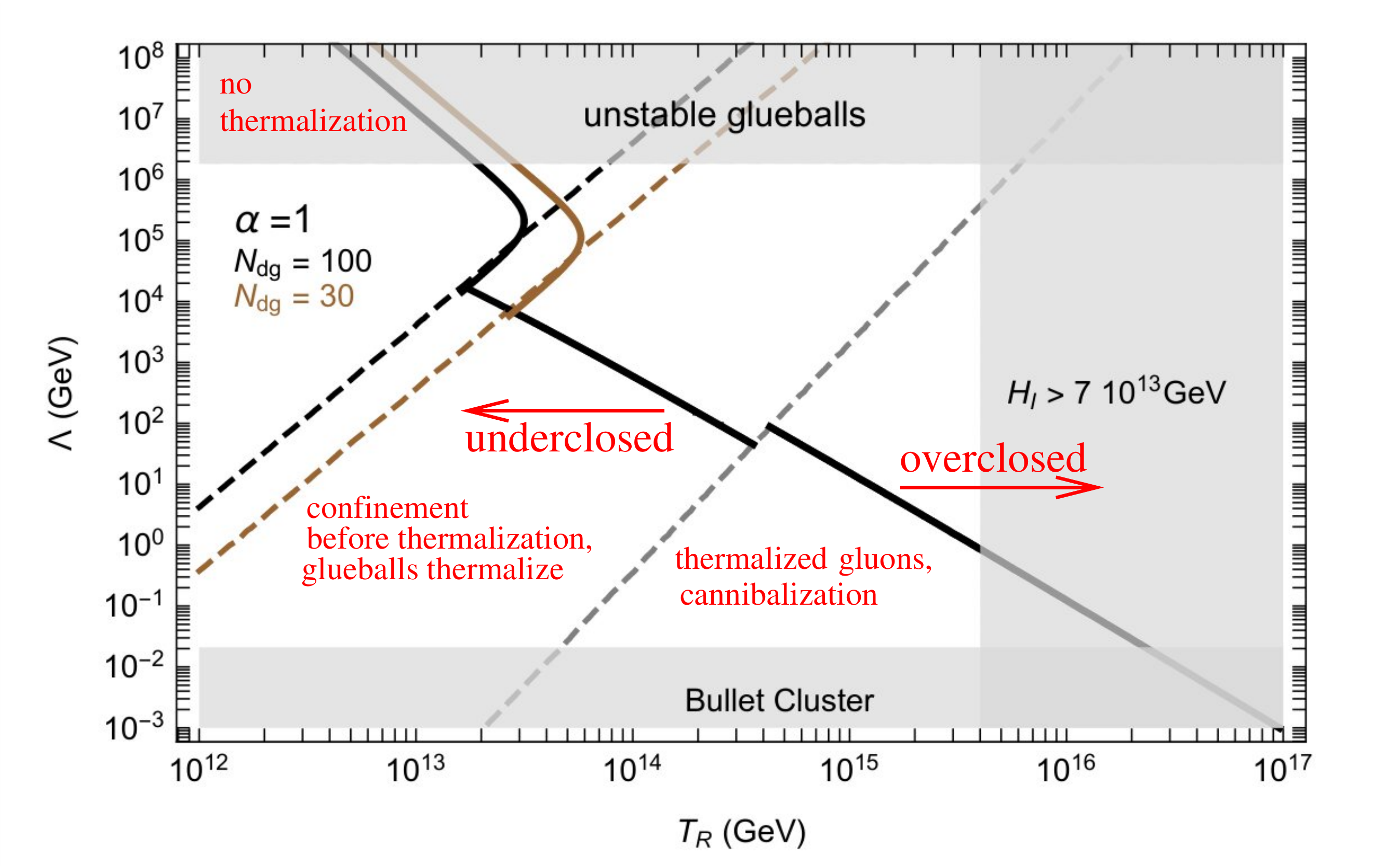}
\includegraphics[width=0.42\linewidth]{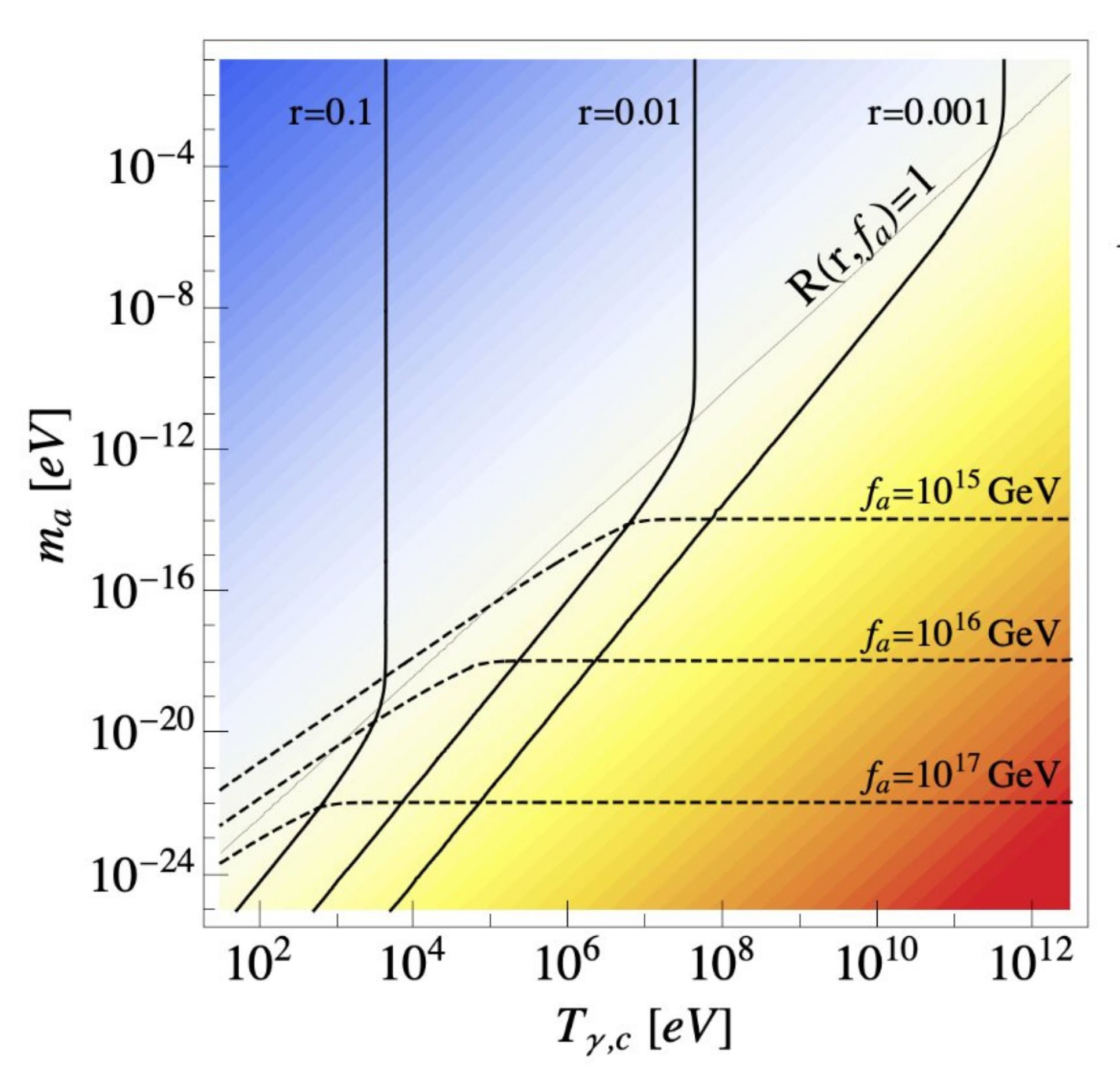}}
\caption{Left: parameter space for gravitational production of dark glueballs, adapted from 
Ref.\ \cite{Redi:2020ffc}.  ``Unstable glueballs'' refers to gravitational
decays (which are unavoidable in any model of dark glueballs)
faster than the present Hubble rate.  $\alpha$ refers to the $3\to 2$
cross section $\langle\sigma v^2\rangle_{3\to 2} \equiv \alpha^3/m_0^3$.
$N_{DG}$ is the assumed number of glueballs produced per gluon in the
phase transition with unthermalized gluons.  Right: regions of 
glueball- (blue) versus axion-dominated (red) DM in the plane of axion mass
versus photon temperature at the time of the confinement phase transition,
from Ref.\ \cite{Jo:2020ggs}.  Here $r=\xi$ and $\xi$, $f_a$ are
adjusted to give observed DM abundance at each point in the plane.
}
\label{fig:redi}
\end{figure}

Eq.\ (\ref{eq:Ogb}) suggests that there would be no production of glueballs
in an inflationary scenario where reheating was purely into SM particles.
But gravity couples to everything, and Ref.\ \cite{Redi:2020ffc} uses
results from conformal field theory to show that
purely gravitational couplings can be sufficient to produce dark glueballs,
depending on the reheat temperature $T_R$.  SM particles annihilating into
an $s$-channel graviton produce dark gluons with a relative abundance going as
$Y' \cong 10^{-6}(N^2-1)(T_R/M_{p})^3$, where $M_p$ is the Planck mass.
One can also compute the relative energy densities in the two sectors.
If $\phi\phi\to\phi\phi$ scattering is fast enough, the dark glueballs will
thermalize before the confinement transition, 
and $3\to 2$ scattering subsequently comes into equilbrium, lowering the 
density.  
However the confinement phase transition could happen before thermalization.
Then the typical gluon is more energetic than $\Lambda'$, and the number of
glueballs produced per gluon $N_{DG}$ depends upon the details of
hadronization, which can affect the final relic density.  These outcomes are
illustrated in Fig.\ \ref{fig:redi} (left).

Ref.\ \cite{Halverson:2016nfq} argues that string theory generically predicts
not just one hidden sector, but many, which can exacerbate the generic problem
that dark glueballs are overproduced unless their sectors are left relatively
unpopulated by reheating after inflation.  An alternative possibility is that
the universe comes to be matter-dominated by moduli and undergoes a second
stage of late reheating by their decays, which could be preferentially into SM
particles \cite{Acharya:2017szw}.  A similar mechanism using domination by
vector-like quarks charged under both SU(N)$'$ and color SU(3) was studied in
Ref.\ \cite{Soni:2017nlm}.

A simple solution to the dark glueball overproduction problem is to include
a coupling of $G_{\mu\nu}$ to axions,
\be
	{g'^2 \over 32\pi^2\,f_a} a\,G_{\mu\nu}\widetilde
G^{\mu\nu}\,.
\ee
The additional interaction allows for a redistribution of abundances to
deplete the glueball density in favor of ultralight axions, resulting in
a two-component DM scenario \cite{Halverson:2018olu,Jo:2020ggs}.
Fig.\ \ref{fig:redi} (right) illustrates the regions of parameter space favoring
glueballs or axions constituting most of the DM.    The dividing line between
glueball versus axion domination is given by the criterion
\be
	R(\xi,f_a) = \left(\xi\over 10^{-2}\right)^2\left(6\times
	10^{13}\,{\rm GeV}\over f_a\right) = 1
\ee
where the temperature ratio $\xi$ is evaluated at time of the confinement
transition.

\begin{figure}[t]
 \centerline{\includegraphics[width=0.7\linewidth]{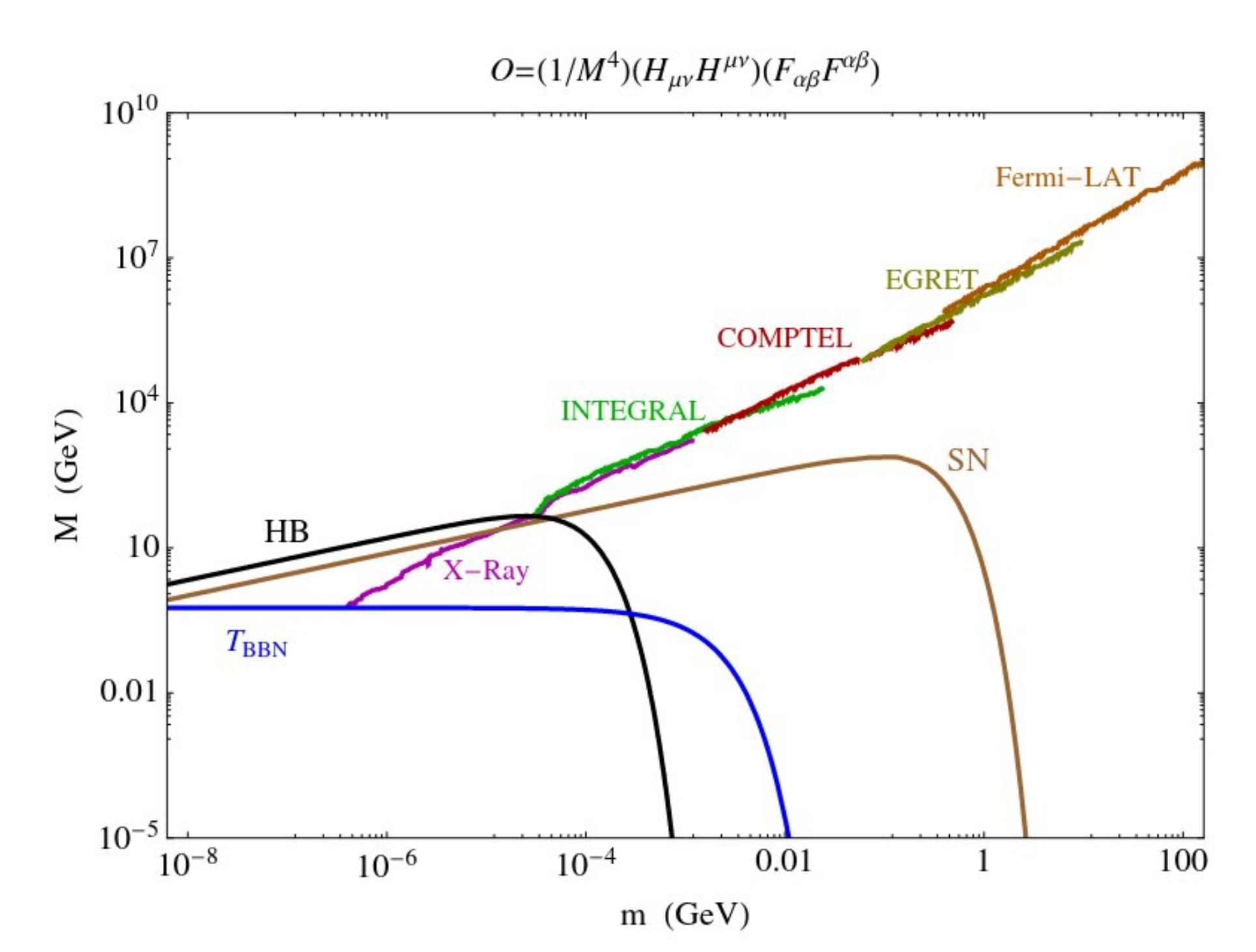}}
\caption{Constraints on the scale $M$ for the $M^{-4}B^2\,{\rm tr}\,{G^2}$
portal versus glueball mass, from cosmic ray 
mononergetic line searches and energy loss in horizontal branch stars
or type II supernovae
\cite{Soni:2016gzf}. }
\label{fig:soni}
\end{figure}

\subsection{Self-interactions and glueballinos}
The glueballs are strongly interacting particles with a geometric cross
section of order $\sigma \sim {4\pi/\Lambda'^2}$;
hence they may be able to address the 
small-scale structure problems of CDM mentioned previously.  Matching to the
desired value of $\sigma/m$ and using the relation between $m$ and $\Lambda'$,
one finds that $\Lambda'\sim 100$\,MeV \cite{Cline:2013zca,Boddy:2014yra}, 
favoring glueballs below the GeV scale.  However these references ignored the
scaling with $N$ and factors of $4\pi$ from NDA.
Taking the potential (\ref{eq:pot}) at face value, one finds a cross section
\be
	\sigma \sim {12\pi^3\over N^4\,m_0^2}
\label{siggpred}
\ee
which would satisfy the criterion (\ref{sigm}) if 
\be
	m_0 \sim 130\,{\rm MeV}\left(3\over N\right)^{4/3}\,.
\ee
Refs.\ \cite{Yamanaka:2019yek,Yamanaka:2019aeq} have computed the
scattering cross section on the lattice for SU(2) glueballs, with
large systematic errors, giving $\sigma = (2-51)/\Lambda^2$.
This is signficantly higher than the prediction (\ref{siggpred}),
which gives $\sigma \cong 0.6/\Lambda^2$.
 
If one takes seriously the indications that DM self-interactions should be
velocity-dependent, glueballs are not the best candidates since they have a
contact interaction leading to constant $\sigma$.  A simple extension is to
include an adjoint fermion $X$ (gluino) which can bind with a gluon to form a
a stable color singlet ``glueballino'' state $\tilde\phi$.  Ref.\ \cite{Boddy:2014yra}
considers the case where the fermion mass $m_X\gg\Lambda'$.  Then glueballinos
are heavier than glueballs, and experience velocity-dependent self-scattering
by virtual glueball exchange.  Glueballinos can undergo thermal freezeout by
$\tilde\phi\tilde\phi\to\phi\phi$ annihilation.

The $\tilde\phi$ DM scenario was also studied in Ref.\ \cite{Contino:2018crt} where it was
called ``gluequark DM,'' with emphasis on the fact that there are generally 
two stages of annihilation: first at the constituent level $XX\to gg$, and
again following the confinement transition through 
$\tilde\phi\tilde\phi\to\phi\phi$.  Moreover if glueballs decay into SM
states, this can significantly dilute abundances in the hidden sector.  In
general one must consider all of these effects to determine the 
relic density.  Ref.\ \cite{Smirnov:2019ngs}  showed that the observed
relic density can be achieved even for glueballino masses as 
high as the PeV scale.  This is well above the conventional
perturbative unitarity constraint for the annihilation cross section
\cite{Griest:1989wd}.

\begin{figure}[t]
 \centerline{{\includegraphics[width=0.75\linewidth]{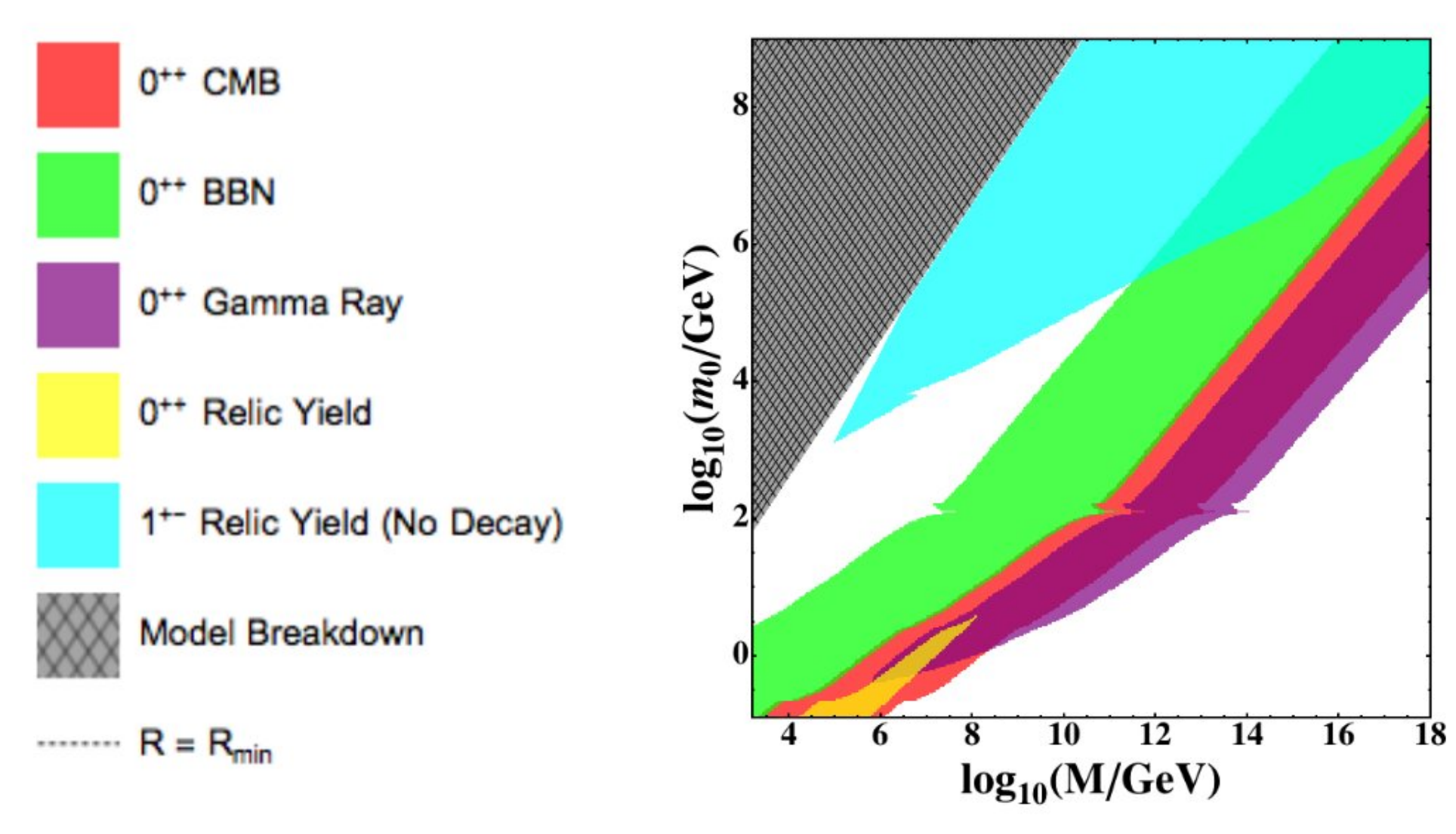}}}
\caption{Example of constraints on $0^{++}$ and $1^{+-}$ glueballs from CMB, BBN
and relic density, assuming a minimal value of the entropy ratio $R =s'/s$ such
that $T'\sim \Lambda'$, conserved dark $C$-parity, and absence of dimension-6
operators in Eq.\ (\ref{eq:ops}), from Ref.\ \cite{Forestell:2017wov}.
``Model Breakdown'' indicates where $m_0>M/10$, and the effective
field theory treatment of the portal interaction may not be valid.}
\label{fig:morris}
\end{figure}

\subsection{Indirect signals}
Various portals connecting glueballs to the SM are possible.  If ${\cal
O}_{SM}$ is a SM gauge singlet operator of dimension $n$, then one can
consider the interaction $M^{3-n}\phi\, {\cal O}_{SM}$ at the effective
field theory level, which mediates glueball decay.  Strong constraints 
on light glueballs arise from
the CMB if $\phi$ can decay into charged particles or photons.  Observations
of cosmic ray photons constrain monoenergetic signals from
$\phi\to\gamma\gamma$ \cite{Soni:2016gzf}, as shown in Fig.\ \ref{fig:soni}.
Moreover
shorter-lived glueball excitations, even if unimportant as DM candidates,
can disrupt big bang nucleosynthesis (BBN) by injecting energy
\cite{Forestell:2016qhc,Forestell:2017wov}.  

Rather than working at scales below the glueball mass, it can be more
theoretically informative to think in terms of portals involving the 
dark nonabelian field strength $G_{\mu\nu}$ and SM U(1) field strength $B_{\mu\nu}$,
Higgs field $H$, or fermions $f$.  Ref.\ \cite{Forestell:2017wov} finds that
the leading operators are
\be
	{1\over M^4}B^2\,{\rm tr}(G^2),\quad {1\over M^4}B_{\mu\nu}
	{\rm tr}(G^3)^{\mu\nu},\quad {1\over M^2}|H|^2 {\rm tr}(G^2)\,.
\label{eq:ops}
\ee
This study emphasized
the relevance of the usually subdominant $1^{+-}$ glueballs, that can be
long-lived and even be the dominant DM. (Unlike the $0^{-+}$ excited
state, the  $1^{+-}$ state is not diluted by coannihilation with 
the ground state, if $C$ is conserved.)
An example of the ensuing constraints 
on glueball mass versus the scale $M$ in Eq.\ (\ref{eq:ops}) is shown in Fig.\
\ref{fig:morris}.  They are sensitive to the initial entropy ratio $R =s'/s$
at the confinement transition
and whether dark $C$-parity is conserved, which would forbid the second
operator in (\ref{eq:ops}).

In addition to constraints, one can address anomalies like the 3.5 keV line
mentioned in section \ref{kevsect}.  Ref.\ \cite{Boddy:2014qxa} noted that
glueballinos with the desired properties for the relic abundance and
self-interactions can also have a hyperfine transition energy of 3.5 keV.
Another interesting example is given by  Ref.\ 
\cite{Jo:2020ggs}, which found that a subdominant, strongly self-interacting glueball
component has the right properties to catalyze early formation of SMBHs,
as discussed in Section \ref{sect:smbh}.

\section{Dark mesons}

If quarks are added to the hidden SU($N$)$'$ sector, then dark 
mesons $\pi'$
become a DM candidate, which can be lighter than the glueballs $\phi$ if the
quark mass $m_{q'}$ is below $\Lambda'$.  Like visible pions, the dark 
$\pi'$ would have quantum numbers $0^{-+}$, and a 
$0^{++}$ glueball could undergo decay as $\phi\to\pi'\pi'$ if 
$2m_{\pi'} < m_\phi$,
leaving $\pi'$ as the sole DM candidate.  Even if decays are kinematically
blocked, annihilations $\phi\phi\to\pi'\pi'$ can greatly deplete the relic
glueball abundance.

With only a single quark flavor and SU(3), one would expect $m_{\pi'}\sim
4\Lambda'$, similar to the $\eta'$ of QCD, which is not light enough to
satisfy  $2m_{\pi'} < m_\phi$, but sufficient for $m_{\pi'}<m_\phi$.  Such a dark
pion would, like the glueball, be unstable to decays into gravitons, with 
amplitude ${\cal M}_{\pi'\to gg} \sim m_{\pi'}^3/M_p^2$ and hence lifetime
\be
	\tau \sim 16\pi\,{M_p^4\over m_{\pi'}^5}\sim \pi\times 10^{17}{\rm s}
	\left(2\times 10^7\,{\rm GeV}\over m_{\pi'}\right)^5\,,
\ee
showing that metastability on cosmological timescales imposes a modest
requirement on the mass.

With more flavors, absolute stability becomes possible, and the pions could
be pseudo-Nambu-Goldstone bosons (pNGBs) from spontaneously broken chiral symmetry. 
The stability criterion (even with quarks that are also coupled to the SM SU(2)) can be formalized analogously to QCD
in terms of $G$-parity: the lightest $G$-odd pion (LGP) is stable 
\cite{Bai:2010qg} for SU($N$)$'$ with $N\ge 3$.  In the SM this would be the
$\pi^0$, in the absence of electromagnetism.  It is not gravitationally
stable, but if couplings to weak interactions were turned off, then $\pi'^\pm$
would become degenerate with $\pi'^0$, and the former would be completely stable.

In addition to pseudoscalar mesons, there will be heavier vector mesons.
In some circumstances they could be the primary DM candidates.
Ref.\ \cite{Hambye:2009fg} presents a model with SU(2)$'$ and a 
complex scalar doublet $\phi$ that can have a Higgs portal coupling 
$\lambda|H|^2|\phi|^2$.  This allows the lighter scalar pion $\phi
^\dagger\phi$ to decay into
SM particles, but leaves the vector stable since a vector cannot mix with the
Higgs.  Even if the vector mesons are not DM, they can play an important role
in the freezeout process, as we will see.

\begin{figure}[t]
 \centerline{{\includegraphics[width=\linewidth]{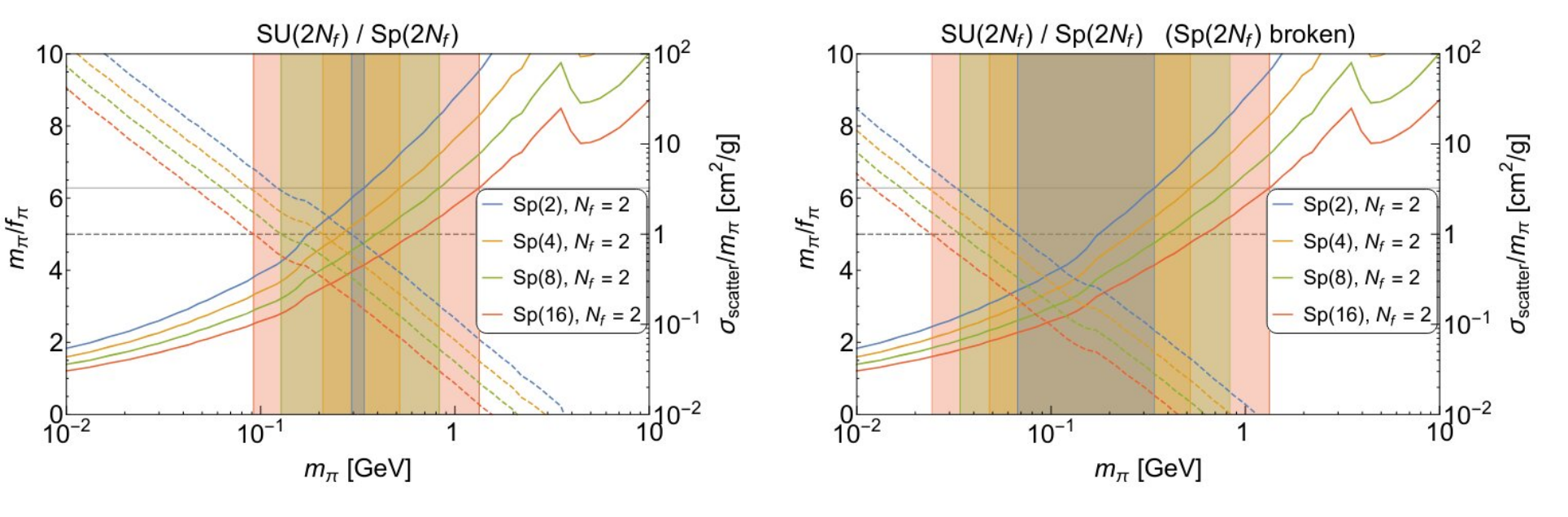}}}
\caption{Allowed ranges of mesonic DM produced by the SIMP mechanism, for
different confining Sp($N$) gauge groups, from Ref.\ \cite{Hochberg:2014kqa}.   Left: unbroken flavor symmetry
(degenerate quark masses); right: broken flavor symmetry (lifting degeneracy
of $\pi'$ masses).  }
\label{fig:hoch}
\end{figure}

\subsection{Relic density}
 
Chiral Lagrangians (see for example Ref.\ \cite{Georgi:1984zwz}) are the appropriate effective theory for dark mesons
that are pNGBs like the SM pseudoscalar octet.  They are constructed from
the matrix $\Sigma = \exp(i\pi'/f)$, where $\pi' = \pi'^a T^a$ and $T^a$ are the 
generators of the flavor symmetry that is spontaneously broken by
$\langle\Sigma\rangle_{ij} = \delta_{ij}$, which is proportional to the matrix of quark
condensates $\langle \overline q'^i q'^j\rangle$ in flavor space.  Including the
symmetry-breaking quark mass matrix $M$, the leading terms in the chiral
Lagrangian are
\be
	f^2\,{\rm tr} (\partial_\mu\Sigma^\dagger\partial^\mu\Sigma) -
f^3\, {\rm tr}(M\Sigma + {\rm H.c.})\,,
\label{chiral-lag}
\ee
giving the pions a mass $m_{\pi'}^2 = fM$.  $f$ is known as the pion
decay constant ($f\sim m_\pi$ in the real world): the hadronic matrix element of the axial 
quark currents can be parametrized as
\be
	\langle 0|\bar q \gamma^\mu\gamma_5T^a q|\pi'^b\rangle \sim 
	f p^\mu \delta_{ab}\,.
\ee
This assumes that chiral
symmetry breaks to SU(N); if it breaks to Sp(N) then 
$\langle\Sigma\rangle_{ij} = J_{ij}$, where $J$ is a symplectic
matrix. 

\subsubsection{Thermal $2\to 2$ freezeout}

The early reference \cite{Bhattacharya:2013kma} considered the portal
interactions
\be
	\lambda_h(|H|^2-v^2){\rm tr} (\partial_\mu\Sigma^\dagger\partial^\mu\Sigma)
	\quad+\quad {\lambda_v\over f} B^{\mu\nu} {\rm tr} 
(M\Sigma\partial_\mu\Sigma^\dagger\partial_\nu\Sigma + {\rm H.c})
\ee
to the Higgs and the hypercharge field strength.  The Higgs portal
enables annihilations $\pi'\pi'\to WW,\,HH$ for thermal freezeout.  The
hypercharge portal would allow for $Z\to \pi'^0\pi'^+\pi'^-$, for example, but
this is kinematically forbidden if the $2\to 2$ processes are allowed.
Such heavy pions would be incompatible with strongly self-interacting DM
(SIDM); see below.

Ref.\ \cite{Cline:2013zca}, motivated by SIDM to consider $m_{\pi'}\sim 30\,$MeV, suggested annihilation 
$\pi'\pi'\to Z'Z'$ into light $Z'$ gauge bosons via a 
$F'_{\mu\nu}F'^{\mu\nu}
{\rm tr}(\partial\Sigma^\dagger\partial\Sigma)$ interaction.  However one needs to keep $Z'$
in thermal equilibrium with the SM for standard freezeout.
Using kinetic mixing to allow for $Z'\to e^+e^-$ leads to conflict
with CMB constraints because of late $\pi'\pi'\to Z'Z'\to 4e$ annihilations.
Taking the $Z'$ to be massless with sufficiently small kinetic mixing
can overcome these problems.

Confining SU(2) models are special since the ``baryons'' are scalars
like the mesons, and differ only in terms of which conserved
quantum numbers assure their stability.  Ref.\ \cite{Buckley:2012ky} considered SU(2) with two
flavors of quarks, $Q_u$ and $Q_d$, assigned equal and opposite 
SM hypercharge.  The baryons $N = Q_u Q_d$ and $\bar N = \overline Q_u \overline Q_d$ are 
stable (and neutral) DM candidates, while
the mesons $\pi'^0$, $\pi'^\pm$ can be made unstable to decays into SM
fermions $f$.   Then $N\bar N\to \pi' \pi'$ along with $\pi'\to f\bar f$
can achieve the desired relic density through conventional freezeout.

\subsubsection{Thermal $3\to 2$ freezeout}

A qualitatively different means of freezeout was proposed in Ref.\ 
\cite{Hochberg:2014kqa}, based on the Wess-Zumino-Witten (WZW) interaction
\be
	{2 N\over 15\pi^2 f_{\pi'}^5}\epsilon^{\mu\nu\rho\sigma}
	{\rm tr}(\pi' \partial_\mu\pi'\partial_\nu\pi'\partial_\rho\pi'
	\partial_\sigma\pi')
\label{eqWZW}
\ee
(where $f_{\pi'}\sim f$ up to factors of 2).  It is a topological term, that only
exists in theories where the 5th homotopy group $\pi_5({\cal G}/{\cal H})$ is nontrivial.
These include the gauge groups SU($N$) and SO($N$) if the number of flavors
$N_F\ge 3$, and Sp($N$) if $N_f\ge 2$.  Notice that $N$ must be even
in the case of Sp($N$).
Then the $3\to 2$ process
(cannibalization), previously discussed for glueballs, becomes possible.
This is an example of the SIMP mechanism introduced in Ref.\
\cite{Hochberg:2014dra}.  This mechanism assumes that the DM is in thermal
equilibrium with the SM at the time of freezeout, but how this is accomplished
for the dark pion model is not discussed in Ref.\ \cite{Hochberg:2014kqa}.

The $3\to 2$ cross section from the WZW interaction scales as
\be
	\langle\sigma v^2\rangle_{3\to 2}\equiv {\alpha_{\rm eff}^3\over m_{\pi'}^5}
	\sim {N^2\,N_f^5\, m_{\pi'}^5\over f_{\pi'}^{10}}\,.
\ee
  In place of a
dimensionless coupling, the ratio $m_{\pi'}/f_{\pi'}$ determines the strength of the
interaction, and chiral perturbation theory breaks down for
$m_{\pi'}/f_{\pi'}\gtrsim 2\pi$.  This puts an upper limit on $m_{\pi'}$ for which the
relic density is small enough.  On the other hand the self-interactions
(discussed below) put a lower limit on $m_{\pi'}$.  This gives rise to somewhat
narrow ranges for $m_{\pi'}\sim 30-1000$\,MeV, depending on the numbers of colors
and flavors, and also on whether the flavor symmetry is exact or broken.
These ranges are illustrated in Fig.\ \ref{fig:hoch}.

\begin{figure}[t]
 \centerline{\includegraphics[width=0.5\linewidth]{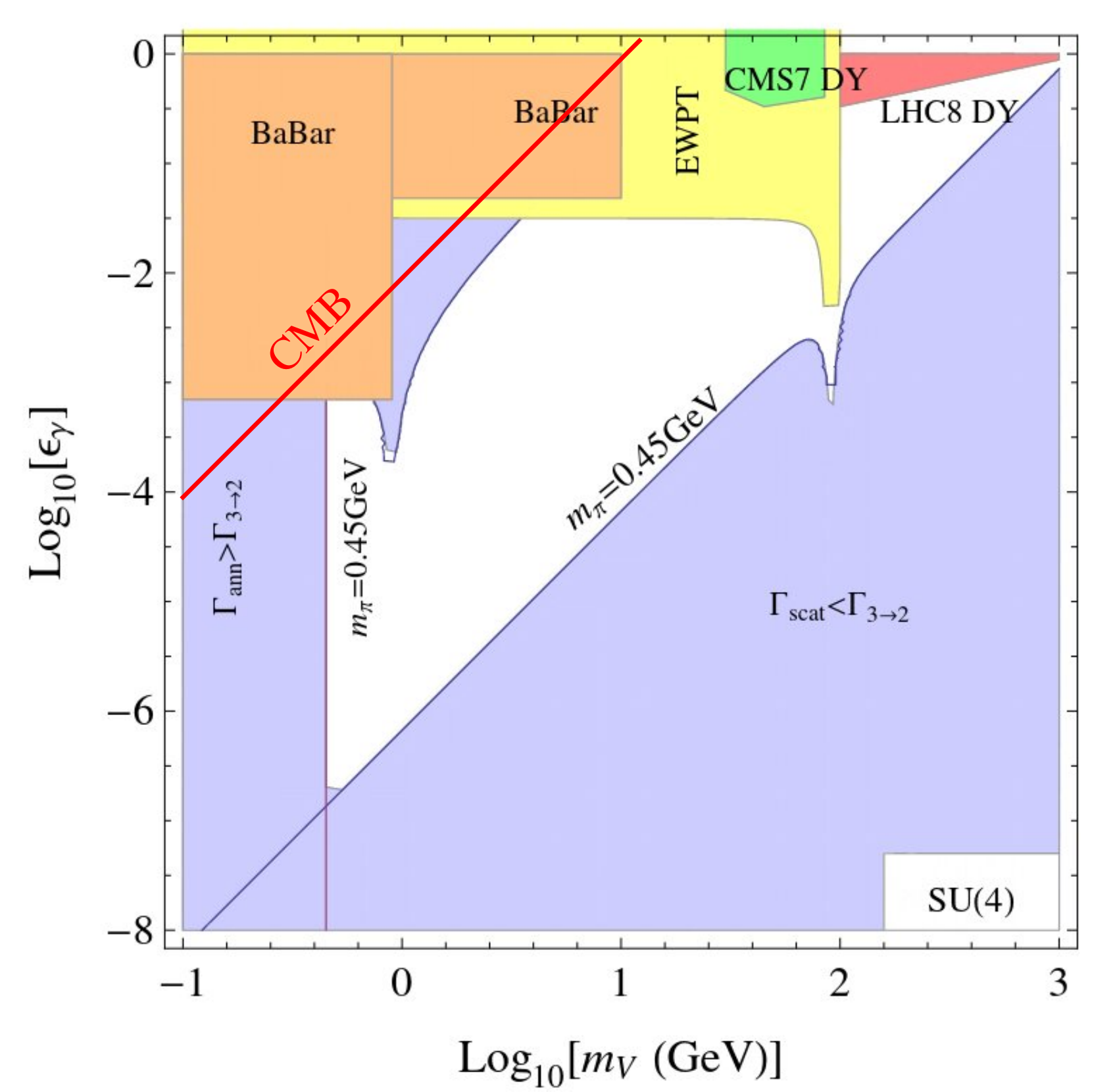}
\includegraphics[width=0.61\linewidth]{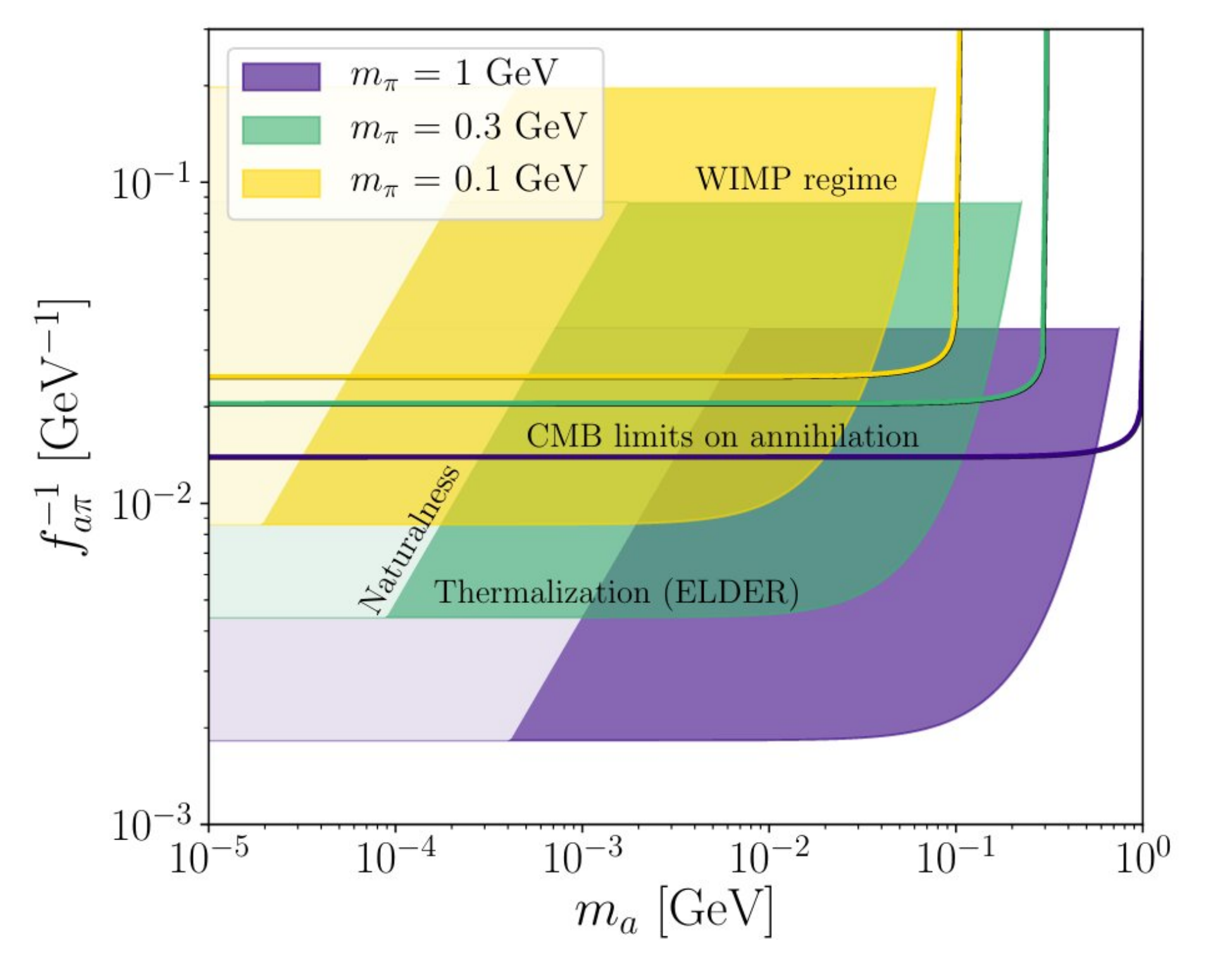}}
\caption{Left: Allowed ranges (unshaded) of kinetic mixing $\epsilon$ versus dark photon
mass $m_{Z'}$ for keeping dark pions in equilibrium with the SM during SIMP
production, adapted from Ref.\ \cite{Lee:2015gsa}. The U(1)$'$ gauge coupling
is taken to be $\alpha'=1/4\pi$.  Right: allowed regions (shaded) for
axion-mediated thermalization of SIMP-produced dark pions, from Ref.\ 
\cite{Hochberg:2018rjs} }
\label{fig:lee}
\end{figure}

\subsubsection{Portals for thermal equilibration}

To study the effect of a mediator to maintain kinetic equilibrium between the
$\pi'$ and SM sectors in the SIMP scenario, Ref.\ \cite{Lee:2015gsa} charged
the quarks under a dark U(1)$'$, assuming $N_f=3$ flavors and charge matrix
$Q = {\rm diag}(1,-1,-1)$, chosen to cancel mixed anomalies of the AVV
type between the global axial and vector flavor currents.  This suppresses
the decay of  the $\pi^0$- and $\eta^0$-like mesons into $Z'Z'$.\footnote{See
the discussion in Sect.\ \ref{sect:VM}.}
  Kinetic mixing of the $Z'$ with the SM hypercharge
can keep the two sectors in equilibrium.  The interactions of $Z'$ with the
pions is obtained from chiral perturbation theory by covariantizing
the derivatives, $\partial_\mu\Sigma\to \partial_\mu\Sigma +
ig'[Q,\Sigma]Z'_\mu$, yielding standard U(1) couplings to the charged
pion currents and seagull terms.  Ref.\ \cite{Lee:2015gsa} analytically
estimated the $\pi'$ abundance from $3\to 2$ freezeout, obtaining the observed
value for pion masses
\be
	m_{\pi'} = 0.03\,\alpha_{\rm eff} (T^2_{eq}\, M_P)^{1/3} \sim 35-350\,{\rm
MeV}
\ee
where $T_{eq} = 0.8\,$eV is the matter-radiation equality temperature, and the
range of $m_{\pi'}$ is from taking $\alpha_{\rm eff} = 1-10$.  The $Z'$ is taken
to be heavier than $m_{\pi'}$ so that $\pi'\pi'\to SM$ annihilations are suppressed
by $Z'$ propagators, as well as kinetic mixing $\epsilon$, and can be
subdominant to $3\to 2$ annihilations for small enough $\epsilon$.  Yet
$\epsilon$ must be large enough to maintain kinetic equilibrium between the
two sectors through $\pi'$-SM elastic scattering. This leads to allowed regions in the plane of $\epsilon$ and
$m_{Z'}$ like in Fig.\ \ref{fig:lee} (left).  The CMB bounds (see for example Ref.\
\cite{Galli:2011rz}) were not considered in Ref.\ \cite{Lee:2015gsa}, but
my estimate (red line) shows that they are less constraining than BaBar.

Another means of thermal equilibration is through an axion coupling to the
dark quarks.  Ref.\ \cite{Hochberg:2018rjs} extended the earlier model of
\cite{Hochberg:2014kqa}, noting that the $a\pi'^3$ coupling vanishes in Sp(2N)
theories, avoiding semi-annihilation processes $\pi'\pi'\to\pi' a$, but
the $\pi'^2 a^2$ interaction exists.  Depending on the axion mass $m_a$ and its
coupling to pions,  $\sim(m_{\pi'}/f_{a\pi'})^2$, $\pi'\pi'\to a a$ annihilation can be subdominant to 
$\pi'\pi'\pi'\to\pi'\pi'$, while equilibration with the SM can be maintained if the
axion-photon coupling $f_{a\gamma}^{-1}$ is large enough.  Allowed regions of
the $f_{a\pi'}^{-1}$-$m_a$ parameter space are shown in Fig.\ \ref{fig:lee} 
(right).  The CMB constraint is relevant here, reducing the allowed region
for light $m_{\pi'} \sim 0.1\,$GeV.

\subsubsection{Role of vector mesons}
\label{sect:VM}

Models with dark pions inevitably have heavier vector meson states $V$ 
as well,
which can play a role in freezeout.  Ref.\ \cite{Choi:2018iit} showed that
vector exchange in the $3\to 2$ annihilations can be near resonance, which
allows for a higher range of possible $m_{\pi'}\lesssim 1\,$GeV through the SIMP mechanism,
without having to resort to nonperturbative couplings in which the chiral
perturbation expansion is breaking down.

\begin{figure}[t]
 \centerline{\includegraphics[width=0.75\linewidth]{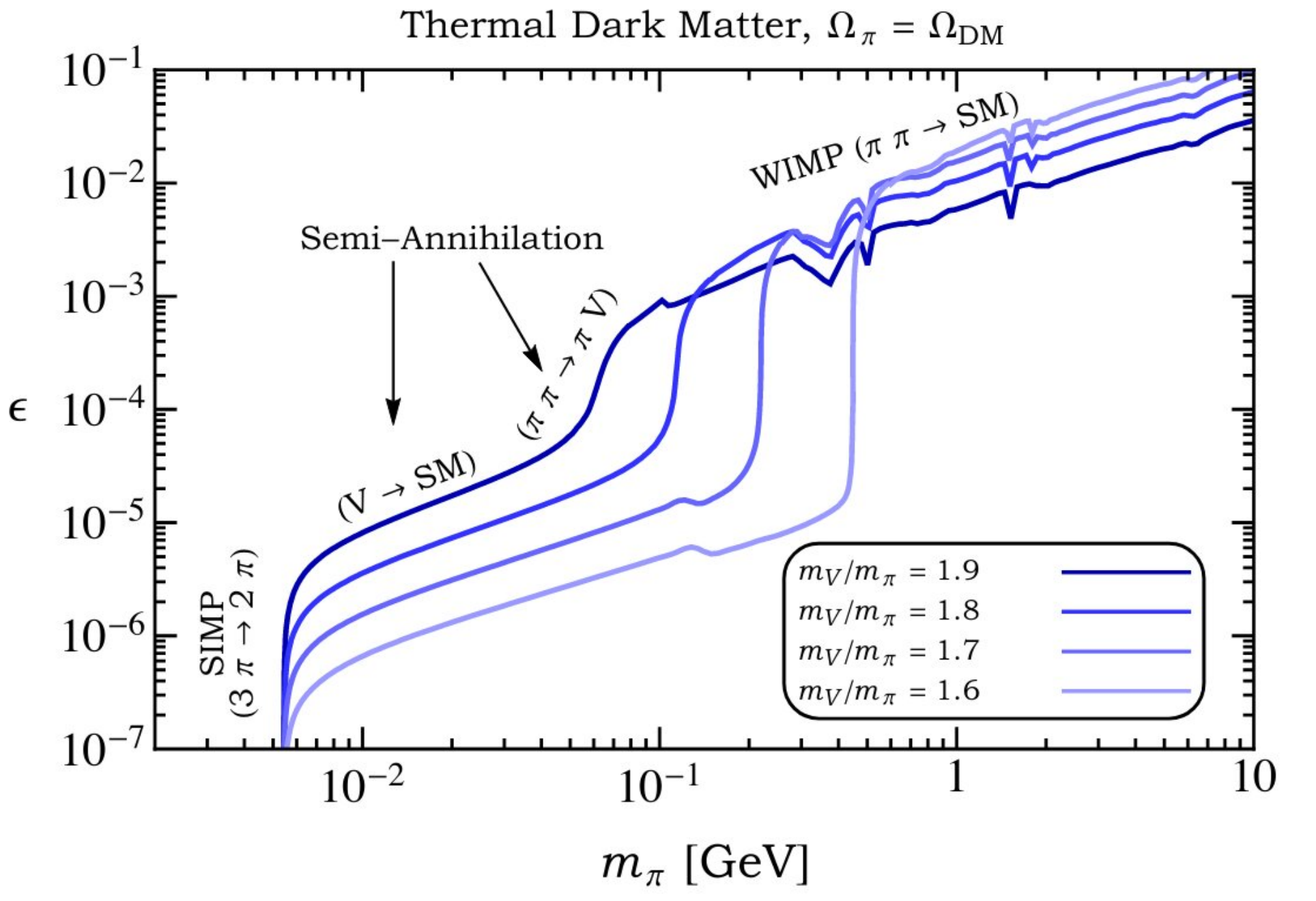}}
\caption{Required value of kinetic mixing versus $m_{\pi'}$ to obtained the
observed DM density, taking account of vector meson effects, from Ref.\ 
\cite{Berlin:2018tvf}.}
\label{fig:berlin}
\end{figure}

Ref.\ \cite{Berlin:2018tvf} observed that the semi-annihilation process
$\pi'\pi'\to \pi' V$ can often dominate over $3\to 2$ annihilation, followed
by the vector decaying into SM particles. They use the same setup as in
Ref.\ \cite{Lee:2015gsa},  but take into account the effects of the vector
mesons, whose mass is expected to be $m_V\sim 4\pi f_{\pi'}/\sqrt{N}$, and
can mix with the $Z'$, similar to  $\rho$-$\gamma$ mixing in the SM. 
Generically $m_V$ could be less than $2 m_{\pi'}$, in which case $V\to 2\pi'$
is blocked, and $V$ will instead decay to SM fermions through its mixing
to $Z'$ and kinetic mixing $\epsilon$ of $Z'$ with the photon.  Ref.\
\cite{Berlin:2018tvf} also emphasized that the cancellation of the chiral
anomaly by the choice of quark charges $Q = {\rm diag}(1,-1,-1)$ is not
sufficient for stability of $\pi'^0$, so that only the charged states $\pi'^\pm$
will be the stable DM.  In the parameter space of $\epsilon$ versus
$m_{\pi'}$, the $3\to 2$ mechanism for thermal freezeout is seen to occupy a
relatively small region in  Fig.\ \ref{fig:berlin} when the effects of the
vector are taken into account.  At large $\epsilon$, freezeout is
dominated by $\pi'^+\pi'^-\to Z',V\to $ SM ($f\bar f$).  The latter cross
section is $p$-wave suppressed and therefore does not lead to strong CMB
constraints.

\subsection{Composite Higgs models}
\label{sect:ch}
Composite Higgs models provide a compelling motivation for dark mesons as DM,
in contrast to a secluded hidden sector.  Analogously to QCD, techniquarks with
an approximate flavor symmetry ${\cal G}$ that breaks to ${\cal H}$ when the
confining technicolor interaction creates a techniquark condensate, give rise
to pNGBs corresponding to the broken generators of ${\cal G}/{\cal H}$.
Some of these should correspond to the Higgs boson, and if there are
additional ones, they can be DM candidates \cite{Ryttov:2008xe}.

A related example is the gauge group SU(2) with $N_f=2$ Dirac flavors
\cite{Lewis:2011zb}.  For massless quarks, this has the flavor
symmetry SU(4), since each Dirac field has two chiralities, and
lattice studies show it breaks to Sp(4), giving 5 Goldstone bosons. 
Three of these can be used for electroweak symmetry breaking (EWSB),
leaving two as scalar DM candidates. This model does not actually have
a composite Higgs (not obviously); this is why two rather than one of
the extra Goldstone bosons are DM, and since they appear as components
of a complex scalar, they can be asymmetric DM.

\begin{figure}[t]
 \centerline{\includegraphics[width=0.55\linewidth]{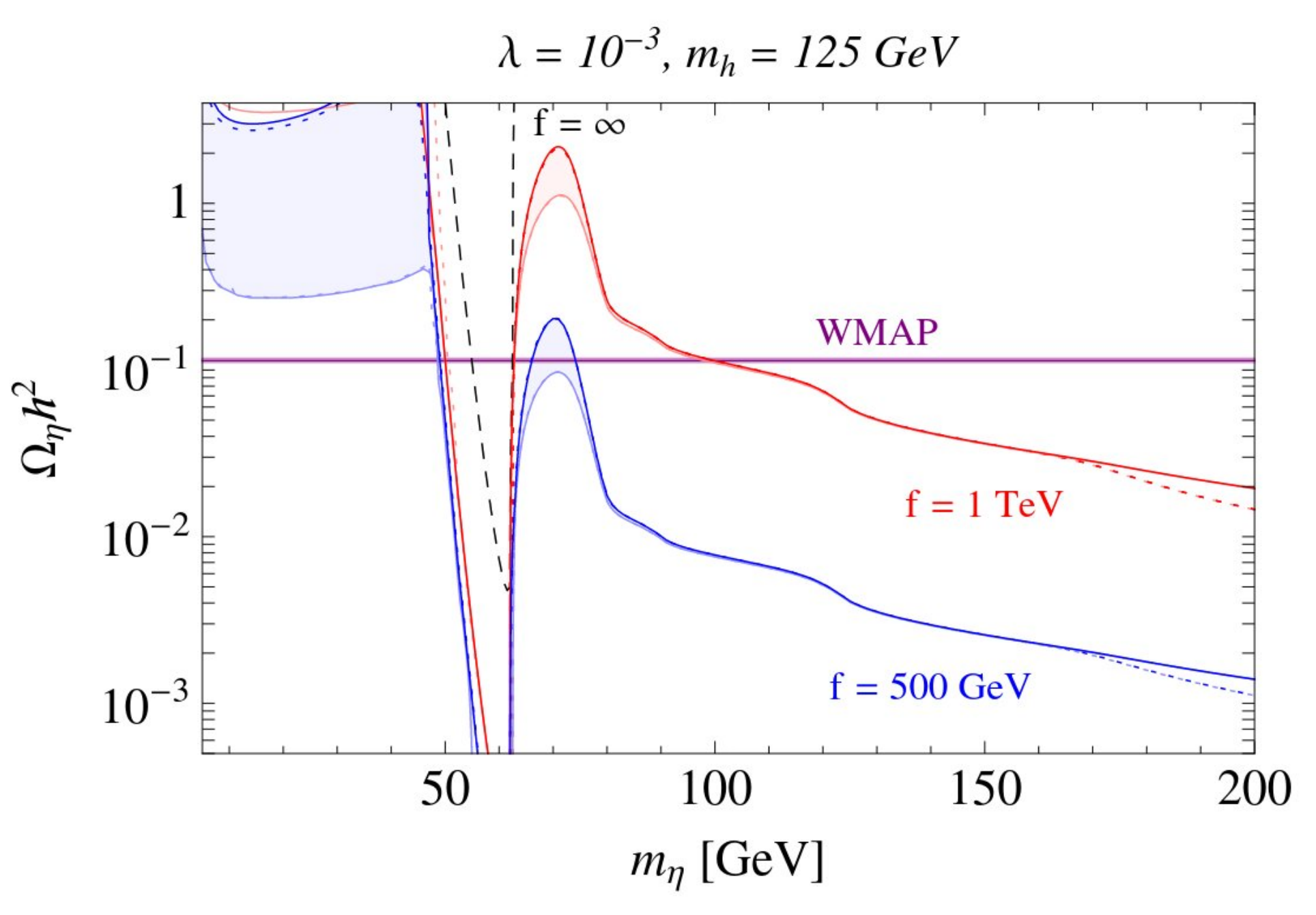}
 \includegraphics[width=0.55\linewidth]{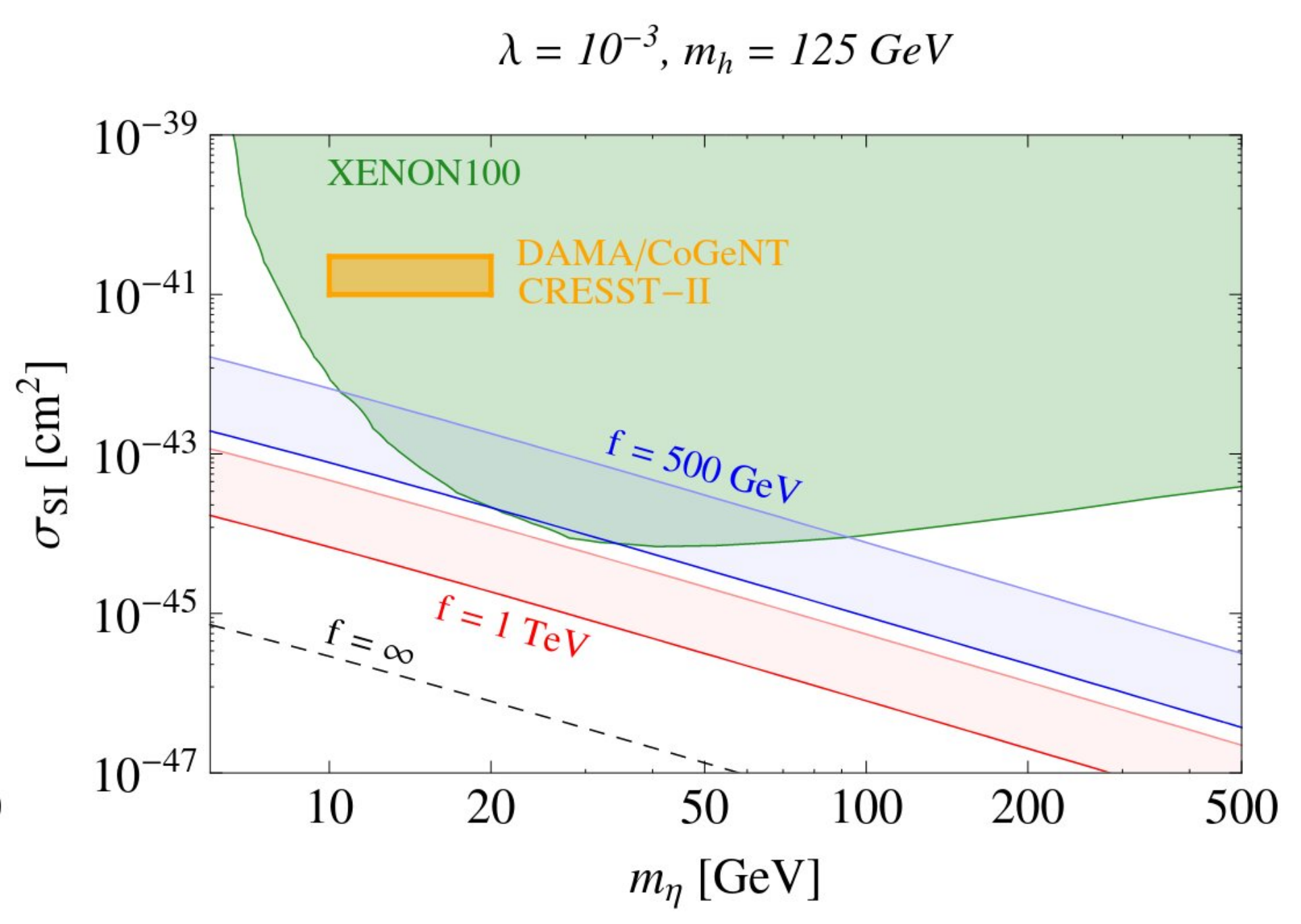}}
\caption{Predictions for the relic density and direct detection for a
dark meson $\eta'$ from 
a SO(6)$\to$ SO(5) composite Higgs model, from Ref.\ 
\cite{Frigerio:2012uc}.}
\label{fig:frig}
\end{figure}

Another popular global symmetry breaking pattern is SO(6)$\to$SO(5), 
which also has five Goldstone bosons, four of which are identified with
the complex Higgs doublet, leaving one as a DM
candidate.\footnote{However it is not generally stable, without
additional global symmetries, due to the WZW interaction (analogous to
that for $\pi^0\to\gamma\gamma$ decay)  which allows
it to decay into electroweak gauge bosons.  This can be overcome by 
taking the coset structure SO(7)/SO(6) \cite{Balkin:2017aep}.  The
resulting DM is stabilized by a dark U(1) symmetry.}  Unlike the 
simpler minimal composite Higgs model \cite{Agashe:2004rs} which has
SO(5)$\to$SO(4), it can be  UV-completed in a techniquark setting
\cite{Frigerio:2012uc}.  In this model the DM meson $\eta'$ has derivative
couplings to the Higgs, $\partial_\mu\eta'^2\,\partial^\mu |H|^2/f^2$, 
standard Higgs portal couplings $\lambda\eta'^2|H|^2$ and couplings to 
SM fermions, $\sim(\eta'/f)^2 y_f\overline Q_f H f$, that allow for
$\eta'\eta'\to f\bar f,$ $HH$ to give thermal freezeout in two different
mass regimes: 50-70\,GeV (with Higgs resonance from the $\lambda v h
\eta'^2$ interaction dominating) 
and 100-500\,GeV (with derivative couplings dominating), illustrated in Fig.\
\ref{fig:frig}.  The value of the portal coupling $\lambda$ is not
predicted, and direct detection rules out much wider ranges of $m_{\eta'}$
when $\lambda=0.1$.

\begin{wrapfigure}[20]{r}[0pt]{0.5\textwidth}
\centering
\vskip-0.5cm
\includegraphics[width=0.5\textwidth]{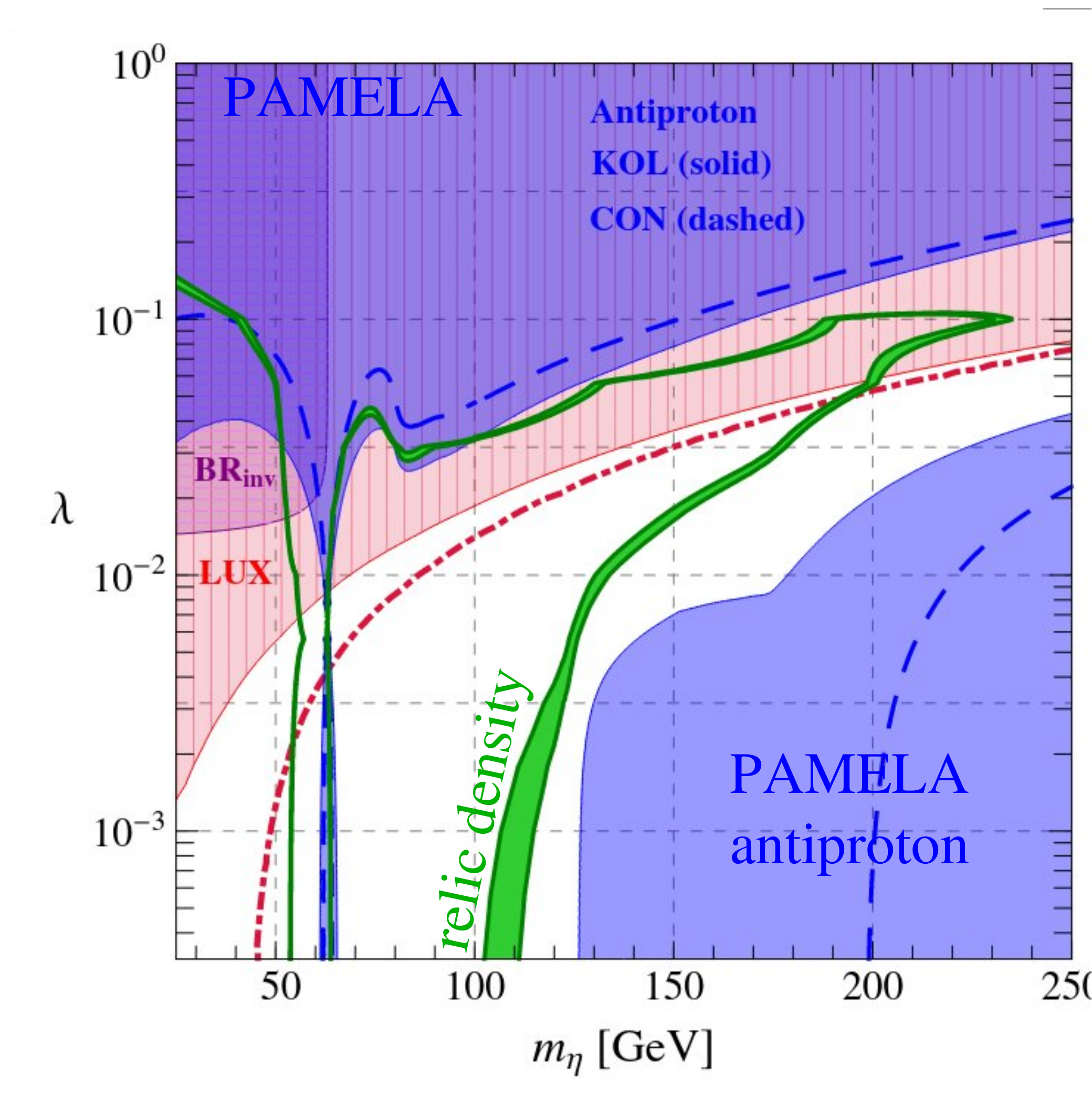}
\vskip-0.5cm
\caption{\label{fig:urb} Combined relic density, direct and indirect
detection constraints, again for the SO(6)$\to$SO(5) model, from Ref.\
\cite{Marzocca:2014msa}.}
\end{wrapfigure}

The same class of models was further examined in Ref.\ 
\cite{Marzocca:2014msa}, focusing on LHC constraints.  Searches for
composite vector resonances, which mix with the SM weak gauge bosons,
exclude low values of the decay constant $f < 800\,$GeV and hence
lower $\eta'$ masses.  For $f=1.1$\,TeV, $m_{\eta'} \sim 100-200$\,GeV is 
predicted, as shown in Fig.\ \ref{fig:urb}.  It is seen that indirect
constraints, in this case production of antiprotons from the primary
annihilation products from $\eta'\eta'$ annihilation in the galaxy, exclude
much of the $\lambda$ versus $m_{\eta'}$ parameter space.

Other possible coset structures for composite dark sectors have been
explored in Ref.\ \cite{Carmona:2015haa}, including
[SU(2)$^2\times$U(1)]/ [SU(2)$\times$U(1)] 
and SU(3)/[SU(2)$\times$U(1)].\\
Their low energy effective descriptions are inert Higgs doublet or
triplet DM models, respectively.  Ref.\ \cite{Wu:2017iji} studied
the SU(4)$\times$SU(4)/SU(4) model, which has 15 Goldstone bosons, and
predicts composite DM with mass 500-1000\,GeV.

\subsection{Self-interactions}
One of the first motivations for dark mesons was to account for strong
DM self-interactions for structure formation.  Ref.\ 
\cite{Cline:2013zca} computed
the $\pi'\pi'$ elastic scattering cross section from the chiral
Lagrangian (\ref{chiral-lag}), with a different normalization
$F_{\pi'} = 2f$ (such that $F_{\pi} = 93\,$MeV for QCD), to find
\be
	\sigma = {m_{\pi'}^2\over 32\pi F_{\pi'}^4}\left(2N_f^4-25 N_f^2+ 90
	-65/N_f^2\over N_f^2 -1\right)
\label{sigmapi}
\ee
for $N_f$ flavors.  To relate $m_{\pi'}$ and $F_{\pi'}$ to the more
fundamental parameters $N$ and $\Lambda'$, lattice gauge theory
calculations would be required \cite{Kribs:2016cew,DeGrand:2019vbx}.
Desired values of $\sigma/m$ can be attained for a range of masses
$m_{\pi'} =30$--100\,MeV, for $N_f = 2$--6.  More generally, Bullet
Cluster constraints put a lower bound on mesonic DM masses of this
order.  

Ref.\ \cite{Hochberg:2014kqa} discusses the analogous
result to (\ref{sigmapi}) for the case where flavor symmetry is
strongly broken by the quark masses so that there is a single lightest
state that dominates the scattering, obtaining $\sigma =
a^2\,m_{\pi'}^2/(32\pi f_{\pi'}^4)$ (note the different normalization of
$f_{\pi'}$, as in Eq.\ (\ref{eqWZW})), where $a\sim 2$ for SU($N$) and
O($N$) gauge theories, and $a\sim 1$ for Sp($N$).

\begin{figure}[t]
 \centerline{\raisebox{3cm}{\includegraphics[width=0.25\linewidth]{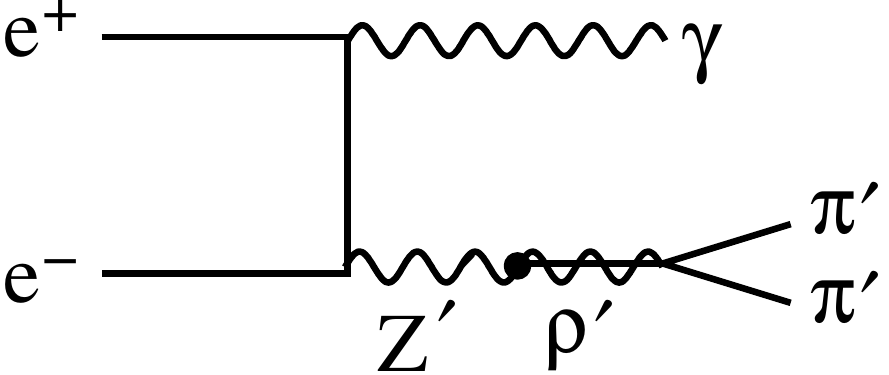}}
\includegraphics[width=0.75\linewidth]{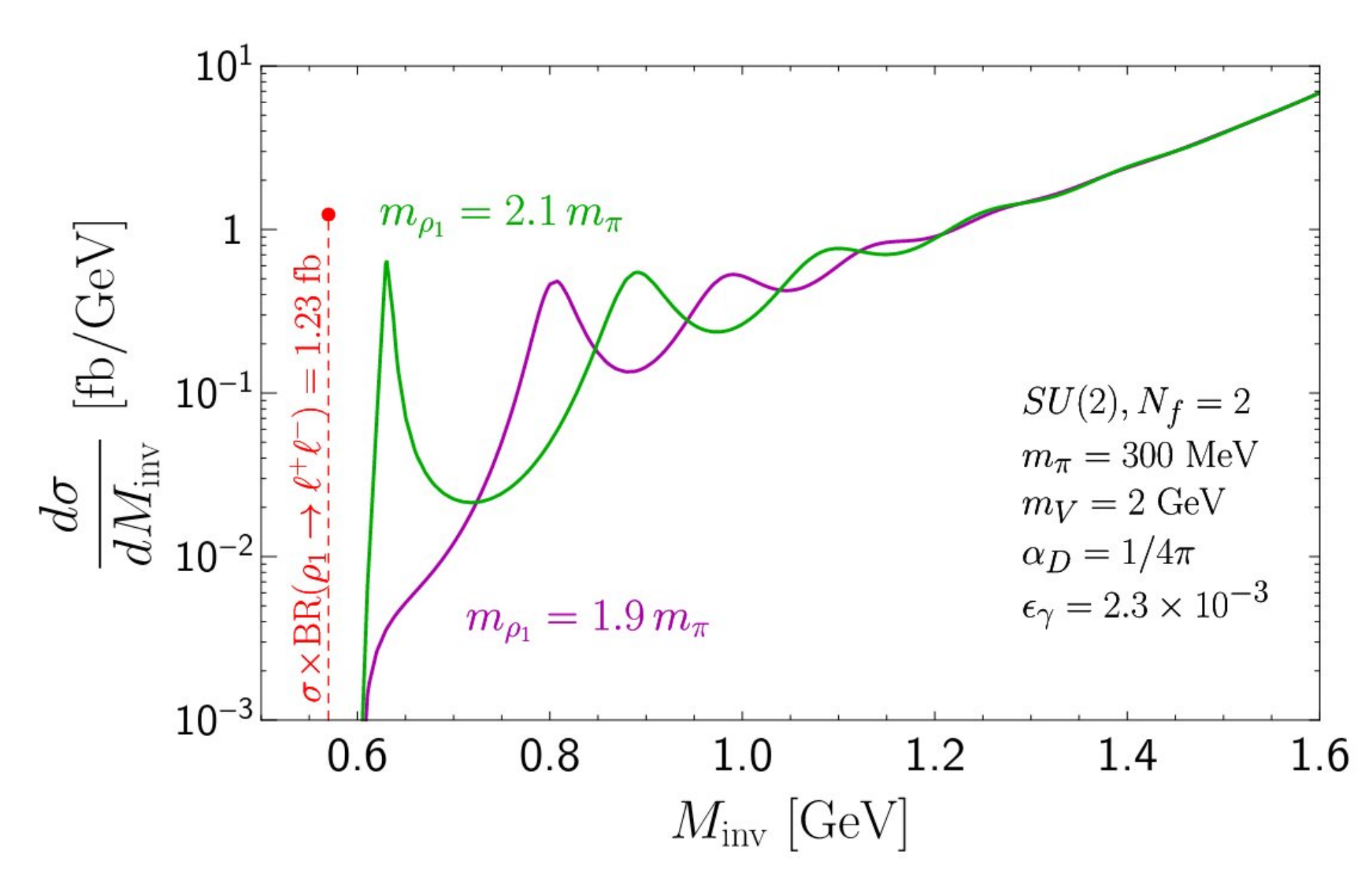}}
\caption{Left: diagram for producing a photon and invisible dark pions
from $e^+e^-$ scattering.  Right: predicted spectrum for invisible
invariant mass,  from Ref.\ 
\cite{Hochberg:2015vrg}.}
\label{fig:simp}
\end{figure}

\subsection{Detection}

\begin{wrapfigure}[15]{r}[0pt]{0.5\textwidth}
\centering
\vskip-1cm
\includegraphics[width=0.5\textwidth]{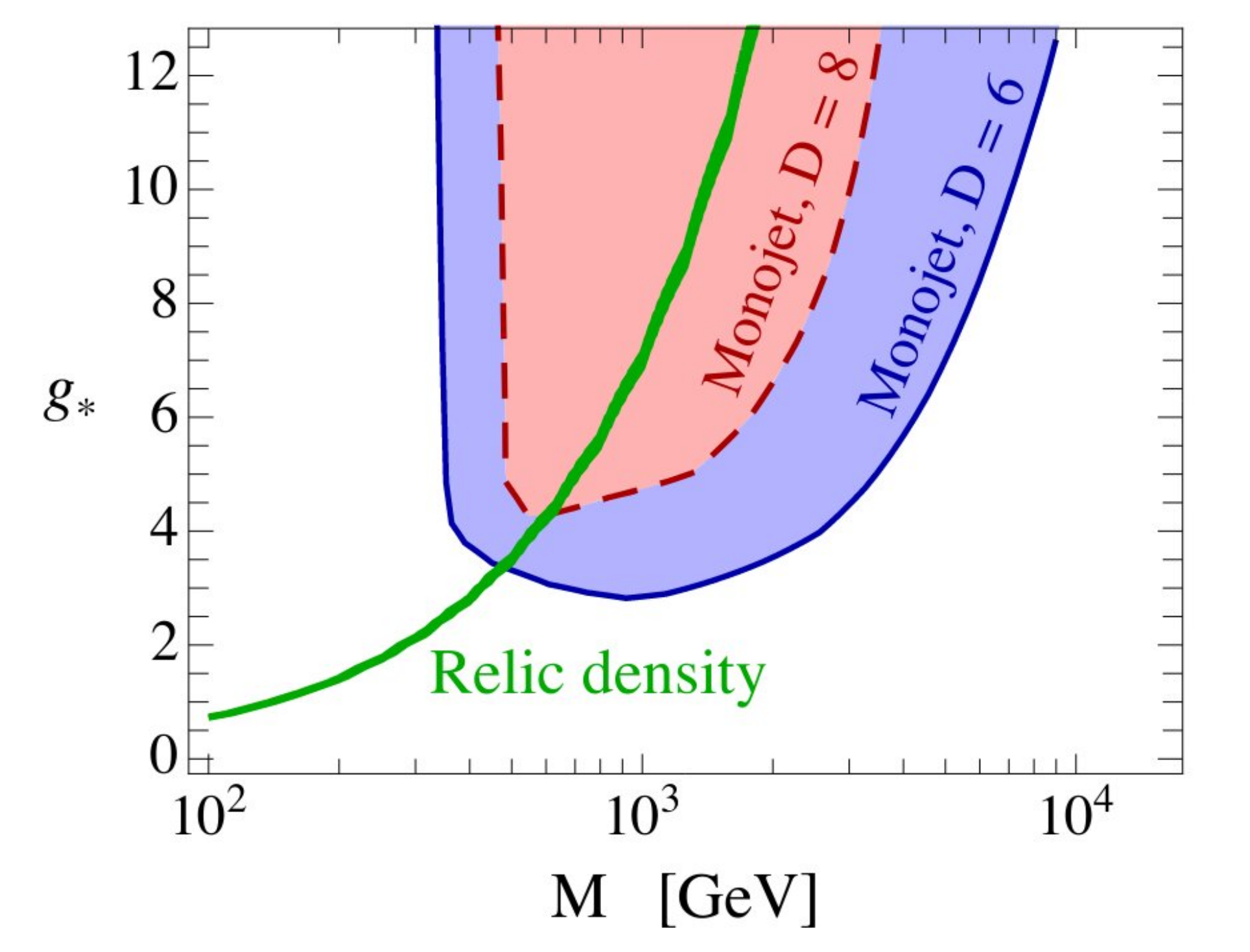}
\vskip-0.5cm
\caption{\label{fig:urb2} from Ref.\
\cite{Bruggisser:2016ixa}; see text.}
\end{wrapfigure}
We have seen in the previous descriptions several examples of direct
and indirect detection of dark mesons, or collider constraints on the
model due to resonant production of the associated dark vectors. 
Because of its scalar nature, the Higgs portal is a common 
interaction for dark mesons,
which leads to scattering on nuclei by Higgs exchange
\cite{Pasechnik:2014ida,Davoudiasl:2017zws}.  Composite Higgs models can also have direct
dimension-6 couplings of the dark meson to SM fermions
\cite{Frigerio:2012uc,Wu:2017iji}.  Light metastable dark mesons 
that can decay to electrons or photons are constrained by the CMB
to have lifetimes $\tau\gtrsim 10^{25}\,$s \cite{Slatyer:2016qyl}.

A distinctive signal of dark mesons, ``SIMP spectroscopy,'' was
suggested in Ref.\ \cite{Hochberg:2015vrg} for $e^+e^-$ collisions.
The Feynman diagram is shown in Fig.\ \ref{fig:simp} (left): it produces a
visible photon and invisible dark pions, through the kinetically mixed 
$Z'$ portal.  The $Z'$ mixes with the dark vector meson $\rho'$ to produce
$\pi'\pi'$.  Through the kinematics, the invariant mass of the invisible 
particles $M_{\rm inv}$ is determined by the beam energy $\sqrt{s}$ and
the observed photon energy: $M^2_{\rm inv} = s - 2E_\gamma\sqrt{s}$.  The spectrum of vector excitations,
expected to go as $m^2_{\rho_n}\sim 4n$ in an AdS-QCD approach
\cite{Karch:2006pv}, can be
observed through the resonances in the differential cross section
$d\sigma/d M_{\rm inv}$, an shown in Fig.\ \ref{fig:simp} (right).

\begin{figure}[t]
 \centerline{\includegraphics[width=\linewidth]{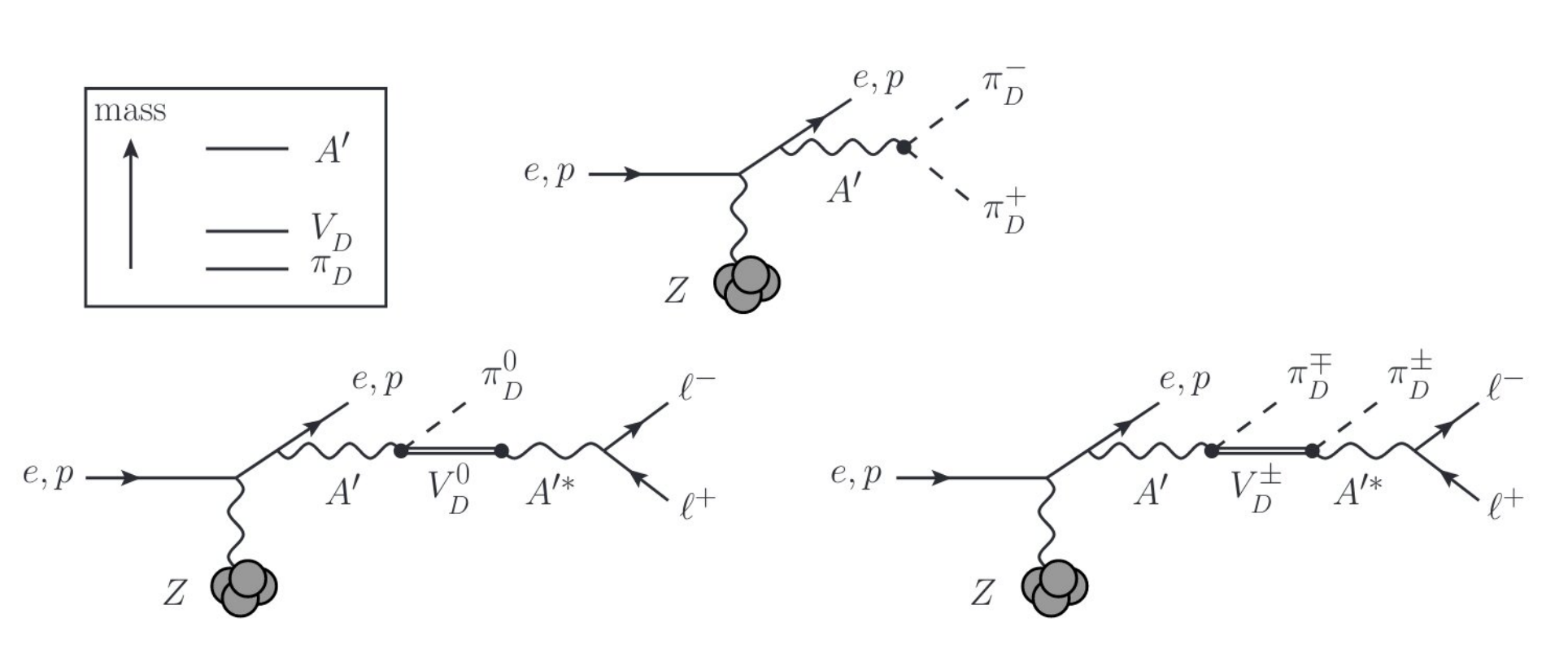}}
\caption{Novel signals from dark mesons coupled to dark photons
for fixed target experiments, from Ref.\ 
\cite{Berlin:2018tvf}.}
\label{fig:toro}
\end{figure}

Ref.\ \cite{Berlin:2018tvf} emphasized the opportunities for fixed-target
experiments to observe similar novel effects connected with production
of dark mesons with interactions to a light (below 10 GeV) $Z'$ with kinetic
mixing $\epsilon$ and mixing with the dark vector excitations.  These processes
are illustrated in Fig.\ \ref{fig:toro}.  Searches for these signals
will be able to probe currently allowed regions of the $\epsilon$-$m_{Z'}$ plane
by planned future experiments.

An interesting example of complementarity between the relic density
requirements and detection at colliders was discussed in Ref.\ 
\cite{Bruggisser:2016ixa}.  Under the assumption that the new
confining dynamics respects approximate SM symmetries, including
custodial, flavor, baryon and lepton number, and can be described
by a single new scale $M$ and coupling $g_*$ \cite{Giudice:2007fh}, the leading
dimension-6 and 8 couplings of mesonic DM $\pi'$ to the SM can be
parametrized up to order 1 coefficients;  for example, operators like 
\be
{g_*^2 \over M^{2}}|\partial_\mu\pi'|^2|H|^2,\
{g_*^2 \over M^{2}}\partial_\mu\pi'^*\partial_\nu\pi'\,B^{\mu\nu}.
\ee
The requirement of a thermal
relic density fixes $g_*$ in terms of $M$ as shown in Fig.\ 
\ref{fig:urb2}, where $m_{\pi'} = 5\,$GeV was assumed.  The shaded regions are excluded by ATLAS searches
for monojets \cite{ATLAS:2015qlt}, putting an upper bound on the
scale of confining dynamics $M\lesssim 500\,$GeV in this example.

Inelastic DM scattering in direct searches requires very small mass splittings
$\lesssim 100$\,keV, that can be naturally achieved in composite
models.  Ref.\ \cite{SpierMoreiraAlves:2010err} used the small
hyperfine splitting between a dark scalar and vector meson
$\pi_d$ and $\rho_d$ in an SU(N)$'\times$U(1)$'$ sector with kinetic
mixing to construct such a scenario.

\section{Dark baryons}
The mass density of the visible universe is dominated by baryons,
so the possibility of dark baryons as DM seems particularly natural.
For SU($N$) or SO($N$) theories, these would be $Q^N$ bound states of the dark
quarks $Q$, whose spin could be $N/2$ (if all $Q$'s are of the same
flavor) or possibly lower (if there are several flavors).  In the
SU($N$) case they are complex, admitting the concept of conserved
dark baryon number, while for SO($N$) they are real, but can
nevertheless still be stable \cite{Antipin:2015xia}.
Ref.\ \cite{Kribs:2016cew} notes an advantage of dark baryons: even if they are
not stable, their decays will be mediated by operators of dimension
$d\ge 6$ if $N\ge 3$.  This makes them more easily long-lived on
cosmological timescales.

In the SM, quark masses are much less than $\Lambda_{QCD}$, making it
difficult to compute the detailed properties of baryons.  One can use
results from lattice gauge theory to infer some of the properties of
dark baryons in the case of SU(3) \cite{DeGrand:2019vbx}. A
computationally  simpler regime is where $M_Q\gg\Lambda'$.  Then the 
quarks are nonrelativistic, and their masses and binding energies can
be calculated using familiar quantum mechanical techniques for
nonrelativistic bound states.  The details of freezeout are different
in these two cases, as we will discuss.

\begin{figure}[t]
 \centerline{\includegraphics[width=0.7\linewidth]{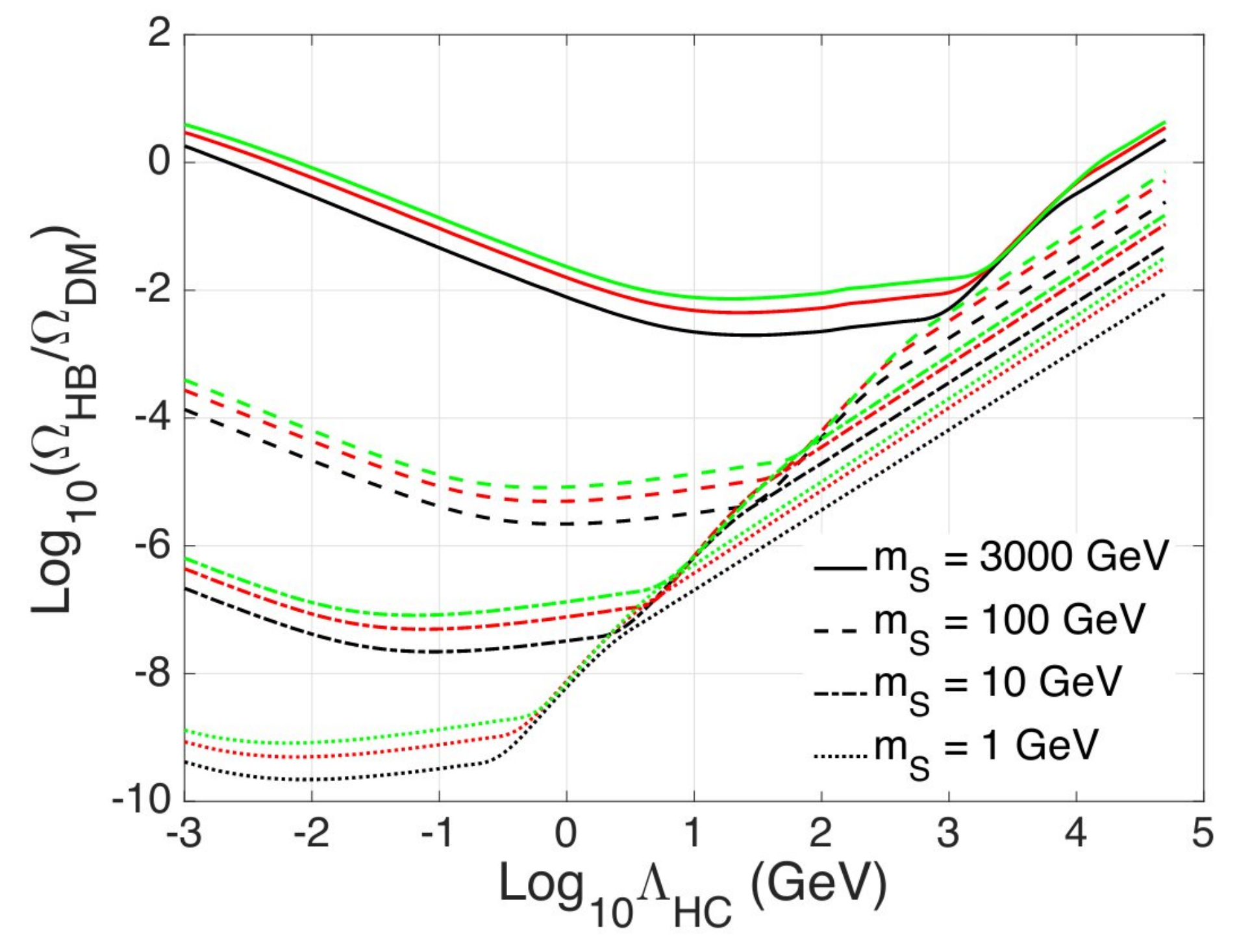}}
\caption{Relic density of dark ``hyperbaryons''
versus the ``hypercolor'' confining scale, for a series of constituent
quark masses $m_S$, and number of colors $N=2,3,4$ (blue, red and
green curves, respectively),  from Ref.\
\cite{Cline:2016nab}.}
\label{fig:weicong}
\end{figure}

\subsection{Relic density}

Like for dark atoms, there is a model-dependent issue as to whether
the dark baryons have an asymmetry or not.  Independently of this
issue, one can address whether their symmetric component can be large
enough to account for all of the DM in a more model-independent way.
In the case $m_Q\ll\Lambda'$ analogous to QCD, one could expect that
the cross section for $p\bar p$ annihilation scales with the baryon
mass as in QCD, $\sigma v \sim 100/m_B^2$.  Matching this to the usual
cross section for thermal freezeout \cite{Steigman:2012nb}, one finds 
that $m_B\sim 200\,$TeV \cite{Antipin:2014qva}, 
close to the unitarity limit \cite{Griest:1989wd,Smirnov:2019ngs}.
In this case $\Lambda'$ is above the freezeout temperature  of the
baryons, $\sim 
m_B/25$, so the details of the confining transition are not important.

\begin{figure}[t]
 \centerline{\includegraphics[width=0.75\linewidth]{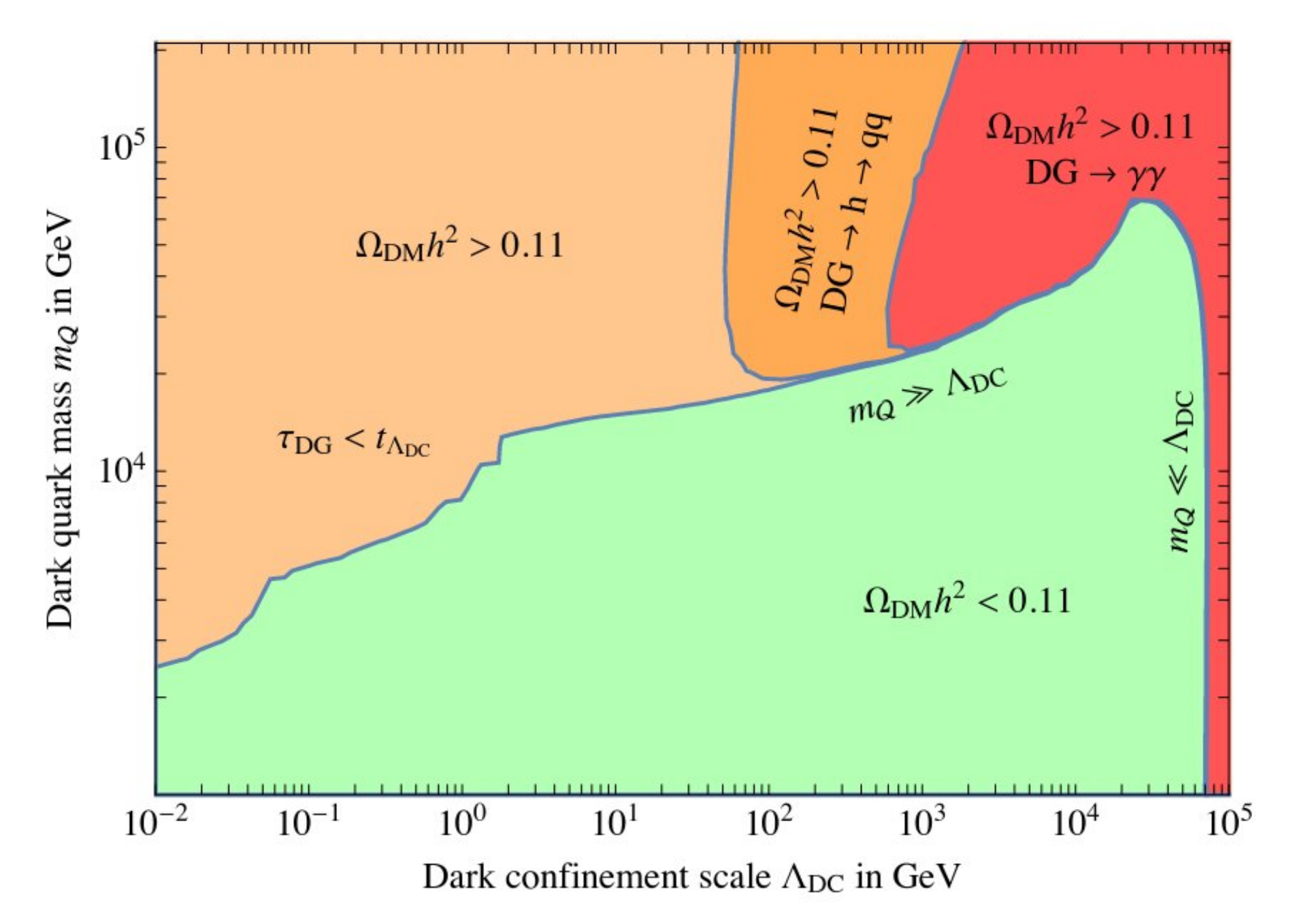}}
\caption{
Regions of dark quark mass $m_Q$ versus confinement scale $\Lambda'$
relevant for the relic density of dark baryons, for gauge group
SU(3), 
from Ref.\
\cite{Mitridate:2017oky}.  Observed abundance
is along the boundary of the green region.  
}
\label{fig:strum}
\end{figure}

In the opposite case of heavy quarks, $m_Q\gg\Lambda'$, annihilation
of $Q\overline Q$ occurs before hadronization, and the details of
hadronization are affected by the residual $Q$ density.  The
qualitative difference between the two scenarios can be seen in Fig.
\ \ref{fig:weicong}
\cite{Cline:2016nab}, which solved the Boltzmann equation in the
general case.  The power-law scaling for $\Lambda'>m_Q$ reflects
the standard relation $\Omega \sim 1/\langle\sigma v\rangle
\sim \Lambda'^2$ for
thermal freezeout: baryons form at an early time, and their final
abundance 
is independent of initial conditions at the confinement
temperature, and only mildly dependent on $m_Q$.   For $\Lambda' < m_Q$, this scaling breaks down
because the initial density of baryons, formed at the confinement
transition, is much higher than their equilibrium abundance at that
temperature, since $m_B\sim N m_Q$, and so 
\be
	n_{Q,\rm eq}\sim e^{-m_Q/T}\gg n_{B,\rm eq} 
	\sim e^{-N m_Q/T}\,.
\ee 
Initially $n_B\sim n_Q/N\gg
n_{B,\rm eq}$ from hadronization at $T'\sim \Lambda'$.  As a result
the relic density has a more complex dependence on $m_Q$ and
$\Lambda'$, sensitive to the dark baryon density at $T'\sim \Lambda'$.
We assumed a geometric cross section for $B$-$\bar B$ annihilation
into dark pions, with a size determined by solving the nonrelativistic
bound state problem with an appropriate potential, and assumed portals
for keeping the two sectors in equilibrium.

These results show for $m_Q\gg\Lambda'$, $m_Q$ above the TeV scale
is favored for getting the observed DM abundance.  This  was further
explored in Ref.\ \cite{Mitridate:2017oky}, which took into
account the effects of the dark glueballs that inevitably also form.
If they are long-lived enough to temporarily matter-dominate the
universe, their decays to DM particles will dilute the $B$ abundance.
The favored values of $m_Q$ versus $\Lambda'$ are shown in Fig.\
\ref{fig:strum} (boundary of green region).

\begin{figure}[t]
 \centerline{\includegraphics[width=0.75\linewidth]{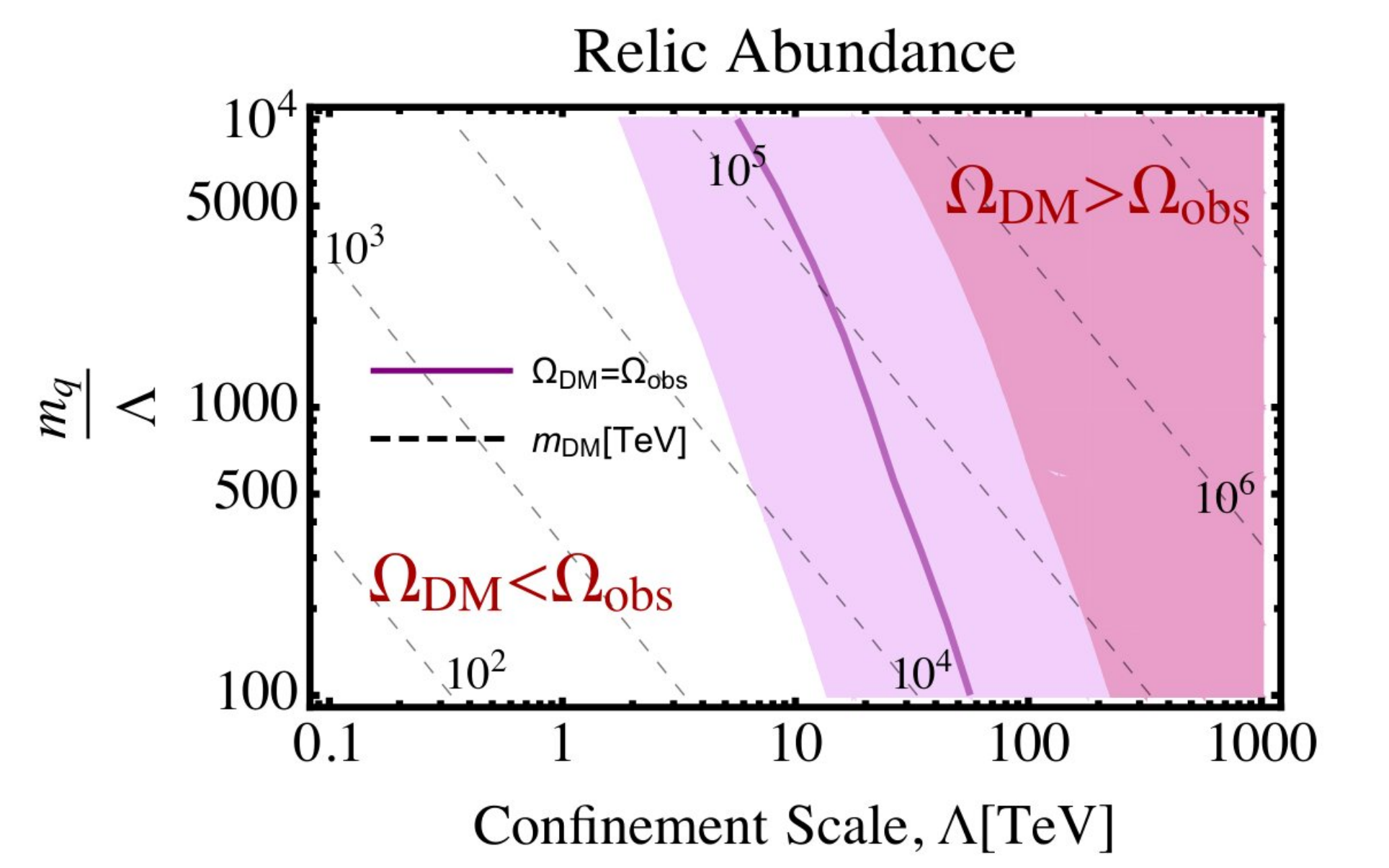}}
\caption{
Regions of  $m_Q/\Lambda'$ versus confinement scale $\Lambda'$
for the relic baryon abundance, accounting for bubble dynamics
from the first order confinement transition, 
from Ref.\ \cite{Asadi:2021yml}.  Dashed lines show contours of
constant $m_B$.  
}
\label{fig:slatyer}
\end{figure}

The two regimes can also be described in terms of weakly coupled
baryons, $m_Q\gg\Lambda'$, in which glueballs provide the thermal bath
of the dark sector, or strongly coupled, $m_Q\ll\Lambda'$, where pions
play that role \cite{Dondi:2019olm}.  The allowed regions for the
relic density can be
expressed in the parameter space of $m_B$ versus $m_{\pi'}$ in the former
case, and in terms of $m_B$ versus the glueball mass in the latter.
The results are sensitive to whether the particles in the bath are
sufficiently long-lived to cause entropy dilution of the baryons by
their decays, and whether they heat up due to $3\to 2$ interactions.
The latter effect can make the dark sector temperature higher than
that of the SM, and the relic $B$ density is enhanced by a factor of
$T'/T$ or $(T'/T)^{3/2}$, depending on which sector is dominant.
This requires larger-than-normal annihilation cross sections for
getting the right $B$ abundance, which in turn enhances the signals
for indirect detection from annihilation in the galaxy, despite the 
large $m_B\gtrsim 10$\,TeV.

However, this is not the end of the story, for the regime where
$\Lambda'\ll m_Q$, since the SU(N)$'$ gauge theory is known to have a
first order confinement transition for $N\ge 3$, and the nucleation of
bubbles can play an important role.  Refs.\
\cite{Asadi:2021pwo,Asadi:2021yml} show that the quarks are kept
outside of the bubbles of confined phase because of the energetic cost
of having a free quark.  After the bubbles percolate, the quarks get
squeezed into small pockets of residual deconfined phase, where they
mostly annihilate away.  But since in each such pocket there is a
statistical $\sqrt{N_q}$ imbalance ($N_q$ being the number of quarks in the
pocket) between quarks and antiquarks, some small asymmetry
is guaranteed to remain, that hadronizes into baryons, which can then
escape to the confined phase.  This leads to a much smaller yield of
dark baryons than in the previous estimates, as shown in Fig.\ 
\ref{fig:slatyer}.  Instead of $m_B\sim 100$\,TeV, values of
$10^4$--$10^5$\,TeV are needed to get the observed abundance. This
suppression is only ameliorated in the $\Lambda'\gtrsim m_Q$ regime,
where the phase transition weakens into a smooth crossover
\cite{Alexandrou:1998wv}.  Although this process is dubbed
``accidentally asymmetric dark matter,'' there is no global asymmetry,
since the sign of the asymmetry from each pocket is random. As for any
such model with conserved dark baryon number, the strong constraints
can be circumvented by introducing a primordial asymmetry.  Another
loophole is the SU(2) case \cite{Francis:2018xjd}, which has a second order transition
\cite{McLerran:1981pb}.

\begin{figure}[t]
 \centerline{\includegraphics[width=0.53\linewidth]{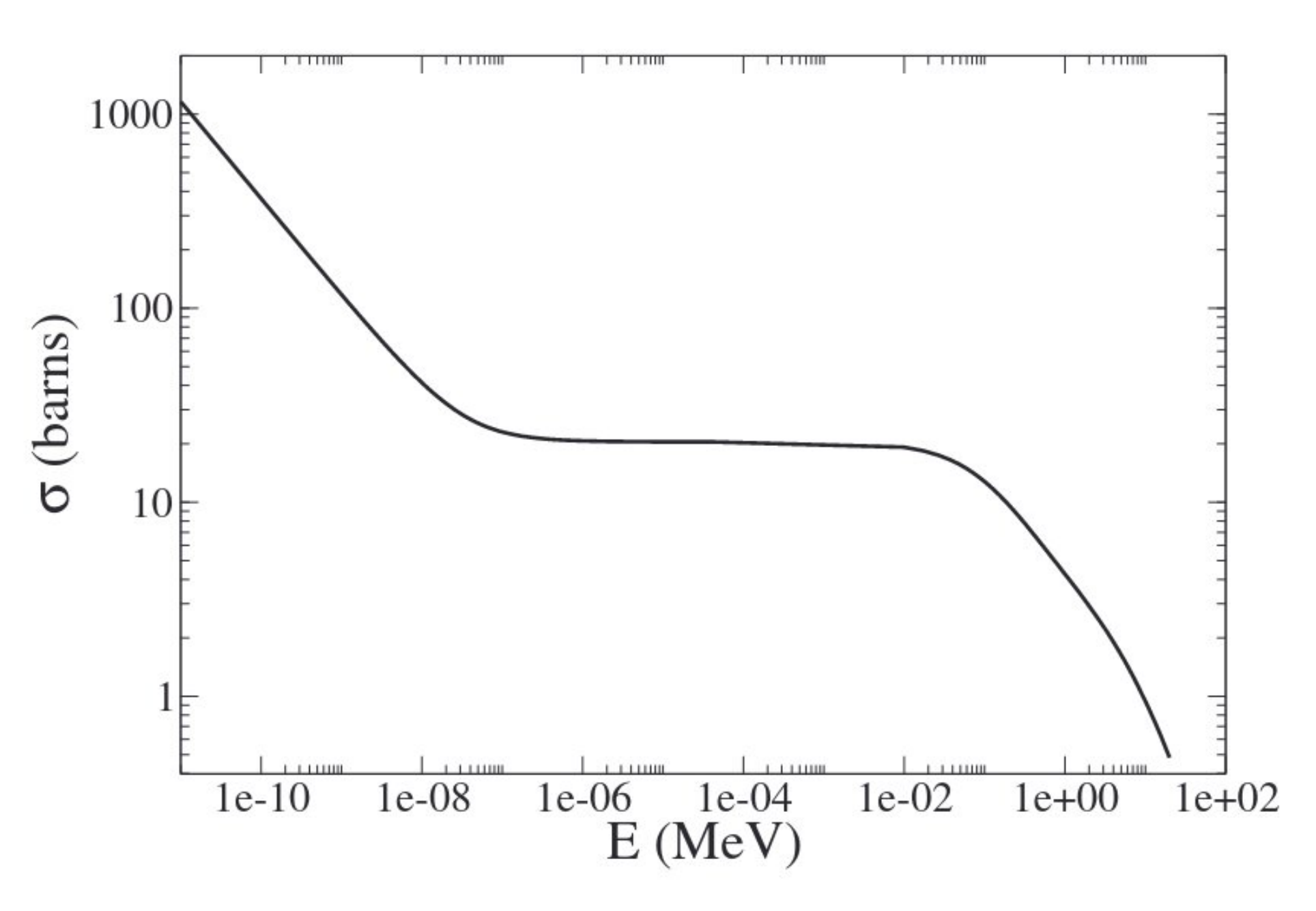}
\includegraphics[width=0.5\linewidth]{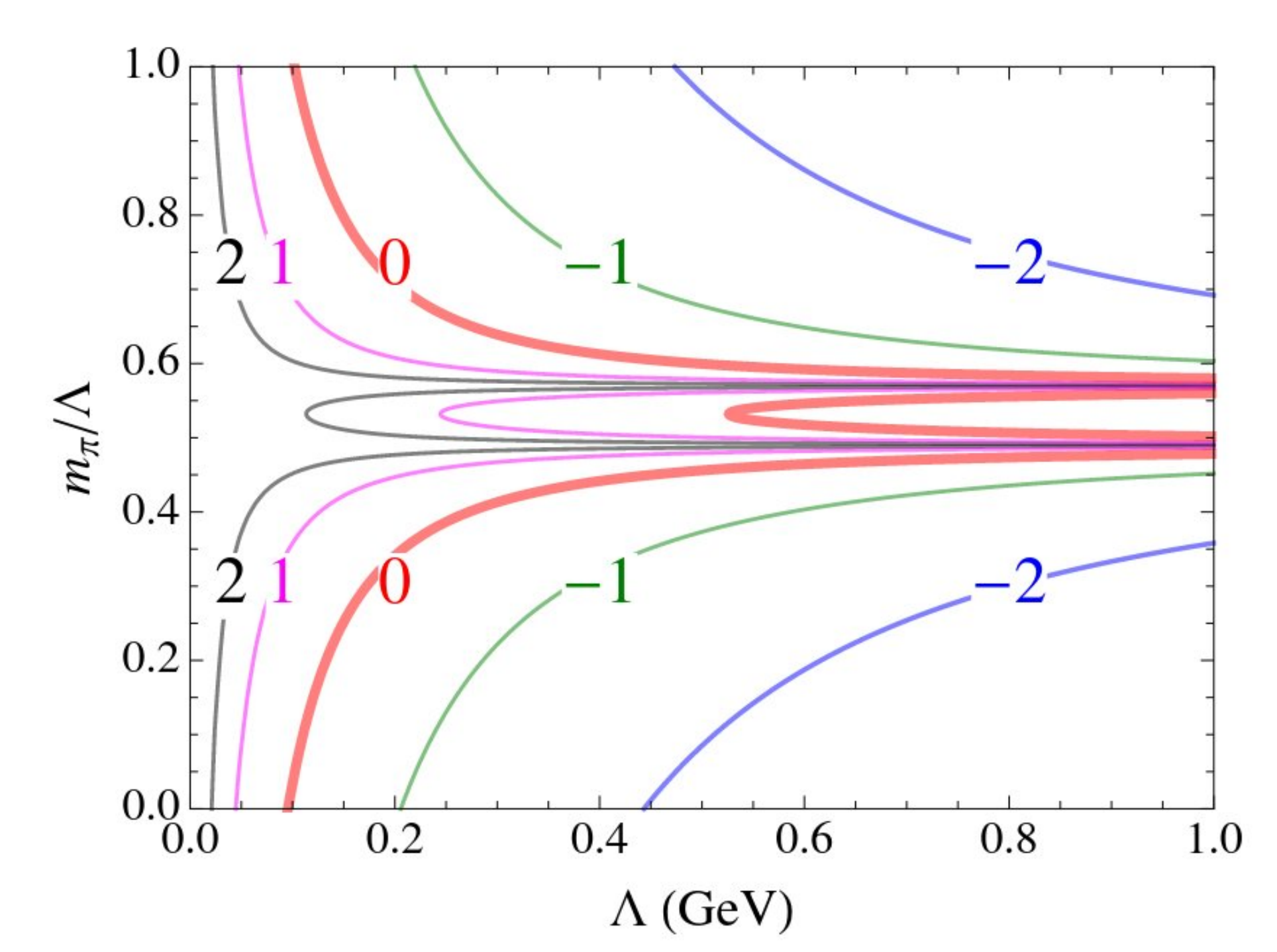}
}
\caption{
Left: actual neutron-proton scattering cross section as a function of 
center of mass energy.  Plateau is the constant cross section of
Eq.\ (\ref{siga}); at lower energies the electromagnetic interaction
dominates.  Right: contours of $\log_{10}[\sigma/m]/[0.6\,{\rm
cm^2/g}]$ in the plane of $m_{\pi}/\Lambda$ versus $\Lambda$ for SU(3)
gauge theory,
from Ref.\ \cite{Cline:2013zca}.
}
\label{fig:npscatt}
\end{figure}

\subsection{Self-interactions}
The elastic scattering cross section for baryons can be estimated
by large-$N$ and NDA to be of order
\be
	{\sigma_{BB}\over m_B} \sim {4\pi\over N\Lambda'^3}
\label{nda-est}
\ee
since $\sigma_{BB}\sim 4\pi/\Lambda'^2$ 
\cite{Witten:1979kh,Manohar:1998xv}
and $m_B\sim
N\Lambda'$.  Witten showed that the amplitude for $B$-$B$
scattering scales as ${\cal M}\sim N$, but the cross section  is
$\sigma_{BB}\sim |{\cal M}|^2/m_B^2$, so the factors of $N$ cancel
out.  Comparing to the actual value of $\sigma_{pn}$ in QCD, the
estimate (\ref{nda-est}) is too small by a factor of 50.  (I choose proton-neutron
scattering here so that the Coulomb interaction that would become
relevant at low energies for $pp$ scattering is not an issue.)  It
turns out that the cross section is resonantly enhanced by the weakly
bound deuteron ($np$) state, whose binding energy is $E_b = 2.2$\,MeV,
and $\sigma \sim 2\pi/(\Lambda E_b)$ gives a better estimate of the
cross section.  The point is that another scale $E_b$ is appearing in
the problem, that cannot be anticipated from order-of-magnitude
arguments.

Ref.\ \cite{Cline:2013zca} noted that one can make quantitative predictions, for 
the case of SU(3), by appropriating results from lattice gauge theory
\cite{Chen:2010yt}.  Lattice gauge theory is computationally expensive
for light quarks, so it is typical for simulations to be done with a
series of decreasing quark masses, for extrapolation to realistically
small values.  This study of the dependence of observables on varying
quark masses can be valuable to the composite model builder.  For the
present case, nucleon scattering amplitudes in the spin singlet and
triplet channels (correlated by Fermi statistics with the isospin
channels) were determined as a function of the pion mass (related to
$m_q$ by $m_{\pi}\sim (\Lambda m_q)^{1/2}$), and expressed as scattering
lengths $a_s$ and $a_t$, defined by the relation
\be
	\sigma = \pi(a_s^2+a_t^2)
\label{siga}
\ee
as $v_{\rm rel}\to 0$.  We fit to the results of \cite{Chen:2010yt} to 
approximate
\be
	a_s\Lambda' \cong {0.58\over m_{\pi'}/\Lambda'-0.57},\quad
	a_s\Lambda' \cong {0.39\over m_{\pi'}/\Lambda'-0.49}\,,
\label{scatlen}
\ee
where the poles indicate the values of $m_{\pi'}/\Lambda'$ at which a
bound state in the $np$ or $nn/pp$ channel is just starting to appear.
Combining Eqs.\ (\ref{siga},\ref{scatlen}) allows one to engineer dark
sectors where the low-velocity baryon self-interactions would match a desired cross
section for small-scale structure problems.  This is illustrated in 
Fig.\ \ref{fig:npscatt} (right), where the thick curves correspond to
a constant cross section of $0.6$\,cm$^2$/g.  This does not take
advantage of the velocity-dependence at high energies to fit galactic
cluster profiles versus smaller halos \cite{Kaplinghat:2015aga,Tulin:2017ara}, which might be worth
investigating.

\begin{figure}[t]
 \centerline{\includegraphics[width=1.2\linewidth]{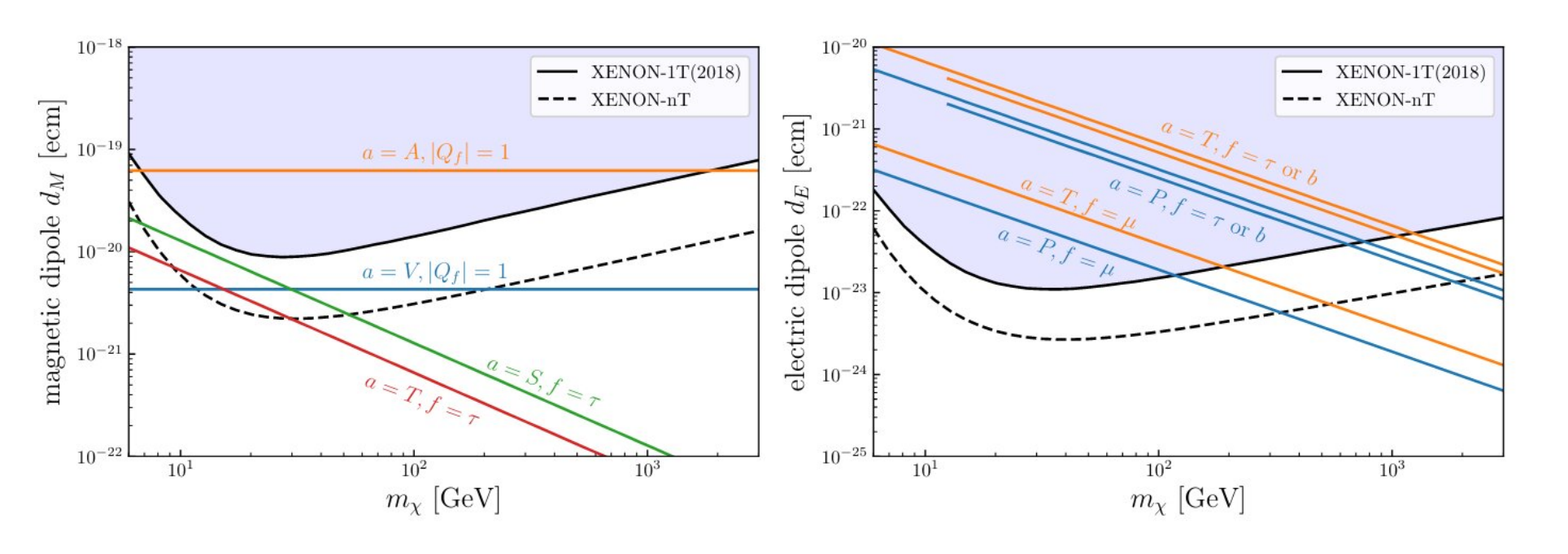}
}
\caption{
Direct detection constraints on dark magnetic dipole moments
(left) and electric dipole moments (right), from Ref.\ \cite{Hambye:2021xvd}.
}
\label{fig:hambye}
\end{figure}

\subsection{Direct detection}

A fully secluded hidden sector is safe from direct detection, but one
often prefers to assume there is a portal to the SM to maintain 
thermal equilibrium, since this facilitates thermal freezeout, and of
course it is more interesting to detect DM than not to detect it. The
dark baryon is typically a bound state of quarks $Q$ that are singlets
under the SM gauge symmetries, although electroweak triplets or
quintuplets (with vanishing hypercharge) are  viable possibilities
\cite{Antipin:2015xia}.  It is also possible to have doublet $Q$
if there is a custodial symmetry that prevents weak neutral current
interactions with the baryons \cite{Appelquist:2015yfa}, dubbed
``stealth'' DM.

Even if $Q$ is a singlet, if it has
interactions with charged particles, {\it e.g.,}
\be
	\lambda \overline Q\Phi\psi\,,
\label{Qpp}
\ee
where $\overline Q\Phi$ is neutral under SU($N$)$'$ and $\Phi\psi$ is
electrically neutral, but $\Phi$ and $\psi$ are electrically charged,
then $Q$ and hence its associated baryon $B=Q^N$ acquire a magnetic moment
at one loop, which is subject to direct constraints.  In more
complicated models, an electric dipole moment could be generated. 

Similarly if $Q$
is an electroweak triplet and $\psi$ is a doublet, both fundamental
under SU($N$)$'$, the interaction
\be
	\lambda \overline Q^i H^\dagger\tau_i\psi\,,
\label{eq:qhp}
\ee
(showing the SU(2)$_{EW}$ index) leads to mass mixing between
$Q$ and $\psi$ when the Higgs gets a VEV.  The mass eigenstate
$Q'$ thus couples to $H$.
Even without the interaction (\ref{eq:qhp}), the neutral $T_3=0$
component of the triplet and
quintuplet models, which is the DM candidate, gets a one-loop coupling
to the nucleons, with the charged components and $W^\pm$ in the loop
\cite{Cirelli:2005uq}.  (The DM coupling to $Z$ vanishes if $Q$ has no
hypercharge.)
Another generic possibility is for $Q$ to be charged under U(1)$'$
that is kinetically mixed with hypercharge.

In the stealth model \cite{Appelquist:2015yfa}, two flavors of vector-like hyperquarks
with even $N\ge 4$ number of colors are introduced, where $Q
=(u,d)_L$, $u_R$, $d_R$ are doublet and singlets respectively under
SU(2)$_{EW}$.  This allows for Higgs couplings to the hyperquarks.
Custodial SU(2)
symmetry forbids neutral weak current interactions of the baryons
$B$, and their even number of constituents forbids magnetic moments,
leaving the Higgs portal as the only means of detection.

\subsubsection{Magnetic and electric dipole moments}
If $N$ is odd, then the $B=Q^N$ baryon can have nonvanishing spin,
which is a necessary requirement for having an electric or magnetic
dipole moment.  Using the quark model, we would estimate that the 
baryon dipole moment is $\mu_B \cong N\mu_Q$ (or less if there are
several flavors of quarks and their spins do not all add).  The one-loop
contribution to the magnetic moment (MDM) $\mu_Q$ from the interaction
(\ref{Qpp}) is of order \cite{Cline:2017aed}
\be
	\mu_Q \sim {e\lambda^2 m_Q\over 64\pi^2 M^2}
\label{eq:muQ}
\ee
where $M$ is the largest mass in the loop.  In a more complicated
theory having several complex couplings with unremovable phases,
the loop diagrams could also give rise to an electric dipole moment
$d_Q$, in analogy to the contribution of a CP-violating $\pi NN$
coupling to the neutron EDM \cite{Crewther:1979pi}.  This occurs in composite Higgs
models (technicolor) \cite{Antipin:2015xia}.  The EDMs are more
strongly constrained than the MDMs because their cross section for
scattering on
nucleons is enhanced by a factor of $1/v_{\rm rel}^2$ compared
to that of MDMs \cite{Barger:2010gv}.  Recent constraints on
dark MDMs and EDMs from direct searches are shown in Fig.\ 
\ref{fig:hambye}.
For example taking SU(3)$'$ and $m_B=2$\,TeV, Eq.\ (\ref{eq:muQ}) 
implies $\lambda/M \lesssim 2/$TeV.

\begin{figure}[t]
 \centerline{\includegraphics[width=0.6\linewidth]{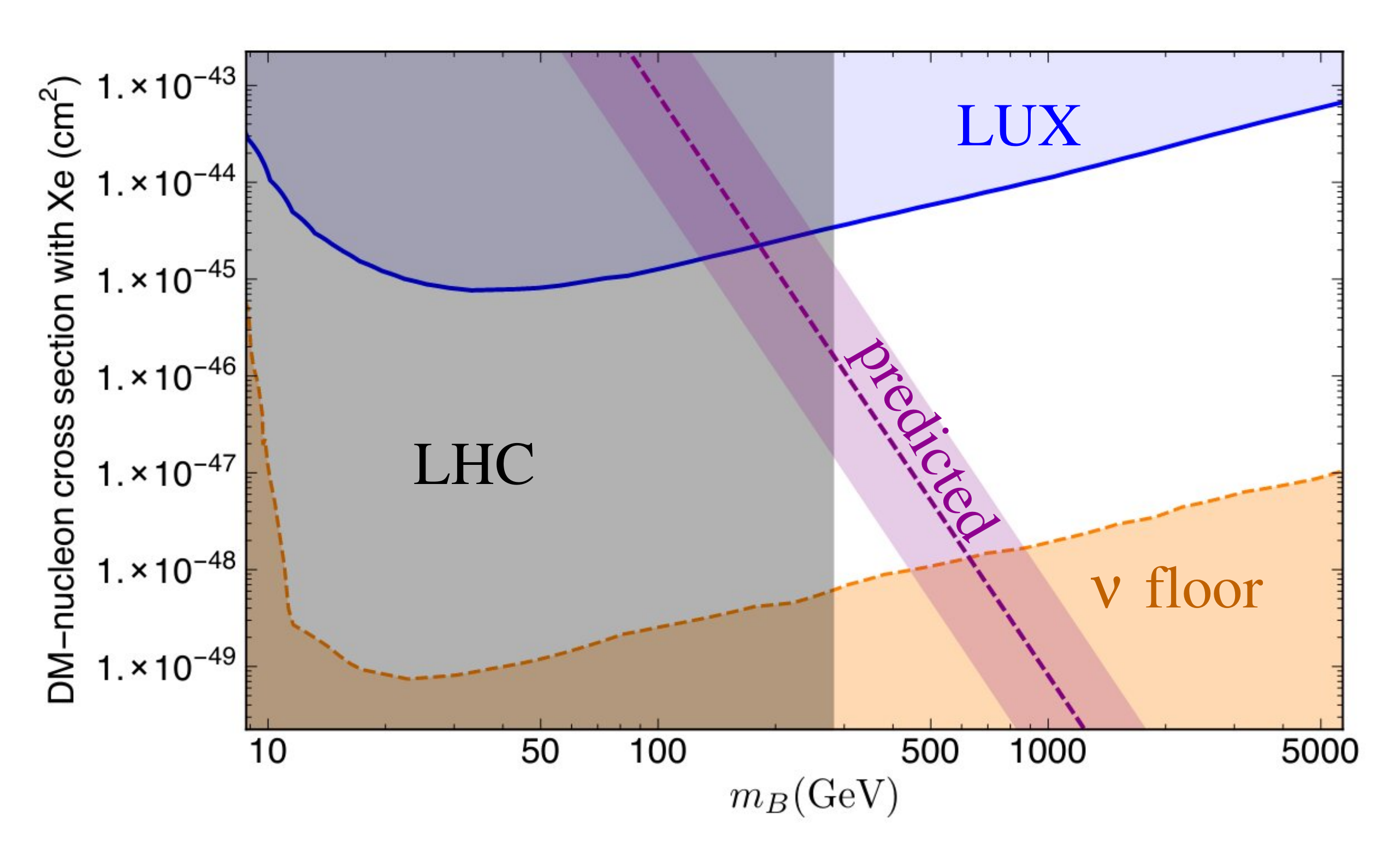}
\raisebox{0.5cm}{\includegraphics[width=0.4\linewidth]{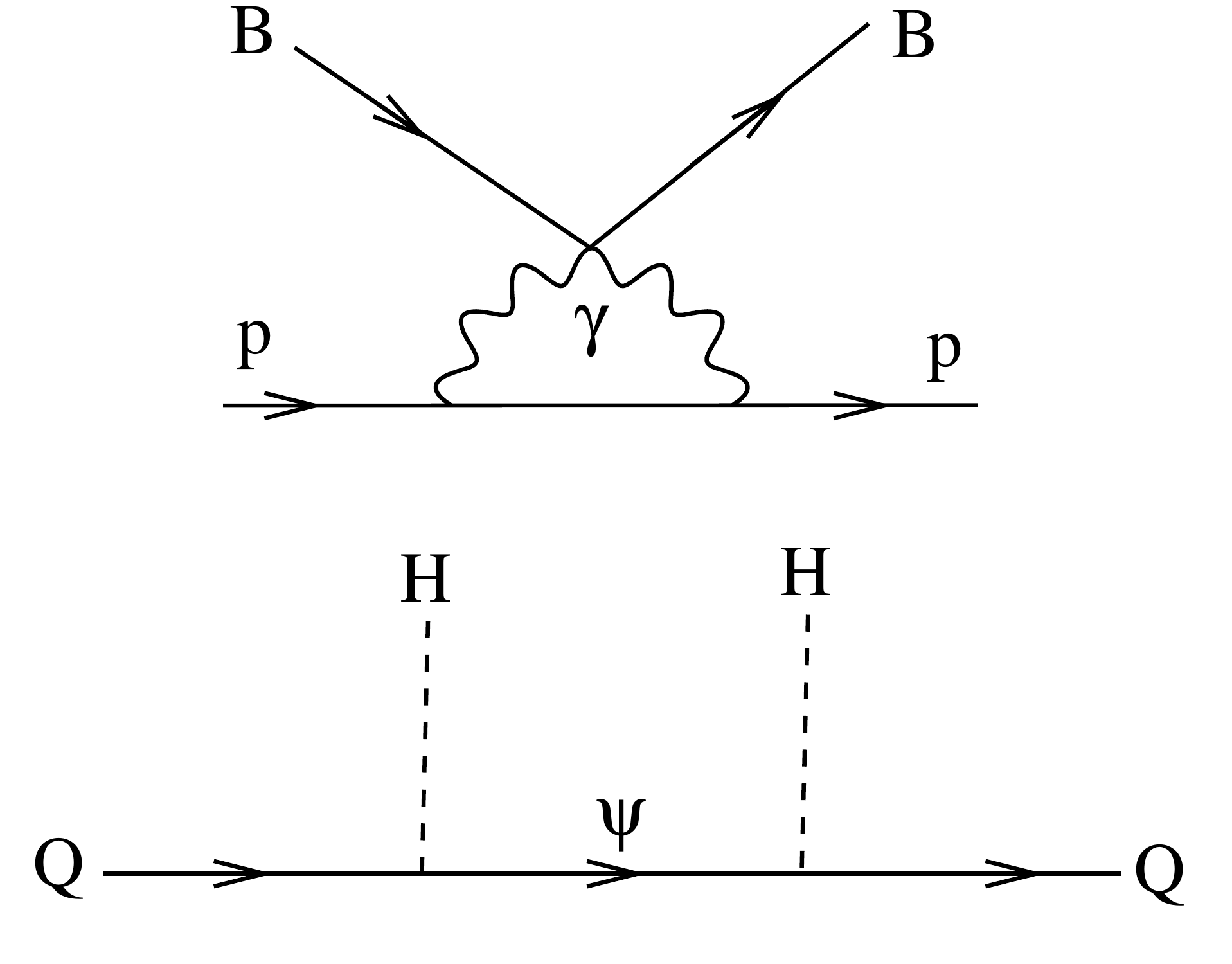}}
}
\caption{Left: constraints on stealth DM (spin 0 baryon) 
scattering on protons induced by electric polarizability, adapted
from Ref.\ 
\cite{{Appelquist:2015zfa}}.
Upper right: dark baryon-proton scattering induced by electric 
polarizability operator of Eq.\ (\ref{eq:polariz}).
Lower right: diagram leading to effective Higgs
portal coupling to a hyperquark $Q$ that is  a triplet under
SU(2)$_{EW}$.  }
\label{fig:hportal}
\end{figure}

In models with even $N$, even though there are no dipole moments,
there can exist higher dimension couplings to
photons---polarizability, which for a scalar baryon $B$ takes the 
form
\be
	C_F\,|B|^2 j_\mu\, F^{\mu\alpha}F_{\alpha\nu}\, j^\nu 
\label{eq:polariz}
\ee 
in an external current $j_\mu$. Interactions of $B$ with protons
then occur at one loop (Fig.\ \ref{fig:hportal} (upper right)).  
Lattice predictions for the
polarizability $C_F$ for SU(4) and ensuing constraints from 
direct searches were carried out in Ref.\ \cite{Appelquist:2015zfa};
see Fig.\ \ref{fig:hportal}.
.

\subsubsection{Higgs portal}

For the SU(2)$_{EW}$ triplet $Q^i$ model, the interaction 
(\ref{eq:qhp}) in the diagram Fig.\ \ref{fig:hportal} (lower right)
leads to the effective Lagrangian 
\bea
	{\cal L} &=&
	{\lambda^2\over m_\psi}\,\overline
	 Q^i\,H^\dagger(\delta_{ij}+i\epsilon_{ijk}\tau_k)\, H\, Q_j
	\nn\\
	&\to& {\lambda^2v\over m_\psi} h \overline Q Q
\quad \equiv\quad y_{\rm eff} h \overline Q Q
\label{yeffeq}
\eea
where $v=246\,$GeV is the Higgs VEV and we assumed
$m_\psi\gg m_Q$ (appropriately, since we want $Q$ to be the dark
matter). To determine the $B$-nucleon scattering cross
section, we need the matrix element
\be
	\langle B|\overline
	 Q^i Q_i | B\rangle 
	\equiv f_B\,,
\ee
where the form factor $f_B$ is a number or order 1 (or perhaps
$N$).  The cross section for scattering on nucleons from Higgs
exchange is
\be
	\sigma_{BN} = {(y_{\rm eff}f_B f_N)^2 m_N^4\over 2\pi v^2
        m_h^4}
\ee
where $f_N\cong 0.3$ for the Higgs-nucleon form factor.  The XENON1T
\cite{XENON:2018voc} (PandaX-4T \cite{PandaX-4T:2021bab}) limits
are
\be
\sigma \lesssim 10^{-48\, (-48.4)}\,\left(m_B\over{\rm GeV}\right){\rm\,cm}^2
\label{eq:xenon}
\ee
for DM mass $m_B$  in the high-mass region, giving
$y_{\rm eff} < 0.06(m_B/{\rm TeV})^{1/2}$.  This gives
$\lambda\lesssim 0.1$ for
$m_B\sim $1\,TeV, $m_\psi \sim 500$\,GeV, for example.

\subsubsection{Kinetic mixing portal}
If the quarks couple to a massive, kinetically mixed $Z'$ with charge
$g'$, the baryon has charge $N g'$, and conservation of the vector
current implies that $Z'$ couples to the current $Ng'\bar B\gamma^\mu B$.
We can use the coupling
(\ref{epscoup}) to protons to compute the $p$-$B$ cross section, assuming 
that $m_A'$ is much greater than the momentum transfer, as
\be
	\sigma_{pB} = {(\mu_{pB} N g' \epsilon e)^2\over \pi\,
m_{A'}^4}\,,
\ee
where $\mu_{pB} = m_p m_B/(m_p + m_B)\cong m_p$ is the reduced mass.
Using the experimental limit (\ref{eq:xenon}) gives
\be
	N g' \epsilon \lesssim 10^{-8}\left(m_A'\over{\rm
GeV}\right)^2\left(m_B\over{\rm TeV}\right)^{1/2}\,.
\ee

\subsection{Masses and wave functions}

Frequently one would like to relate the dark baryon mass and size
to the fundamental parameters of the model.  For relativistic bound
states this would be most reliably done using lattice gauge theory.
For nonrelativistic or mildly relativistic systems, one can use 
quantum mechanics with a model for the two-body potential between
constituents \cite{SpierMoreiraAlves:2010err,
Cline:2016nab,Mitridate:2017oky}.  

If $m_Q\gg\Lambda'$, the consituents are highly nonrelativistic, and
the quark-quark force is dominated by the short-distance Coulomb
contribution \cite{Raby:1979my}
\be
	V_C = -{\alpha'\over 2r}\left(N-{1\over N}\right)\,.
\ee
For faster and hence less deeply bound quarks, the linear confining
part of the potential can become significant,
\be
	V_L = \sigma r \cong 2 (N -1)\Lambda'^2 r\,,
\ee
where the string tension $\sigma$ was estimated by Ref.\ 
\cite{Cline:2016nab} using large-$N$ scaling together with lattice
determinations for $N=3$.

To approximately solve the Schr\"odinger equation for the full potential,
\be
	V = N m_Q + \sum_{i<j}^N 
\left[V_C(r_{ij}) + V_L(r_{ij})\right]\,,
\ee
one can make an ansatz for the ground-state wave function
\be
	\psi \propto \exp\left(-\mu\sum_i^N r_i\right)
\ee
(considering $r=0$ to be the centroid of the baryon)
and use the variational method: compute the total energy 
$E = \sum_i p_i^2/2m_Q + V$ as a function
of $\mu$ and minimize it to find the mass and size $\mu^{-1}$ of the bound
state.  This method can be extended to relativistic systems by using
the kinetic energy $T = \sum_i \sqrt{p_i^2 + m_Q^2}$ \cite{Rai:2002hi}
and working in the momentum basis to evaluate its expectation value.

\section{Conclusion}
Although Occam's razor seemingly makes composite dark matter not the
theorist's 
first choice, nature may well think differently.  The fact that we are
made from composite (visible) matter gives credence to the possibility
of a rich dark sector including gauge interactions.  Whether abelian
or nonabelian, this can give rise to bound states forming the dark
matter.  

In terms of experimental motivation, the hints of strong DM
self-interactions for solving the small-scale structure formation 
problems of standard CDM are perhaps the most persuasive indication
that DM could be composite.  It is intriguing that the target cross
section of $\sigma/m \sim 1$b/GeV is of a similar order of magnitude
to that for nucleons.  One day experiments may provide definitive
evidence that will allow us to narrow the currently vast scope of our 
speculations.

\section*{Acknowledgements}
I thank S.\ Caron-Huot,  M.\ Fairbairn, E.\ Hardy, H.-M.\ Lee,
G.\ Moore,  
D.\ Morrissey, E.\ Neil, J.-S.\ Roux, F.\ Sannino, K.\ Schutz, T.\ Slatyer, N.\ Toro
 and A.\ Urbano for helpful correspondence, and S.\ Heeba for
correcting an error.  I thank M.\ Cirelli
and I.\ Masse
for their encouragement to attend Les Houches in person in 2021.
Thanks also to N.\ Selimovic for proofreading and helpful suggestions
to improve the clarity.  I thank the referees for numerous constructive
suggestions.

\paragraph{Funding information}
This work was supported by the Natural Sciences and Engineering Research Council (NSERC) of Canada.

\bibliography{DA.bib}

\newpage
\appendix
\section{Road bike rides from Les Houches}
Mountain bikers have several options near Les Houches, but road bikers
have just two: up the valley or down the valley.  In either case,
be prepared for significant climbing.  In the direction of Vallorcine,
Col des Montets is the high point and makes for a pleasant climb,
especially coming back from the other side, which is very scenic.
Chamonix now has a nice bike path crossing most of the town, that
allows one to get off the main road (D243) and away from the traffic.
After crossing the Arve on D243, take the first right to cross it
again and join the bike path.  After some kilometers it ends abruptly; turn left to join
D1506 toward Argenti\`ere.
From the Col, one can continue to 
Switzerland if desired.  Vallorcine is a charming destination, just a
few kilometers from the border.

To go down-valley, ignore Google's suggestion to take All.\ des
Diligences (suitable for hikers and mountain bikes only); instead 
take Route de Vaudagne to Route de la Plaine Saint-Jean (D13).
Continue to Passy and turn right at D43 
(or turn a bit sooner at Chem.\ de Perrey toward Maffray and take
Route de Maffrey to join D43) for the climb to Plaine Joux,
a popular cycling challenge, where the maximum grade 
is marked at each kilometer, going up to 8\%.  Save some energy for
the ride back, which entails about the same gain in elevation
and similar grades.

Not highly recommended: D143 toward Lac Vert (ends at parking, far
from the lake, steep grades); and especially not Parc de Merlet:
very crowded, very steep, no rewarding views.

\bigskip
 \centerline{\includegraphics[width=0.95\linewidth]{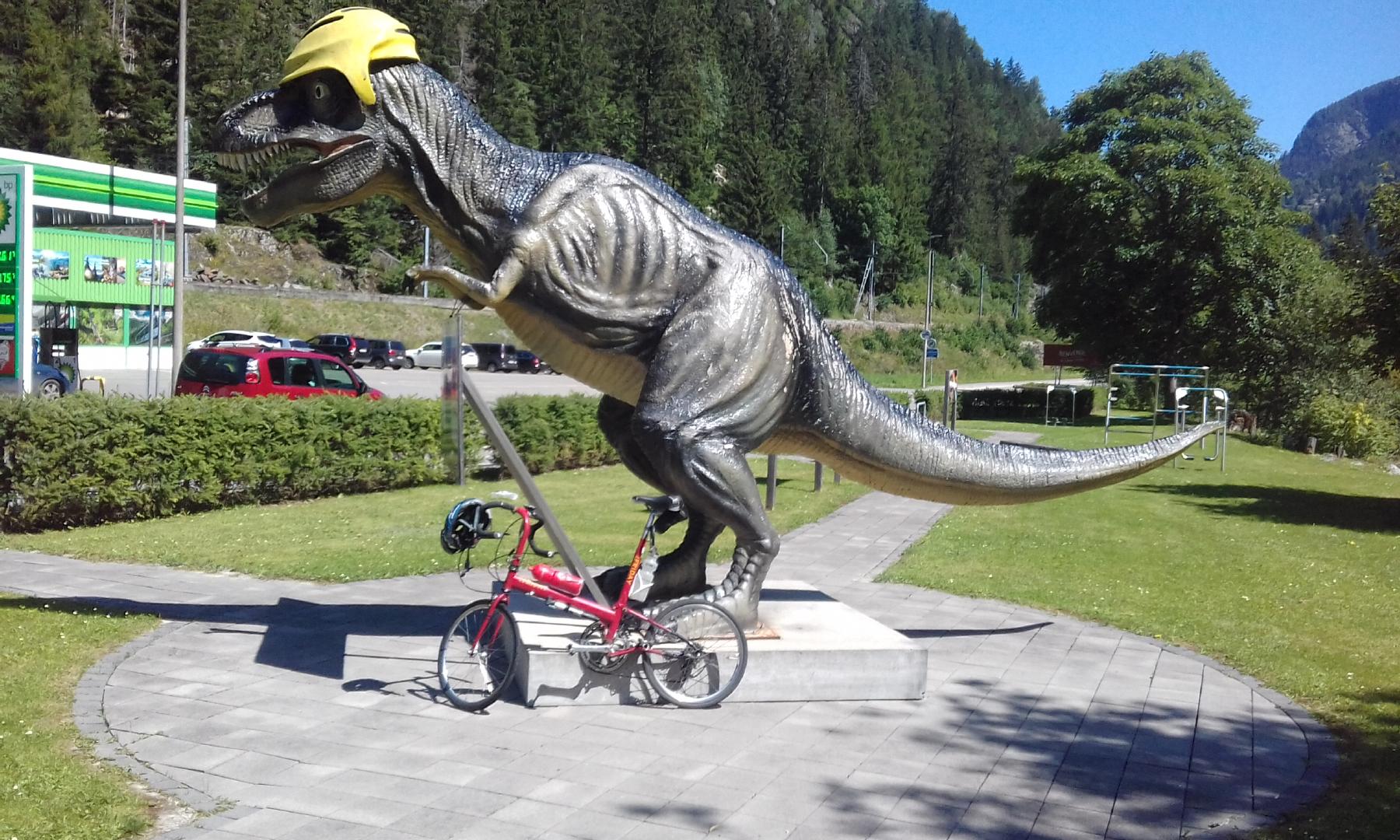}}

\nolinenumbers

\end{document}